\documentclass[11pt, a4paper, aastex]{report}
\usepackage{fancyhdr}
\usepackage{epsfig}
\usepackage[update,prepend]{epstopdf}
\usepackage{amssymb}
\usepackage{natbib}
\usepackage{setspace}
\usepackage{ifthen}
\usepackage{deluxetable}

\newboolean{PAPERS}
\setboolean{PAPERS}{true}

\def\aj{AJ}%
\def\actaa{Acta Astron.}%
\def\apj{ApJ}%
\def\apjl{ApJ}%
\def\apjs{ApJS}%
\def\apss{Ap\&SS}%
\def\aap{A\&A}%
\def\mnras{MNRAS}%
\def\nat{Nature}%
\def\procspie{Proc.~SPIE}%

\begin{document}


\pagestyle{empty}

\begin{figure*}[t]
\vskip -2.5cm
\hskip 3mm
\centerline{\includegraphics{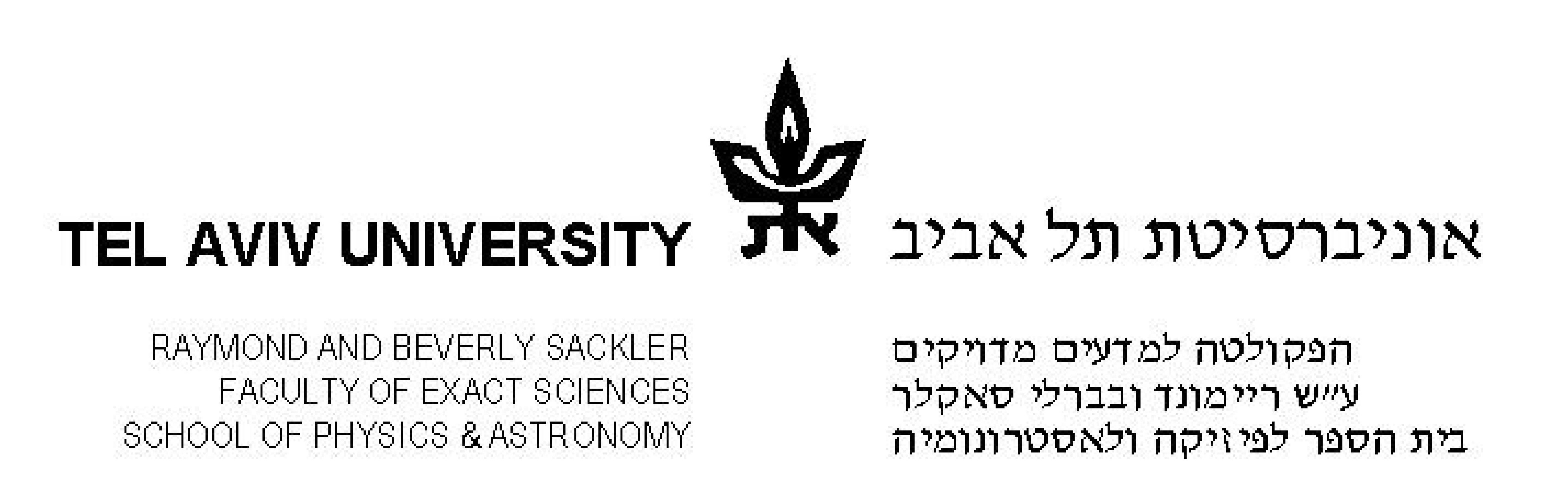}}
\end{figure*}
\vspace*{-1.9cm}
\hskip -0.8cm
\begin{picture}(-20,0)
\thicklines
\put(-50,0){\line(1,0){505}}
\end{picture}
\vglue 1.5cm
\LARGE
\centerline{\bf BEER analysis of {\it Kepler} and {\it CoRoT} light curves:}
\vskip 0.5cm
\centerline{\bf discovering binaries and exoplanets }
\vskip 1.8cm
\Large
\centerline{Thesis submitted for the degree ``doctor of philosophy''}
\vskip 0.4cm
\centerline{by}
\vskip 0.5cm
\LARGE
\centerline{\bf Simchon Faigler}
\vskip 1.8cm
\Large
\centerline{Submitted to the senate of Tel Aviv University}
\vskip 0.4cm
\centerline{February 2016}
\Large
\vskip 1.8cm
\centerline{This work was carried out under the supervision of}
\vskip 0.4cm
\centerline{Professor Tsevi Mazeh}
\vskip 0.4cm
\clearpage

\newpage
\thispagestyle{empty}
\
\newpage

\vglue 0.5cm
\large
\leftline{\bf{ \em To my family}}
\vskip 0.2cm
\cleardoublepage
\clearpage

\newpage
\newpage
\thispagestyle{empty}
\
\newpage

\pagenumbering{Roman}
\pagestyle{plain}

\tableofcontents
\newpage
\thispagestyle{empty}
\
\newpage

\doublespacing

\section*{Abstract}
\addcontentsline{toc}{section}{Abstract}

This thesis consists of seven scientific papers that cover the proof-of-concept, the development, and discoveries made through the use of the BEER (BEaming, Ellipsoidal, and Reflection) algorithm for searching for companions in the light curves from the {\it Kepler} and {\it CoRoT} space telescopes. 

Paper I presents the detection of the ellipsoidal and the beaming effects 
in the {\it CoRoT} light curve of CoRoT-3, a system of a $22$$M_{\rm Jup}$ brown dwarf orbiting an F star with an orbital period of $4.3$ days.
This work served as a proof-of-concept that these effects are detectable in the space light curves of systems with brown-dwarf or planetary secondaries, thus indicating that similar modulations may be detected in the light curves of non-transiting systems.
As a follow-up to the first paper, Paper II presents the BEER algorithm for detecting {\it non-transiting} short-period low-mass companions, through the BEaming, Ellipsoidal and Reflection effects, in {\it Kepler} and {\it CoRoT} light curves. 
The paper also analyzes the expected performance of the algorithm and predicts that it should enable detection of secondaries with masses down to $5$--$10$$M_{\rm Jup}$.

To establish the effectiveness of the BEER method, Paper III, and later Paper VI, present the discovery and radial velocity (RV) confirmation of seven, and seventy, non-eclipsing short-period binaries, in {\it Kepler} and {\it CoRoT} data, respectively.
The papers illustrate that unlike eclipses searches, the BEER algorithm searches for {\it non-eclipsing} companions, and therefore can detect additional systems with lower inclination angles. 

As envisioned by the previous papers, Paper IV presents the discovery of Kepler-76b, a 2$M_{\rm Jup}$ transiting hot Jupiter orbiting an F star with orbital period of $1.54$ days. This planet was first identified by the BEER algorithm, and then confirmed by RV follow-up observations.
The Kepler-76 light curve also showed the first evidence in the {\it Kepler} band for superrotation. This phenomenon involves a phase shift of the planetary thermal-emission modulation, due to equatorial superrotating winds in the planet atmosphere, and was previously observed only in the infrared.
Paper V extends the discovery of superrotation in Kepler-76b to two additional known hot Jupiters, HAT-P-7b and KOI-13b. 
These discoveries illustrate that detailed phase-curve studies, of precise space-surveys light curves, allow the investigation of atmospheric phenomena, such as thermal winds or reflective clouds, in multiple close-in exoplanets.

The last study, Paper VII, demonstrates the different strengths and utility of the BEER search algorithm. 
It presents the discovery of four short-period eclipsing binaries in the {\it Kepler} light curves, consisting of an A-star primary and a low-mass WD secondary (dA+WD).
The systems show BEER phase modulations together with primary and secondary eclipses.
These add to the 6 {\it Kepler}, and 18 WASP, previously known short-period eclipsing dA+WD binaries.
The paper shows that three of the new systems harbor the smallest WDs detected so far in such binaries. These  three binaries extend the previously known population to older systems with cooler and smaller WD secondaries, allowing to test binary evolution theories in a parameter region not observed before.

The seven papers illustrate the effectiveness of the BEER algorithm in finding both common stellar binaries and rare astrophysical objects.
 As such, the BEER tool can serve as an important component in the virtual astronomy toolbox for mining the vast astronomical data produced by current and future photometric surveys. 

\newpage

\chapter{Introduction}
\label{intro}
\pagestyle{plain}
\pagenumbering{arabic}

The {\it Kepler} and {\it CoRoT} space telescopes were launched with the primary mission of detecting extrasolar transiting planets, through the minute periodic transit dips they induce in the photometric light curves of their host stars \citep{borucki10,koch10,rouan98, baglin06, auvergne09}. As such, these missions produced hundreds of thousands of nearly uninterrupted, high precision light curves, with time spans of tens to more than a thousand days, and relative photometric precision of up to $10^{-3}$--$10^{-4}$ per data point, depending on the stellar brightness. This exquisite photometry revolutionized exoplanet research, by enabling the detection of Earth size planets (and white dwarfs) with transit depths of $\sim$$0.01\%$ \citep[e.g.,][]{leger09,batalha11}, a two orders of magnitudes improvement relative to the $\sim$$1\%$ typical sensitivity of ground-based transit surveys \citep[e.g.,][]{butters10,udalski15}.

As of January 2016, analysis of the {\it CoRoT} light curves yielded the discovery of 30 exoplanets
 by the transit method, that were later confirmed by spectroscopic radial-velocity (RV) observations \citep{exoplanetseu}. 
More importantly, the analysis of the {\it Kepler} data has produced more than 4600 planetary candidates, of which more than 1000 have been verified as planets by various methods \citep{batalha13,mullally15,kepexoplanets}.  As a byproduct of this effort, the {\it Kepler} mission has also identified more than 2800 eclipsing binary (EB) systems\footnote{http://keplerebs.villanova.edu/} \citep{slawson11}.

  For EB or transiting planets systems, the orbital period, inclination and the radii of the primary star and the companion, relative to the semi-major axis,  are directly measurable through analysis of the eclipses/transit shape \citep[e.g.,][]{seager03}. Yet, there are additional astrophysical effects that produce flux variations along the orbital phase of a binary system, which depend on, and thus probe, the mass of the companion, be it a stellar, brown dwarf, planet, or a compact object. Such out-of-eclipses phase modulations that are seen also in non-eclipsing systems, are the result of three main effects: reflection/emission, ellipsoidal, and beaming. 
  
The reflection/emission effect (referred to here as the reflection modulation) is a result of each component's light scattered off the facing hemisphere of its companion (``day side''), combined with light absorbed and later thermally re-emitted by the companion atmosphere, at different wavelengths \citep{vaz85,wilson90,maxted02,harrison03,for10,reed10}. As such, this effect causes a modulation at the orbital period, and probes the companion radius relative to the semi-major axis, together with properties associated with the companion atmospheric response to its host-star radiation, such as the Bond albedo, scattered-light geometric albedo, and heat-redistribution parameters, among others. 

The ellipsoidal modulation \citep{kopal59,morris85} is a well-known and well-studied effect in close binaries, that is due to the tidal distortion of the primary star by the gravity of the secondary \citep[e.g.,][]{loeb03,zucker07,mazeh08}, resulting in a phase modulation at half the orbital period. 

The beaming effect, sometimes called Doppler beaming or Doppler boosting, causes an increase (decrease) of the brightness of any light source approaching (receding from) the observer \citep{rybicki79,loeb03}, with an amplitude proportional to the RV of the source. Therefore, the stellar RV modulation due to an orbiting companion will produce an RV like beaming phase modulation at the orbital period.
The beaming amplitude is the result of a bolometric effect of $4V_r/c$, where $V_r$ is the star RV and $c$ is the speed of light, corrected by an $\alpha_{\rm beam}$ factor that compensates for the stellar-spectrum shift into, or out of, the observed bandpass.
  The amplitudes of the beaming and the ellipsoidal modulations both depend, but in different ways, on the masses of the two components, that cannot be probed by the transit method. As such they can provide important complementary information about the basic astrophysical properties of the system.

While the reflection and the ellipsoidal effects are well known and studied in the field of short period binaries, the beaming modulation became observationally relevant only recently. Before the era of space photometry this effect has been
noticed only once, by \citet{maxted00}, who observed KPD 1930+2752, a binary with an orbital period a
little longer than 2 hours, and an RV  semi-amplitude of 350 km/s. The beaming effect of that system, on the order of $10^{-3}$, was hardly seen in the photometric data. It was though anticipated that the high precision light curves of {\it CoRoT} and {\it Kepler} will detect each of the three modulations \citep[e.g.,][]{drake03,loeb03,zucker07}, for binaries and planets alike.
As predicted, once the {\it Kepler} and {\it CoRoT} photometric light curves became available, 
 several studies detected various combinations of the three effects in eclipsing binaries, for which the orbital period was well established from the space-obtained light curves. These detection were in eclipsing binaries with a white-dwarf secondary  \citep{vankerkwijk10,carter11,bloemen11,bloemen12,breton12,rappaport15, faigler15b}, and even in a few transiting brown-dwarfs and planetary secondaries \citep{snellen09,welsh10,mazeh10,shporer11,mazeh12,barclay12,herrero14}.

However, space mission data can yield much more. In addition to eclipse events, The  {\it CoRoT} and {\it Kepler}
data can indicate the binarity of a system based on the beaming, ellipsoidal and reflection
effects themselves. \citet{loeb03} suggested that the beaming 
effect can be used to detect non-transiting exoplanets, and \citet{zucker07} extended this idea to binaries. 
\citet{loeb03} \citep[see also the discussion of][]{zucker07} showed that for relatively long-period orbits, of the order of $10$--$100$ days,
the beaming modulation is stronger than the ellipsoidal and the reflection effects, and therefore could be observed without interference from the other two modulations. 
However, the beaming modulation by itself might not be enough to render a star a good exoplanet candidate, as a pure sinusoidal modulation could be produced by other effects, stellar modulations in particular \citep[e.g.,][]{aigrain04}.
The BEER search algorithm, therefore, searches for stars that show in their space-obtained 
light curves some {\it combination} of the BEaming, Ellipsoidal, and Reflection (BEER) modulations, with amplitudes and phases that are consistent with a low-mass companion.


This thesis consists of seven scientific papers that cover the proof-of-concept, the development, and discoveries made through the use of the BEER algorithm for searching for companions in the light curves of the {\it Kepler} and {\it CoRoT} space telescopes. 

Spectroscopic RV confirmations in Papers III, IV and VII were performed mainly by the Tillinghast Reflector Echelle Spectrograph \citep[TRES;][]{furesz08} mounted on the 1.5-m Tillinghast Reflector at the Fred Lawrence Whipple Observatory operated by the Smithsonian Astrophysical Observatory (SAO) on Mount Hopkins in Southern Arizona. This thesis relies on hundreds of TRES observations of $\sim$180 target stars, led by Dave Latham from the Harvard-Smithsonian Center for Astrophysics, that contributed in a crucial way to the discoveries reported in the thesis papers.  

Paper I \citep{mazeh10} describes the detection of the ellipsoidal and the beaming effects, through the use of a novel cosine-transform based detrending method, in the {\it CoRoT} light curve of CoRoT-3, a system of a $22$$M_{\rm Jup}$ brown dwarf orbiting an F3 star in an orbital period of $4.3$ days \citep{deleuil08}.
This work, together with \citet{snellen09} and \citet{welsh10}, served as a proof-of-concept that such modulations are indeed detectable in the space light curves of systems harboring brown-dwarf or planetary secondaries. These studies thus indicate that similar modulations may be detectable in the light curves of non-transiting systems.

Paper II \citep{faigler11} describes the BEER algorithm for detection of non-transiting short-period low-mass companions, through the beaming, ellipsoidal and reflection effects, in {\it Kepler} and {\it CoRoT} light curves. 
The paper presents the algorithm, including an assignment of a likelihood factor to any possible detection, based on the expected ratio of the beaming and ellipsoidal effects. It then provides two examples of candidates found in the light curves of the first {\it Kepler} quarter, with detected periodic amplitudes as small as 100 parts per million (ppm).

Paper III \citep{faigler12} presents the first discoveries of seven non-eclipsing binaries in the {\it Kepler} light curves through the BEER method, that were confirmed by RV observations. Two of the detected binaries were the two candidates presented in Paper II. This work marks the first detections made through the use of the BEER algorithm.

Paper IV \citep{faigler13} describes the discovery of Kepler-76b, a 2$M_{\rm Jup}$ hot Jupiter orbiting a 13.3 mag F star with orbital period of $1.54$ days. This system that initially appeared in the {\it Kepler} EB catalog, was identified by the BEER algorithm, based on its amplitudes and phases of the three effects, as a system harboring a hot Jupiter. The paper covers the BEER algorithm detection of the hot Jupiter, RV observations confirmation of the companion mass, and first evidence in the {\it Kepler} band for superrotation in the atmosphere of Kepler-76b.
Superrotation involves a phase shift of the planetary thermal-emission modulation, due to equatorial superrotating jets in the planet atmosphere. This phenomenon was predicted by \citet{showman02} and later observed by \citet{knutson07,knutson09} in the infrared for HD 189733.
In addition to TRES observations, part of the RV measurements presented in this paper were performed by the SOPHIE spectrograph \citep{Perruchot08, bouchy09, bouchy13} mounted on the 1.93-m telescope at Observatoire de Haute-Provence, France.

Paper V \citep{faigler15} extends the discovery of superrotation in Kepler-76b to two additional known hot Jupiters, HAT-P-7b and KOI-13b, and presents the Lambertian superrotation BEER model, that enables estimating the masses of hot Jupiters from photometry alone. The paper concludes that hot Jupiter superrotation may be a common phenomenon that is detectable in the visual-band {\it Kepler} light curves.

Paper VI \citep{tal-or15} describes the discovery of seventy non-eclipsing binaries, detected by the BEER algorithm in {\it CoRoT} light curves, and confirmed by RV follow-up observations.
The discoveries included two brown-dwarf candidates on a $\sim$$1$ day period orbit. This was the first time non-eclipsing beaming binaries were detected in {\it CoRoT} data , and the paper estimates that $\sim$$300$ such binaries can be detected in {\it CoRoT} long-run light curves.
RV measurements presented in this paper were performed by the AAOmega multi-object spectrograph \citep{smith04,saunders04} at the Anglo-Australian Telescope (AAT).

Paper VII \citep{faigler15b} presents the discovery of four short-period eclipsing systems in the {\it Kepler} light curves, consisting of an A-star primary and a low-mass WD secondary (dA+WD), through the BEER algorithm. These add to the 6 {\it Kepler}, and 18 WASP, previously known short-period eclipsing dA+WD binaries. Three of the new systems harbor the smallest WDs discovered so far in such binaries. These three binaries extend the previously known population to older systems with cooler and smaller WD secondaries.

\newpage
\newpage
\thispagestyle{empty}
\

\chapter{The Papers}
\label{papers}
\thispagestyle{empty}

\
\ifthenelse{\boolean{PAPERS}}{

\newpage
\thispagestyle{empty}
\

\pagestyle{fancy}

\cfoot{}

\addcontentsline{toc}{section}{Paper I}
\chead{Paper I -- CoRoT-3b}
\addtolength{\headsep}{-1.5cm}

\centerline{\psfig{figure=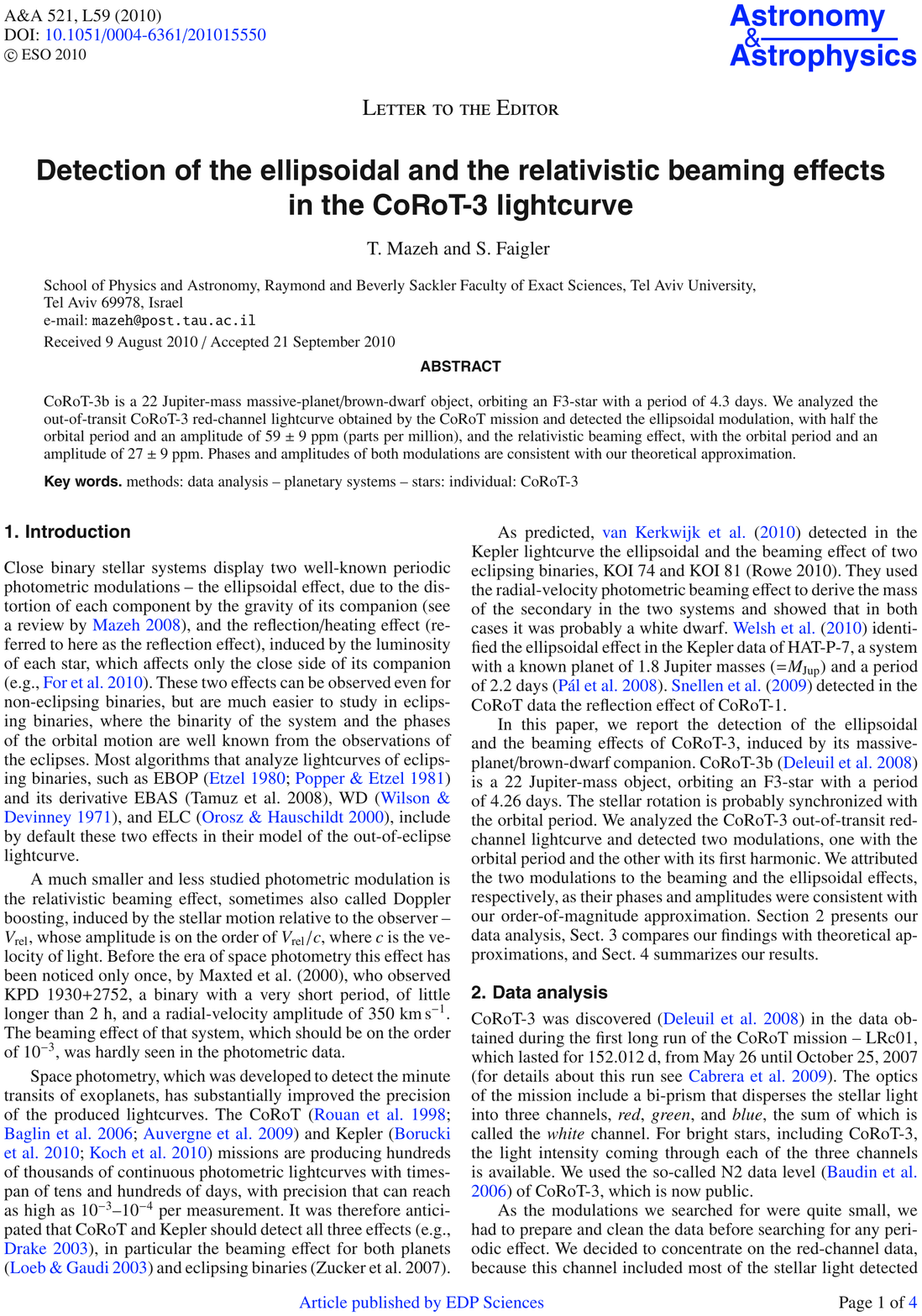,height=10.5in,clip=}}
\chead{Paper I -- CoRoT-3b}
\centerline{\psfig{figure=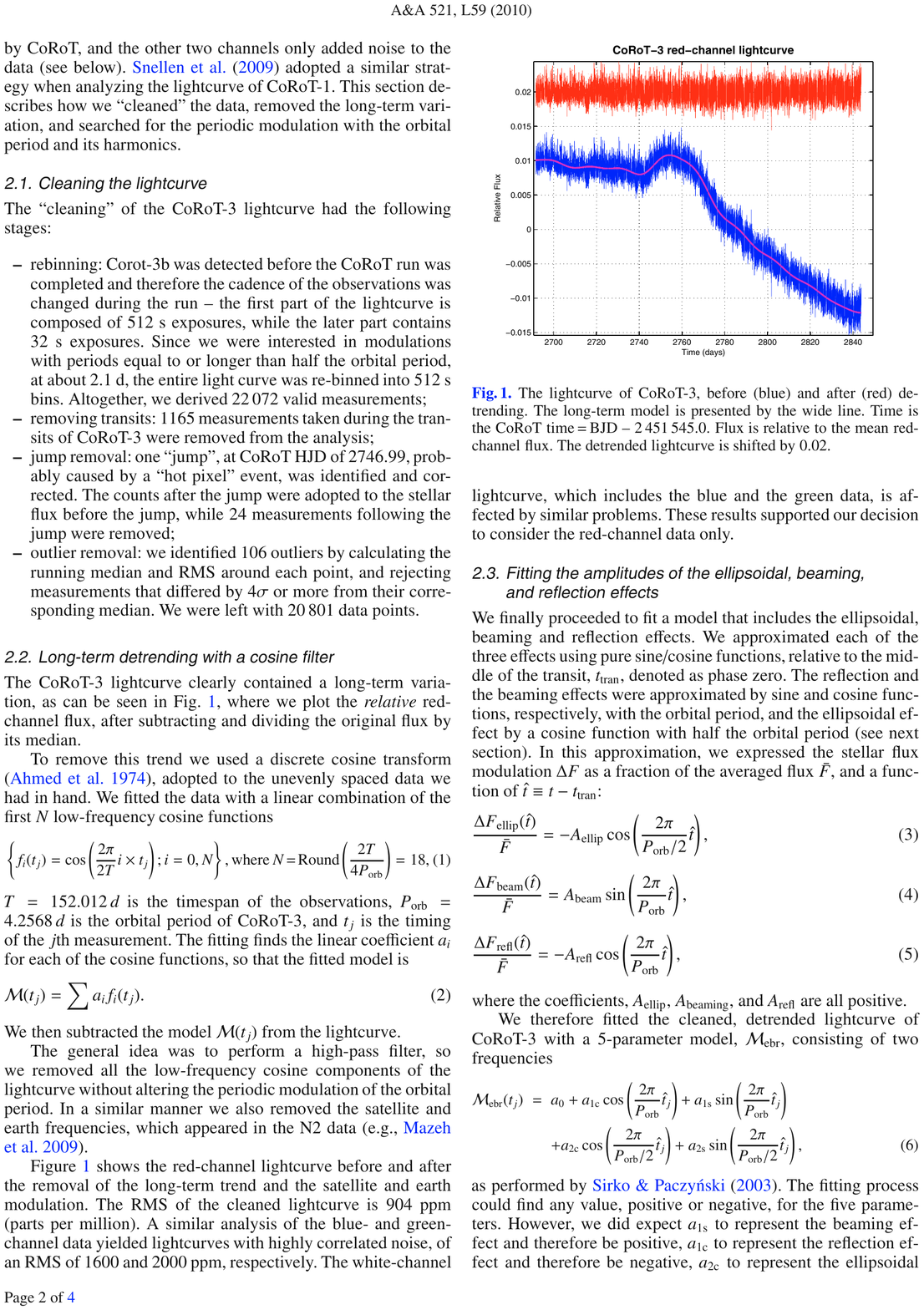,height=10.5in,clip=}}
\centerline{\psfig{figure=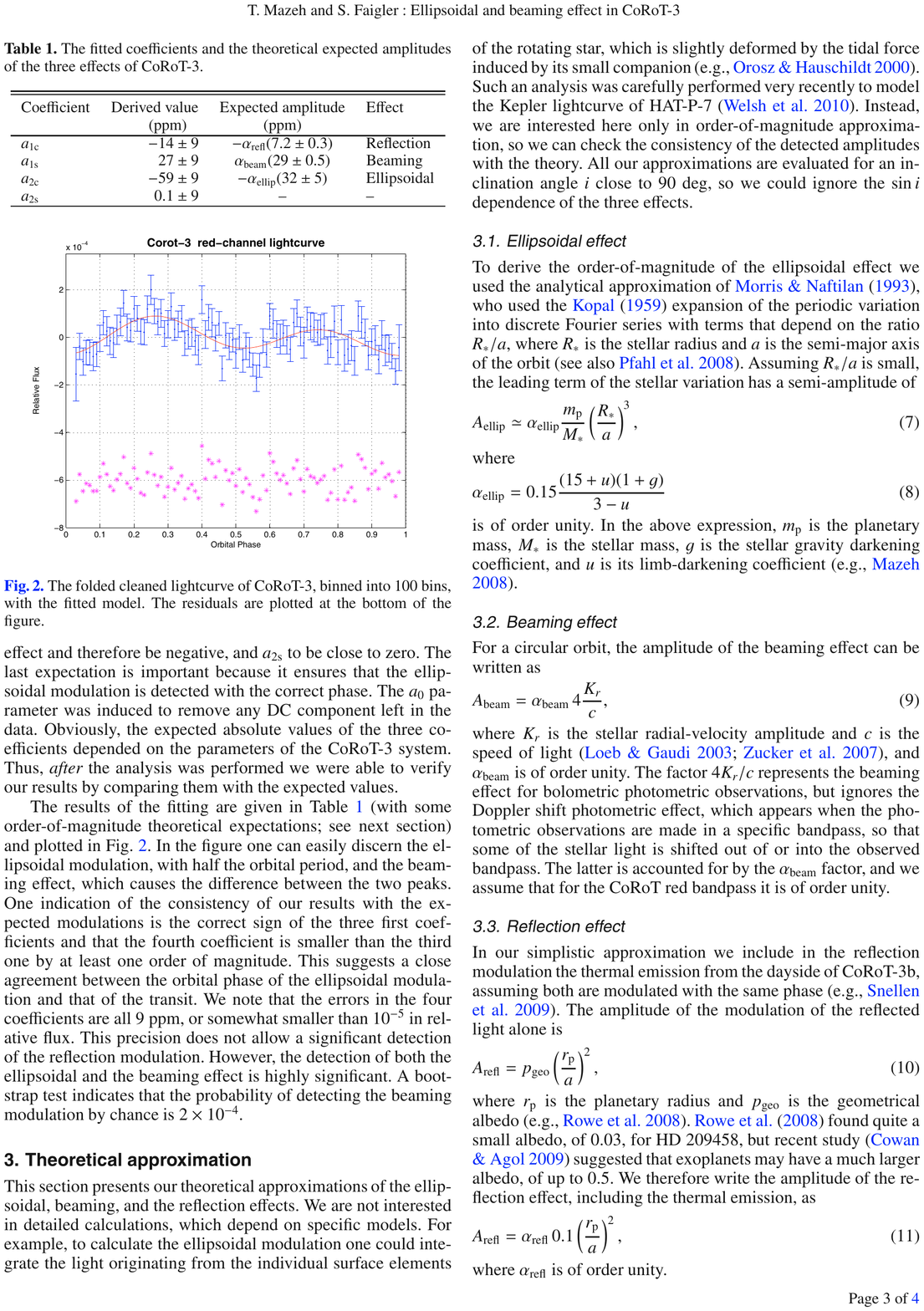,height=10.5in,clip=}}
\centerline{\psfig{figure=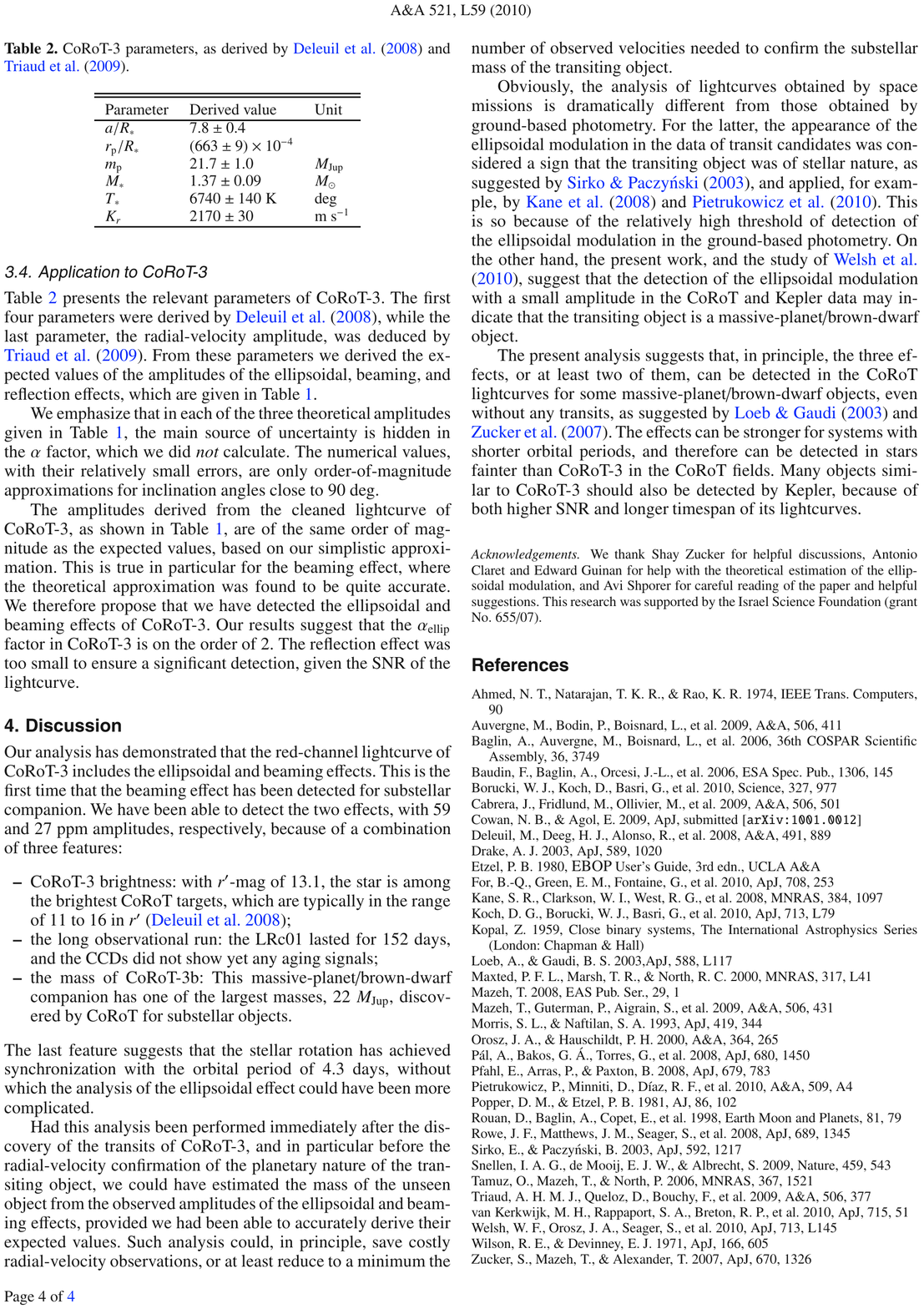,height=10.5in,clip=}}

\addcontentsline{toc}{section}{Paper II}
\centerline{\psfig{figure=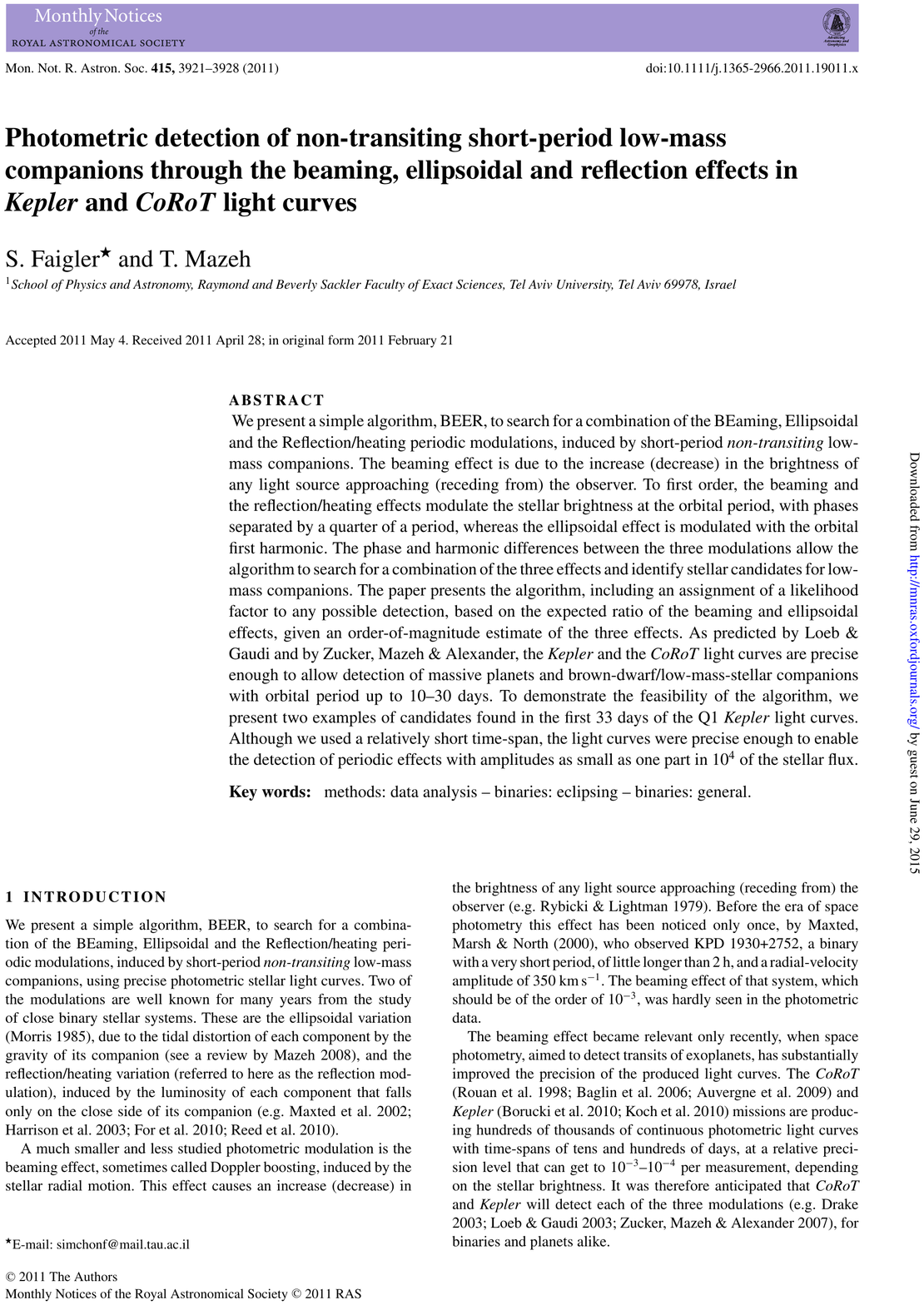,height=10.5in,clip=}}
\chead{Paper II -- The BEER Search}
\addtolength{\headsep}{-1.0cm}

\centerline{\psfig{figure=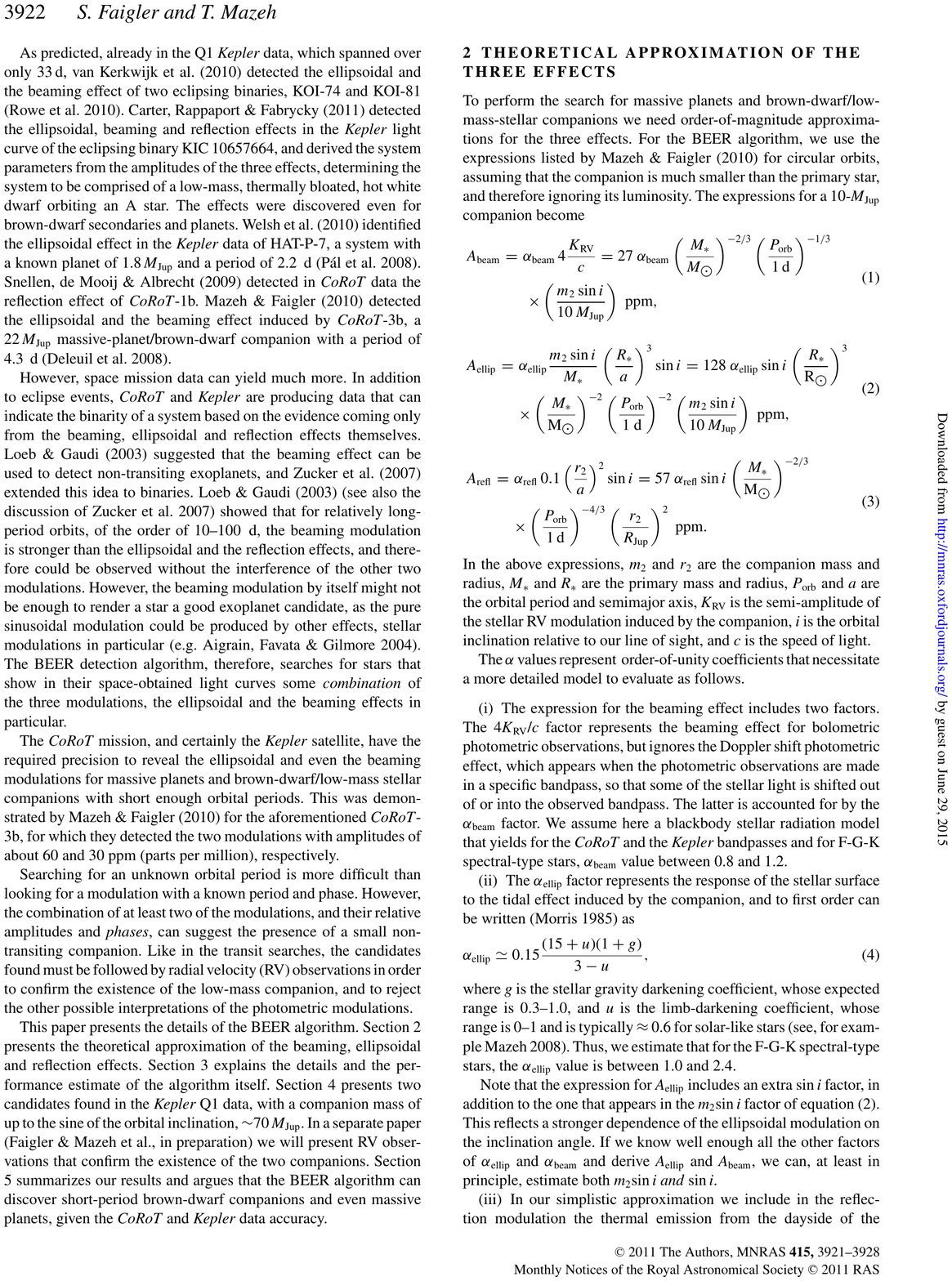,height=10.5in,clip=}}
\centerline{\psfig{figure=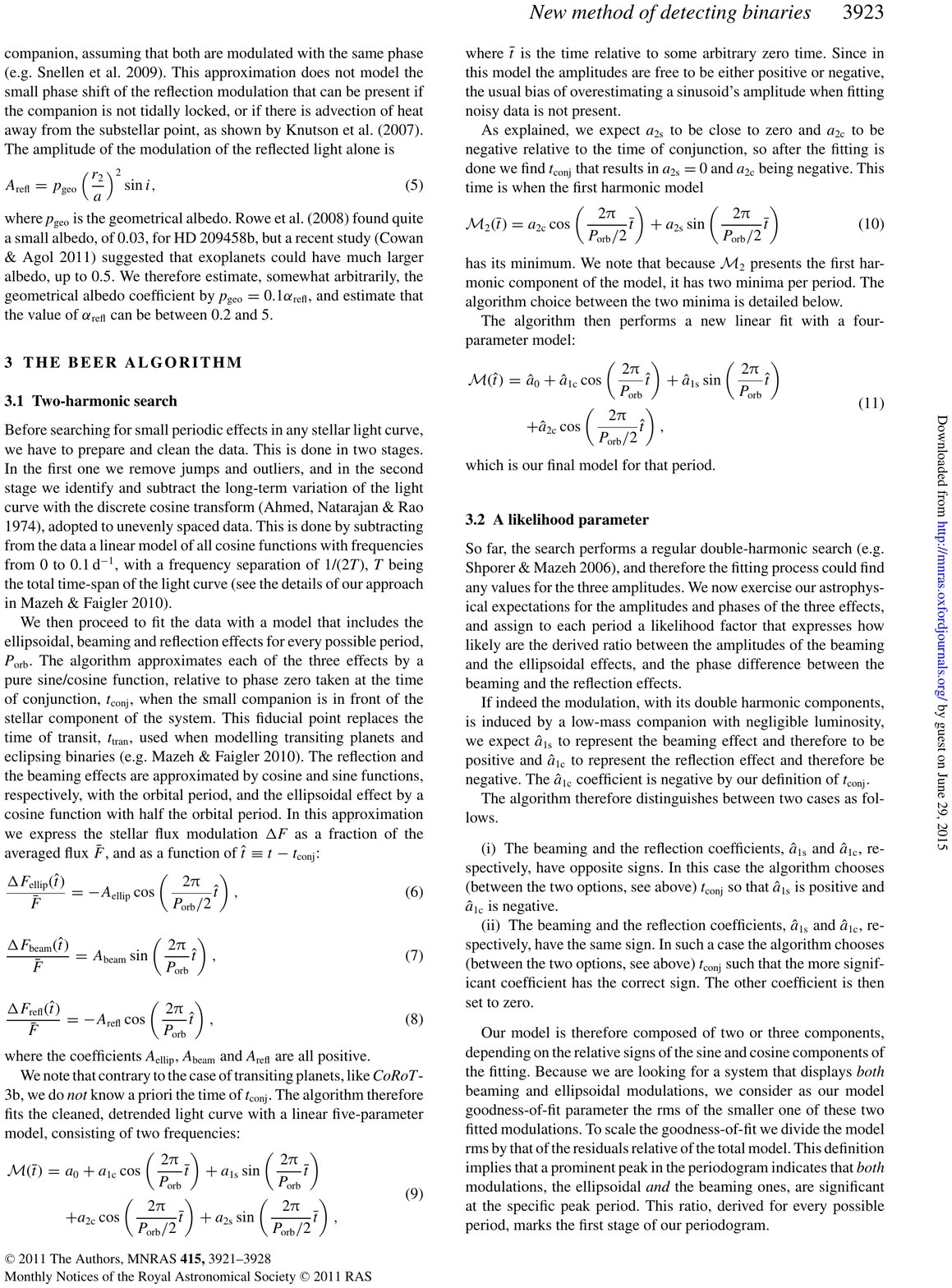,height=10.5in,clip=}}
\centerline{\psfig{figure=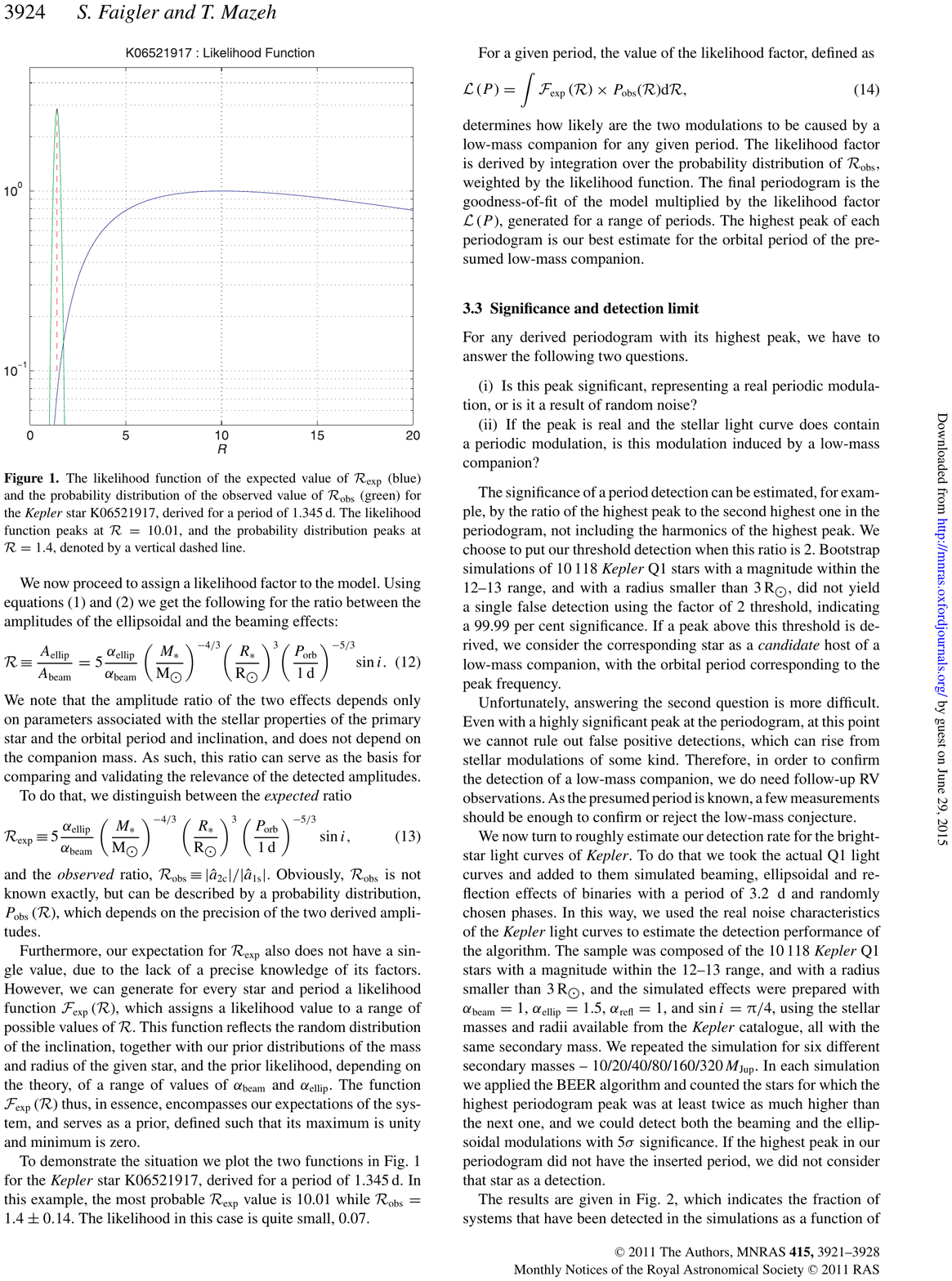,height=10.5in,clip=}}
\centerline{\psfig{figure=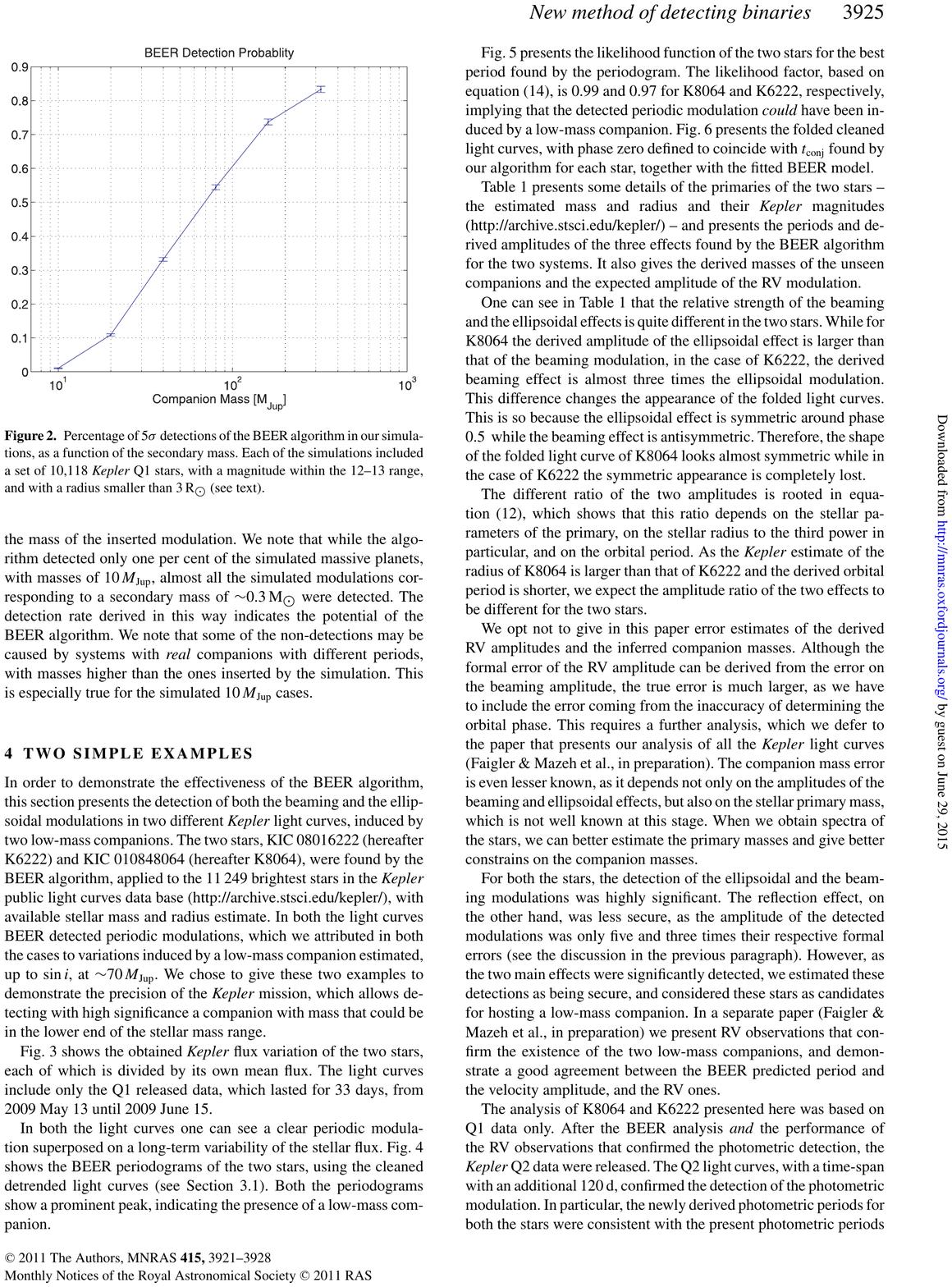,height=10.5in,clip=}}
\centerline{\psfig{figure=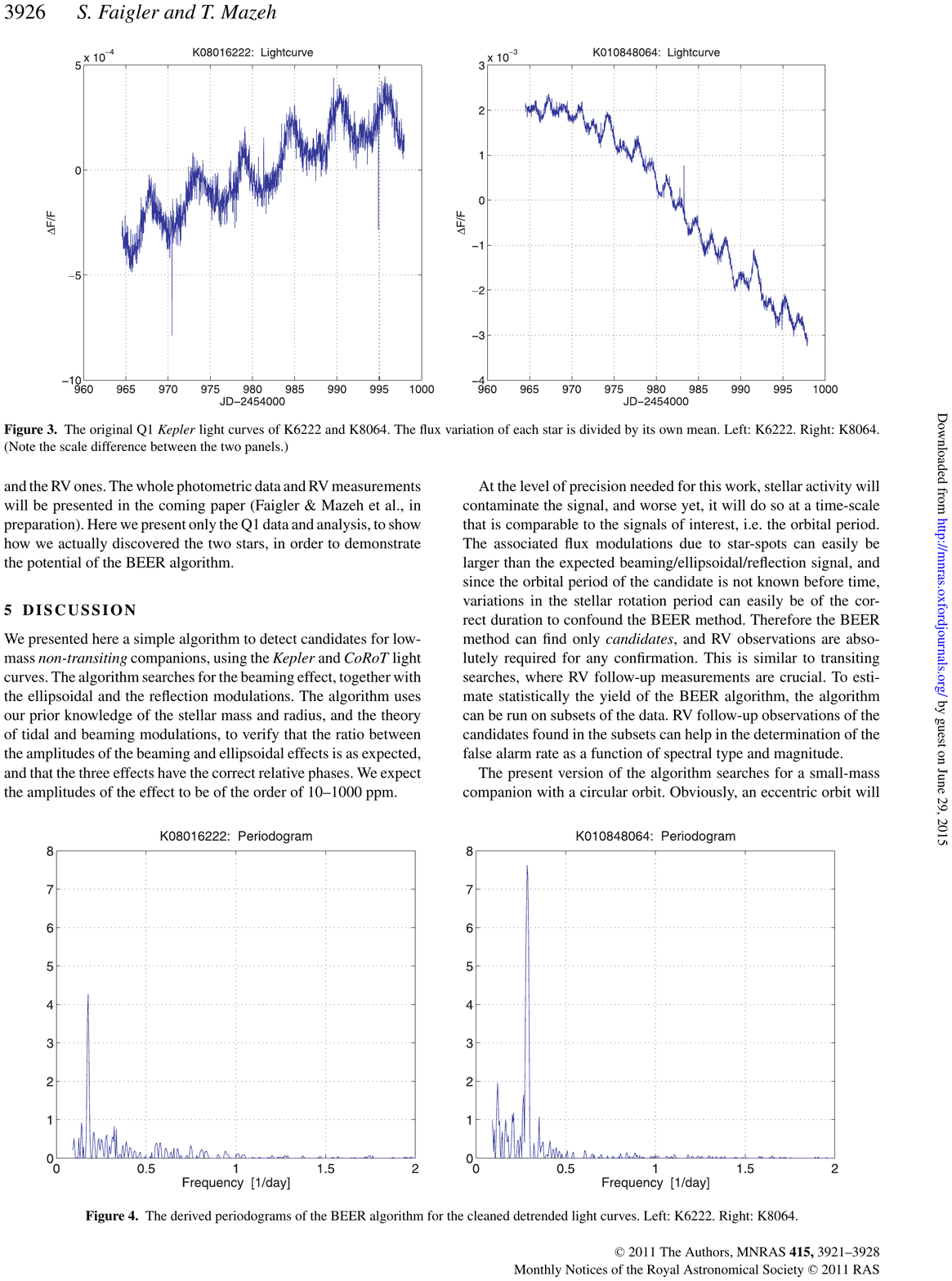,height=10.5in,clip=}}
\centerline{\psfig{figure=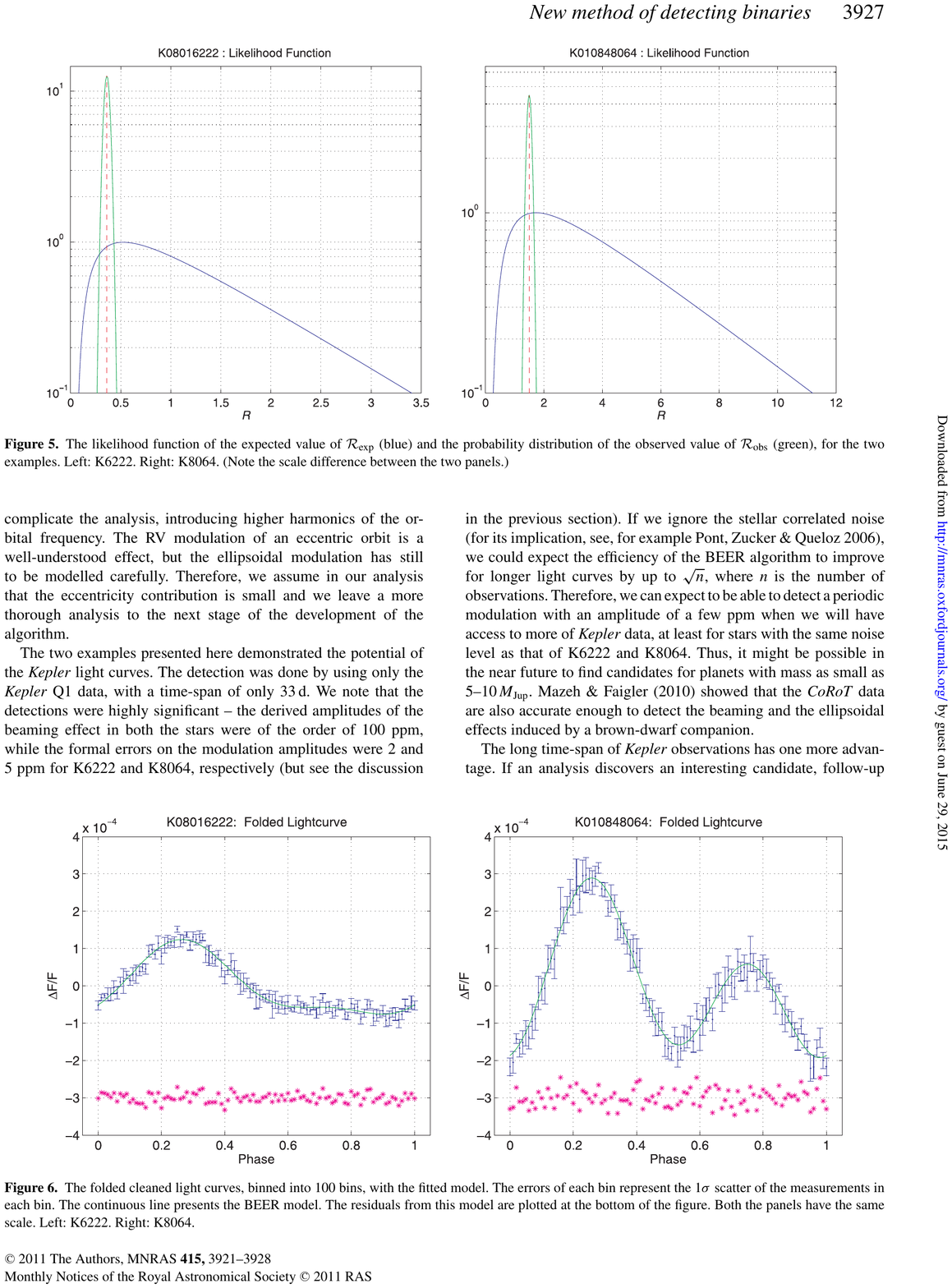,height=10.5in,clip=}}
\centerline{\psfig{figure=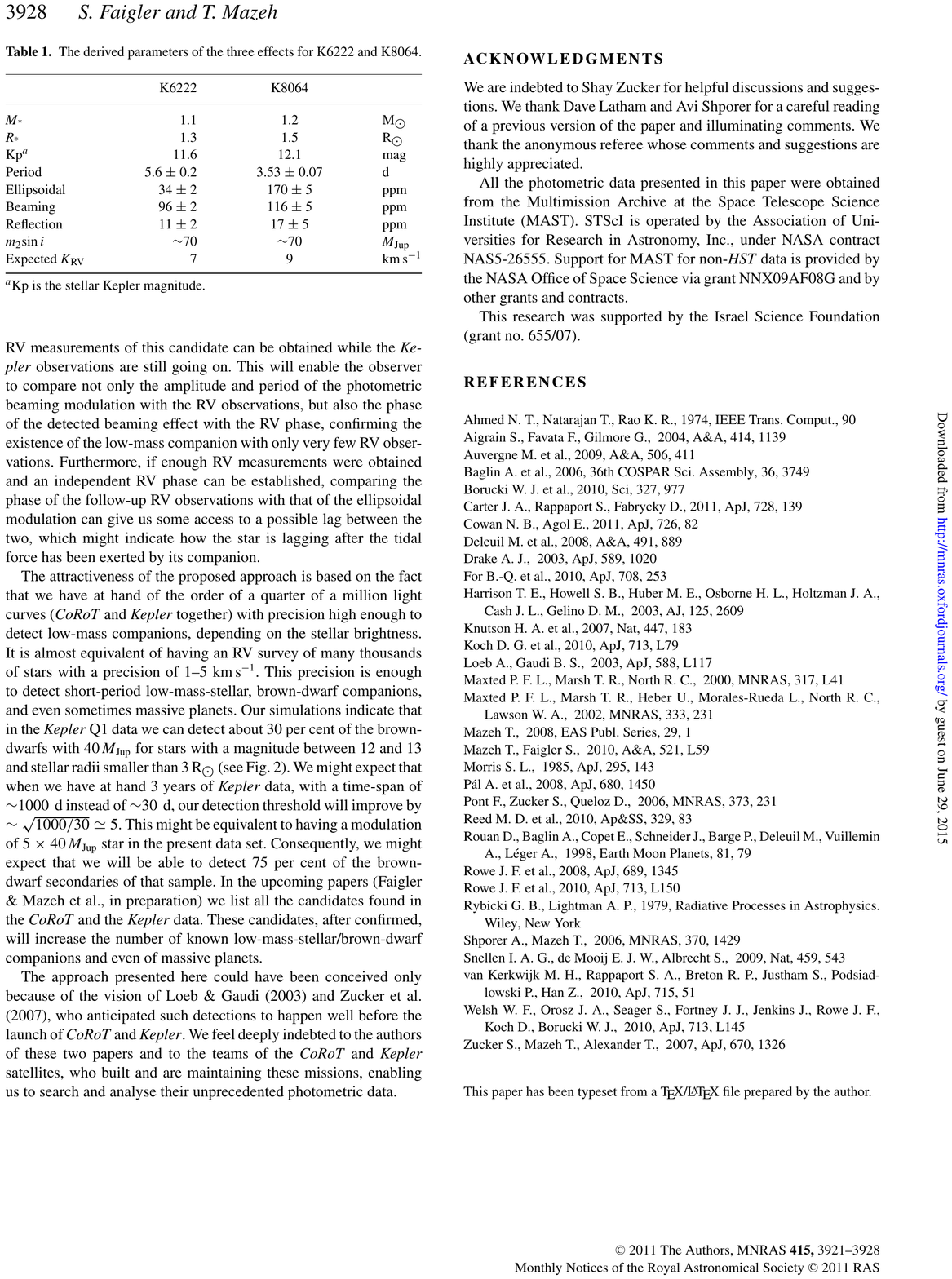,height=10.5in,clip=}}

\addcontentsline{toc}{section}{Paper III}
\centerline{\psfig{figure=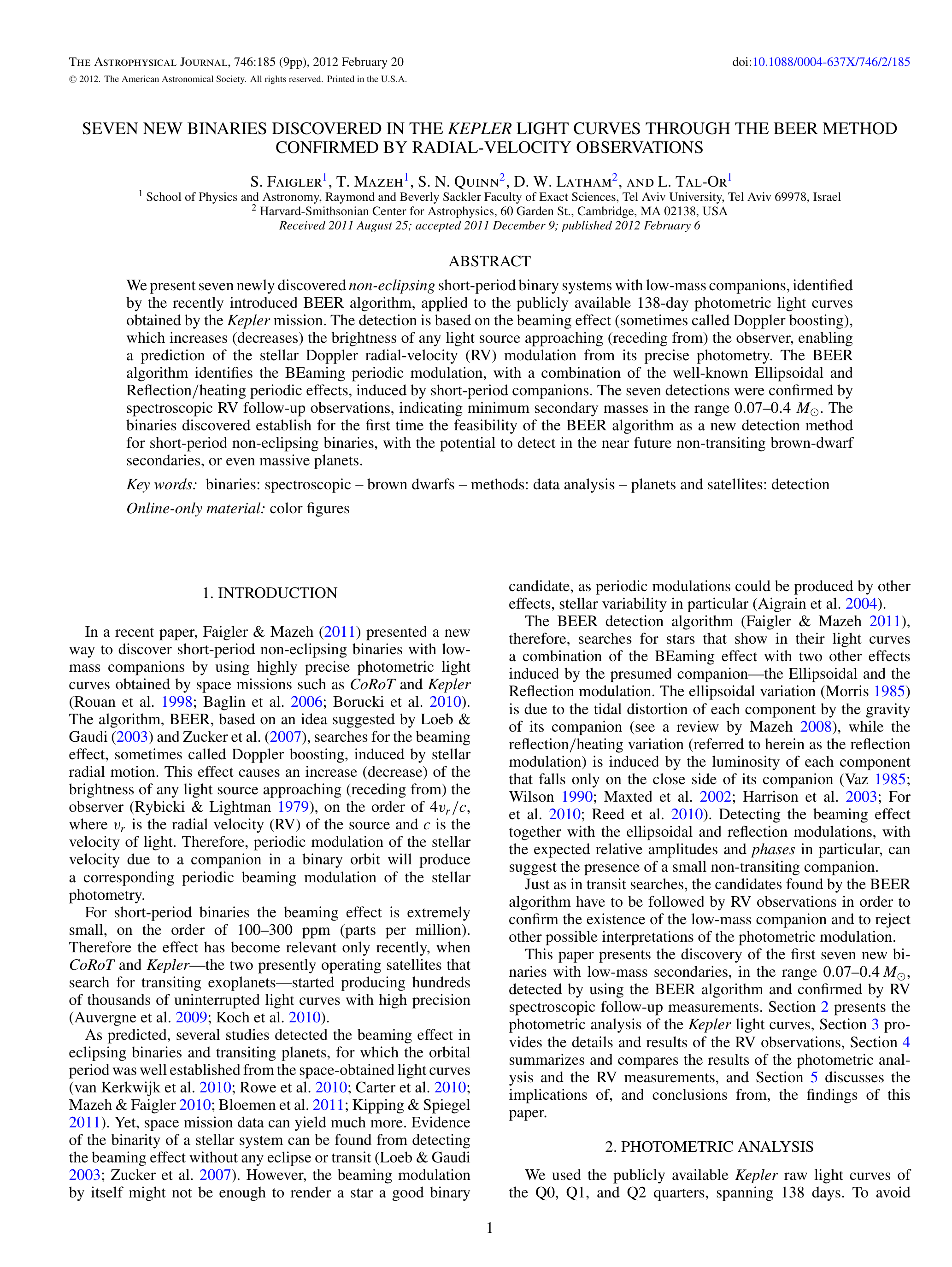,height=10in}}
\chead{Paper III -- Seven BEER binaries}
\addtolength{\headsep}{+1.5cm}

\centerline{\psfig{figure=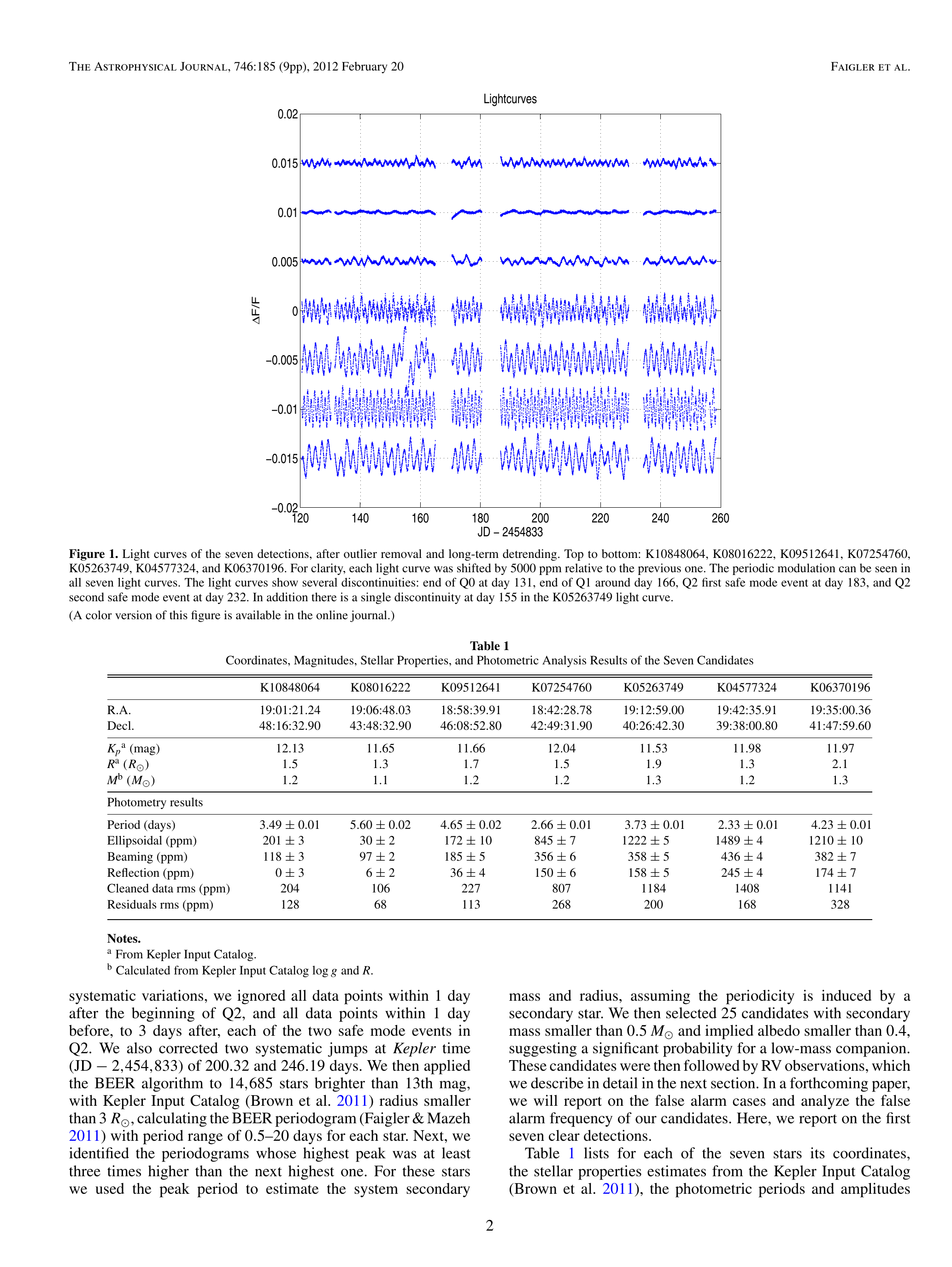,height=10in}}
\centerline{\psfig{figure=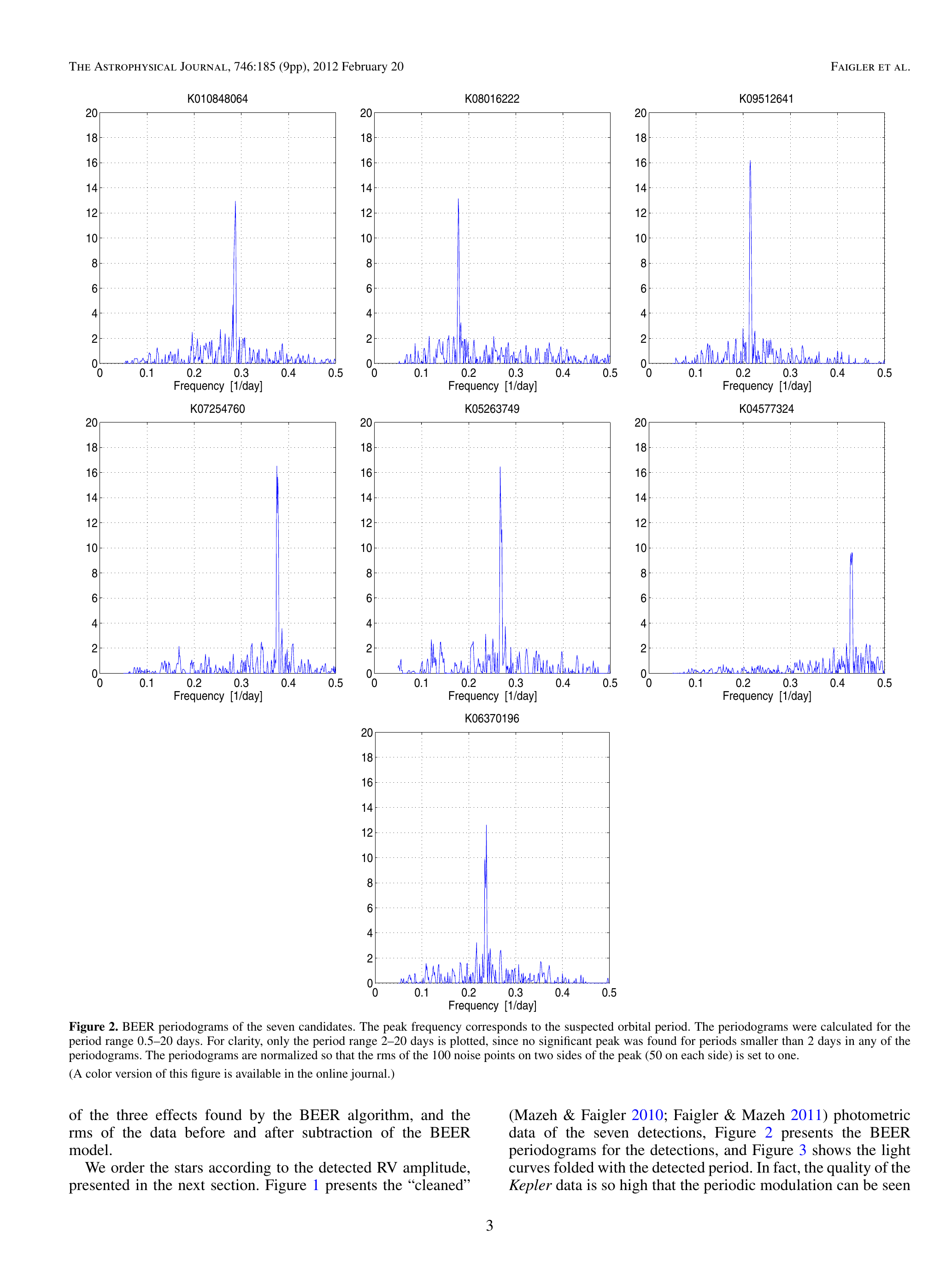,height=10in}}
\centerline{\psfig{figure=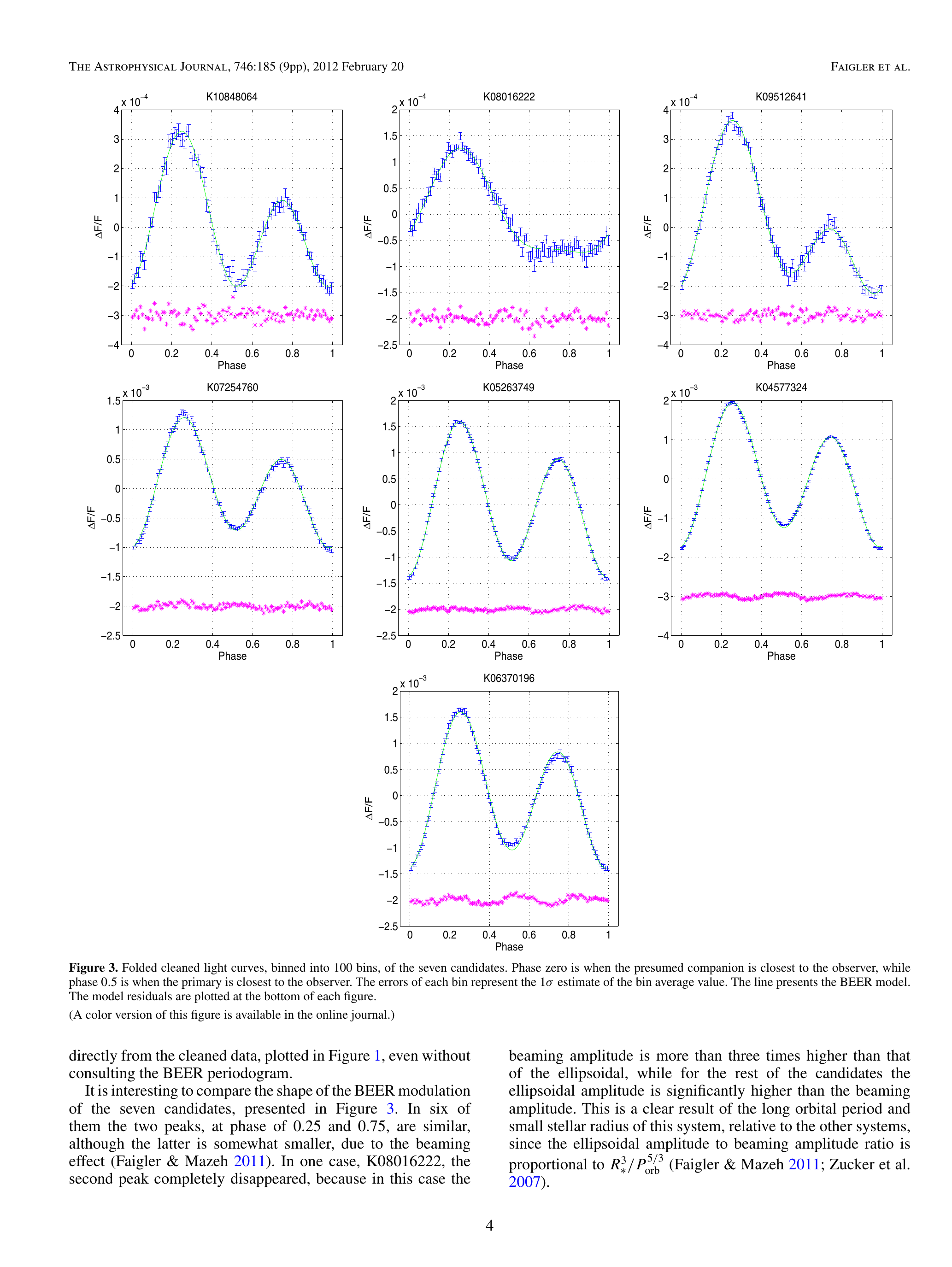,height=10in}}
\centerline{\psfig{figure=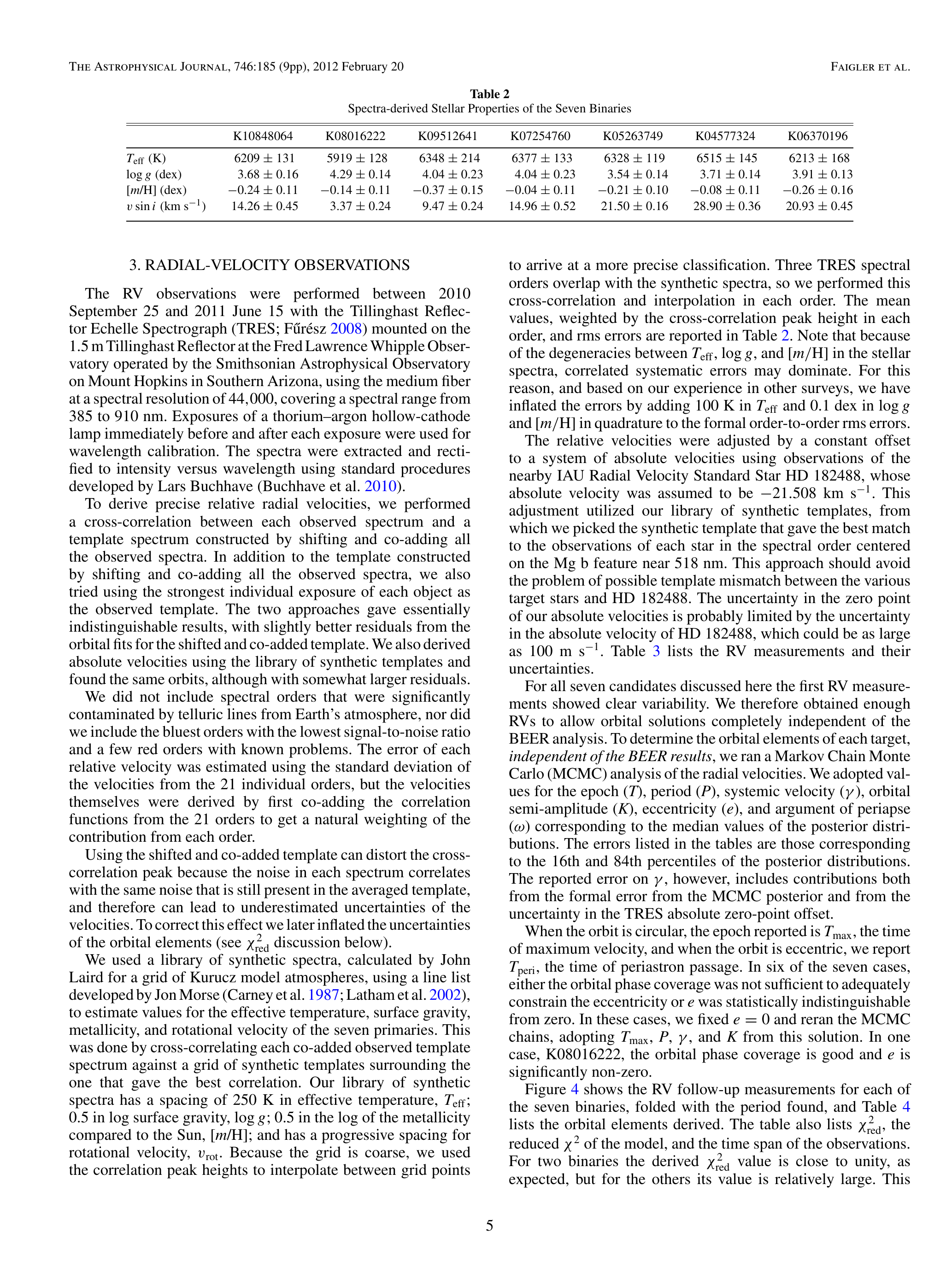,height=10in}}
\centerline{\psfig{figure=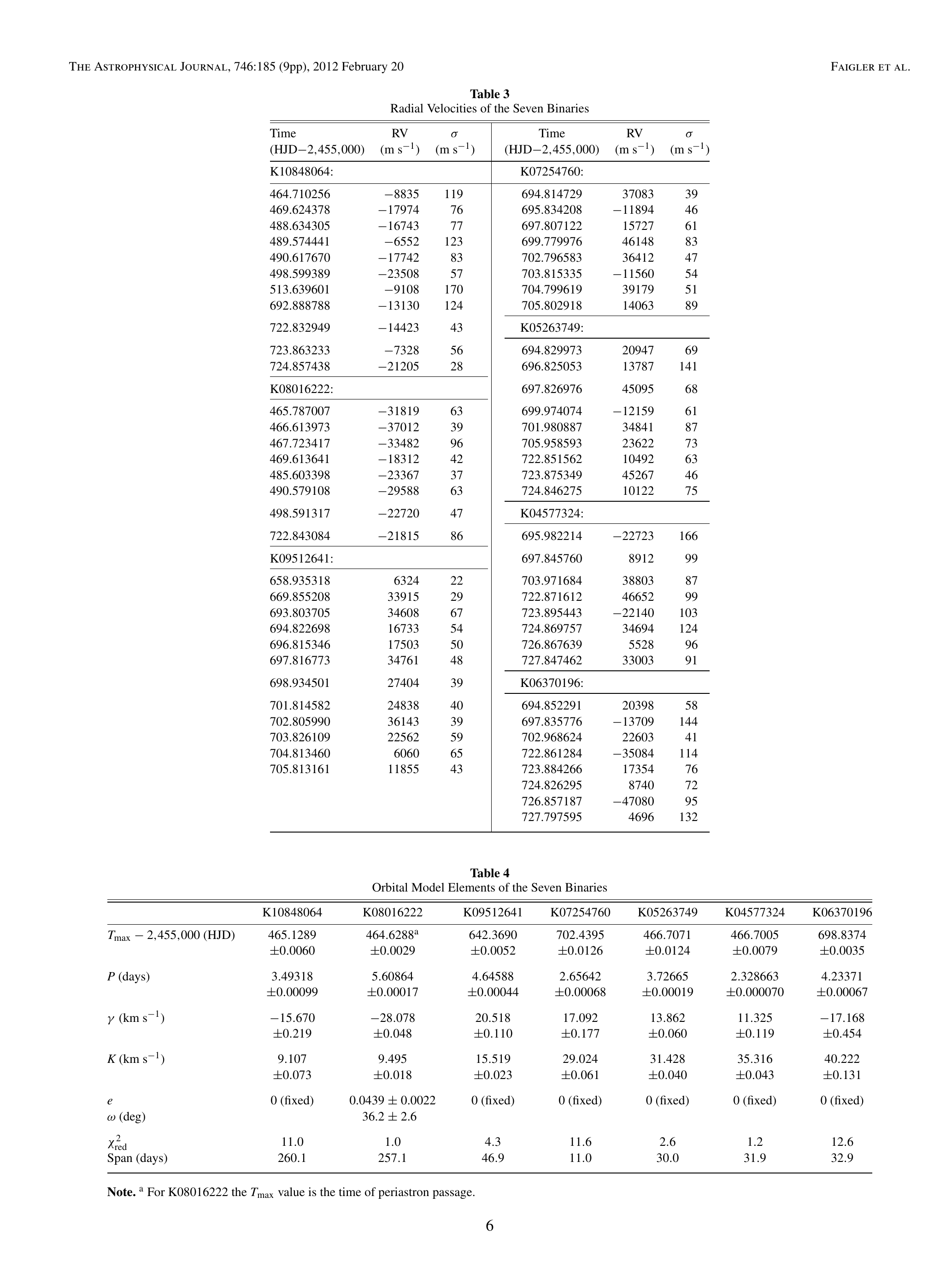,height=10in}}
\centerline{\psfig{figure=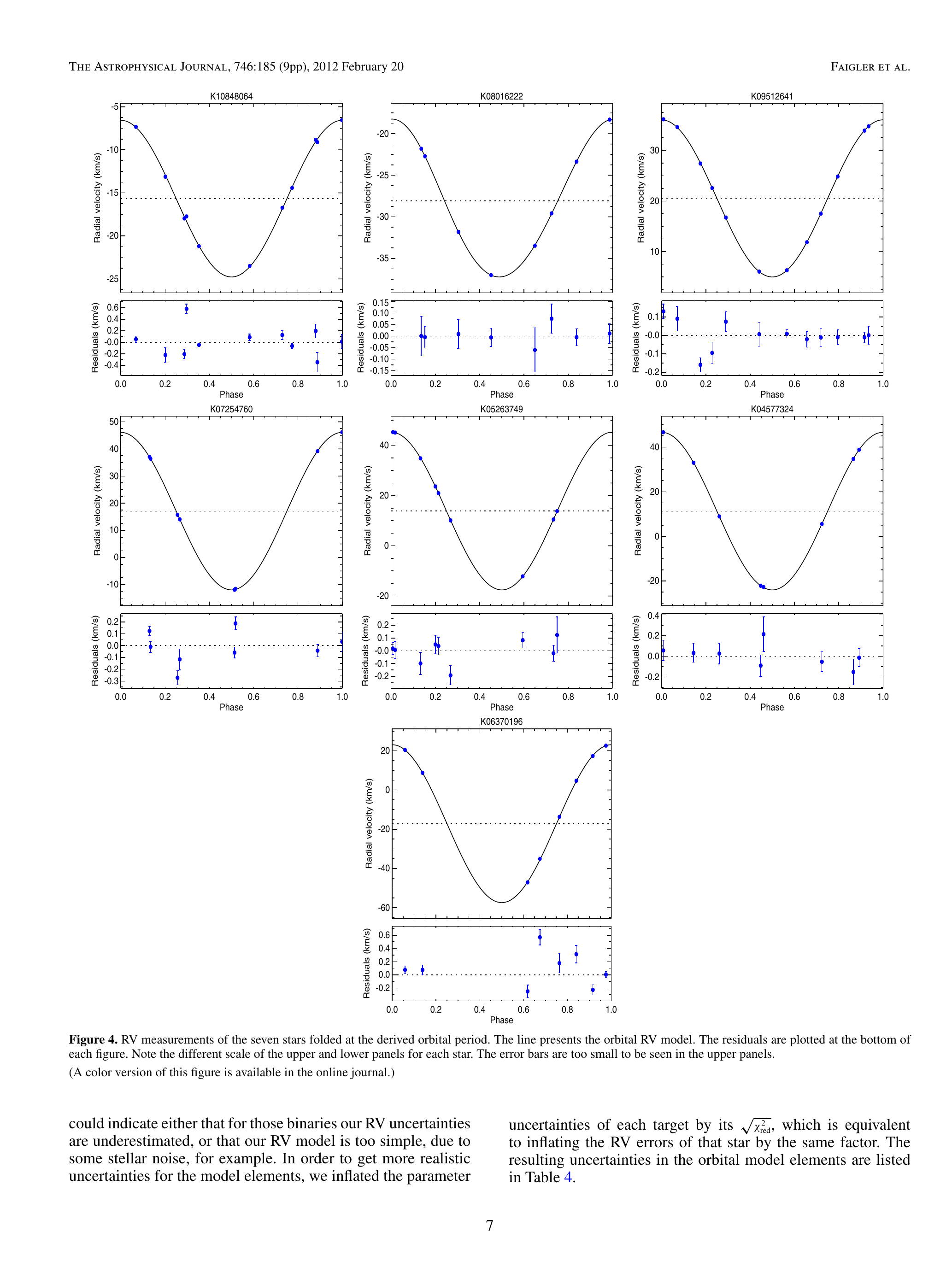,height=10in}}
\centerline{\psfig{figure=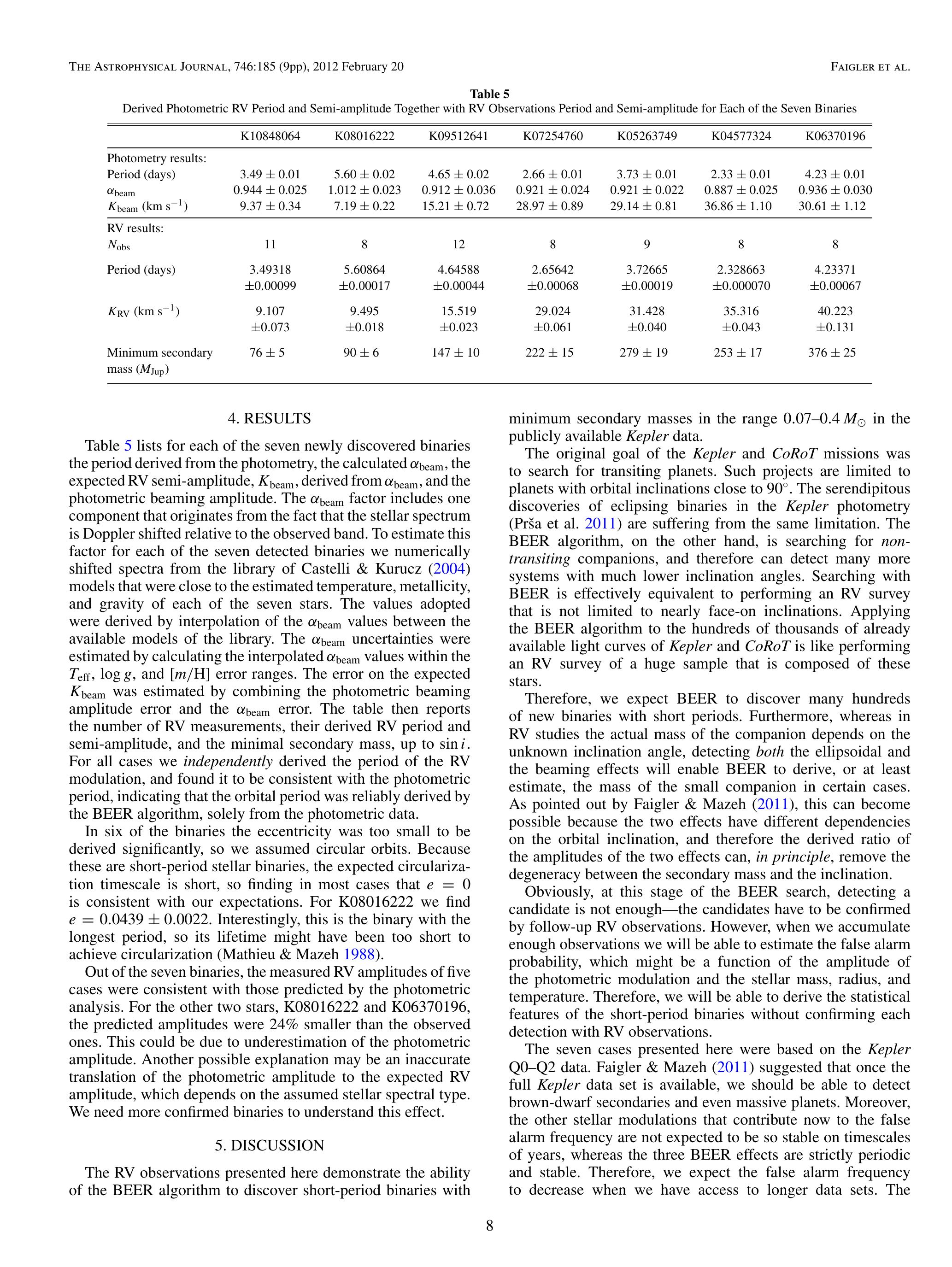,height=10in}}
\centerline{\psfig{figure=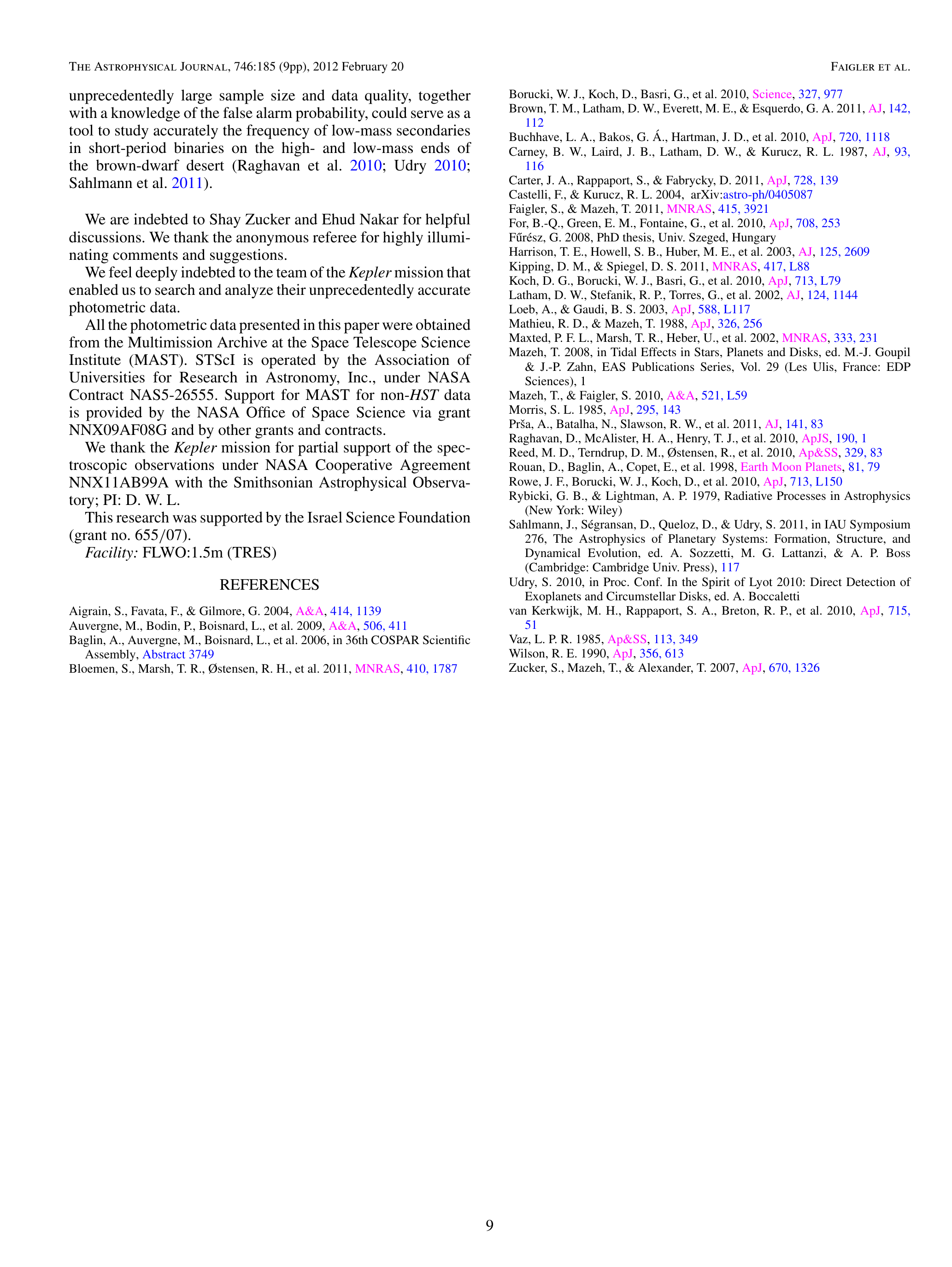,height=10in}}

\clearpage
\thispagestyle{empty}

\addcontentsline{toc}{section}{Paper IV}
\centerline{\psfig{figure=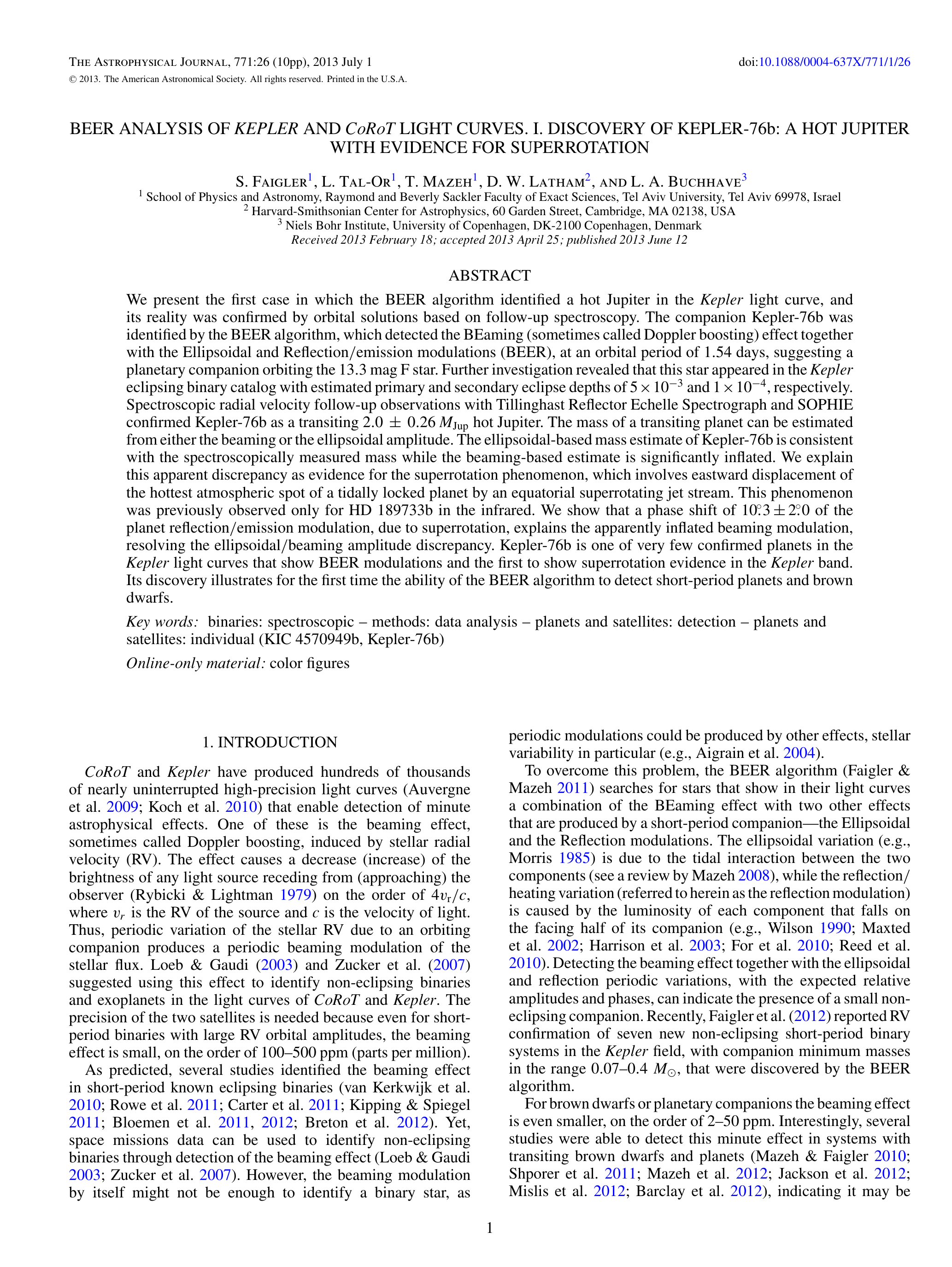,height=10in}}
\chead{Paper IV -- Kepler-76b}
\addtolength{\headsep}{0.0cm}

\centerline{\psfig{figure=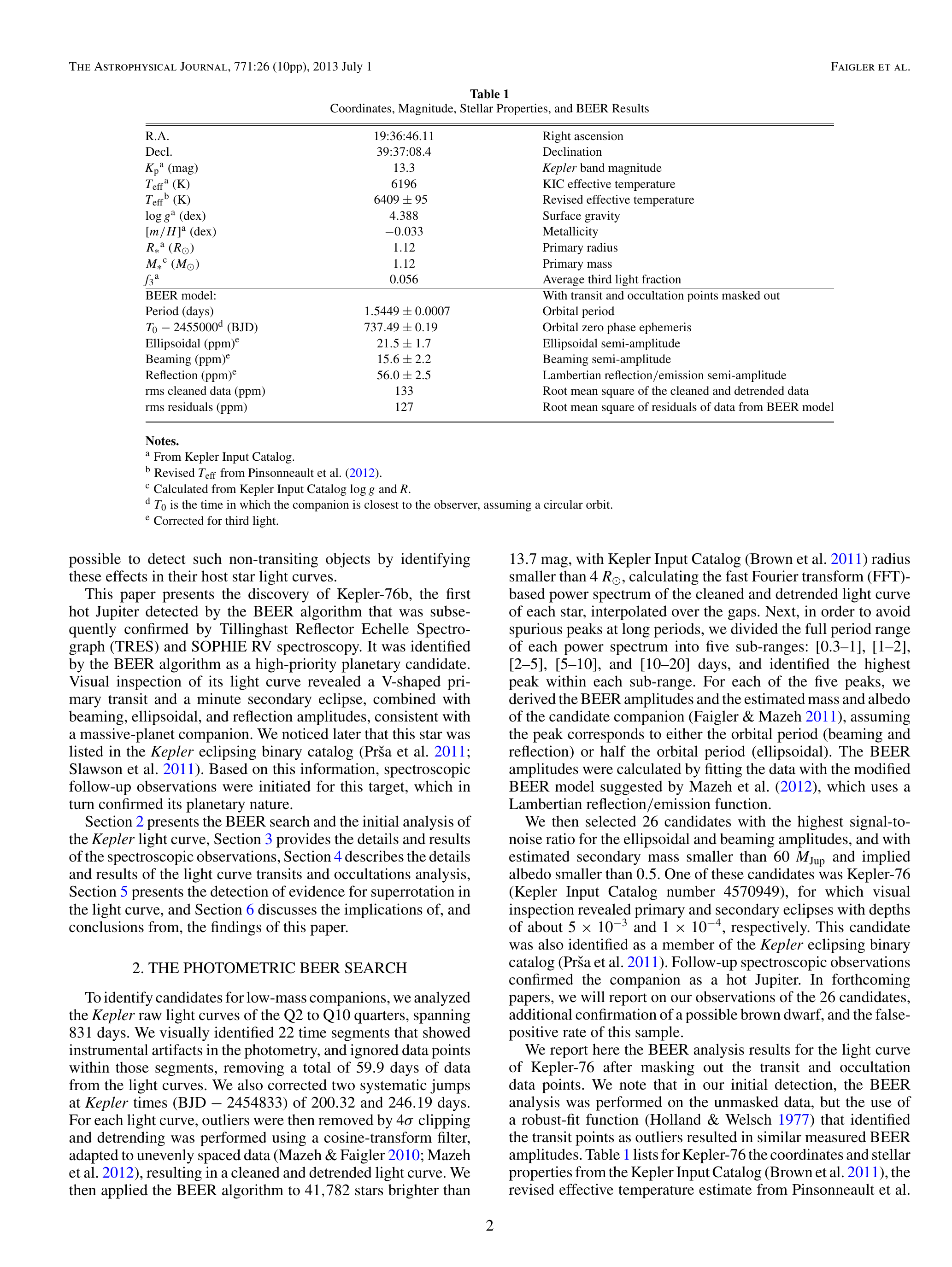,height=10in}}
\centerline{\psfig{figure=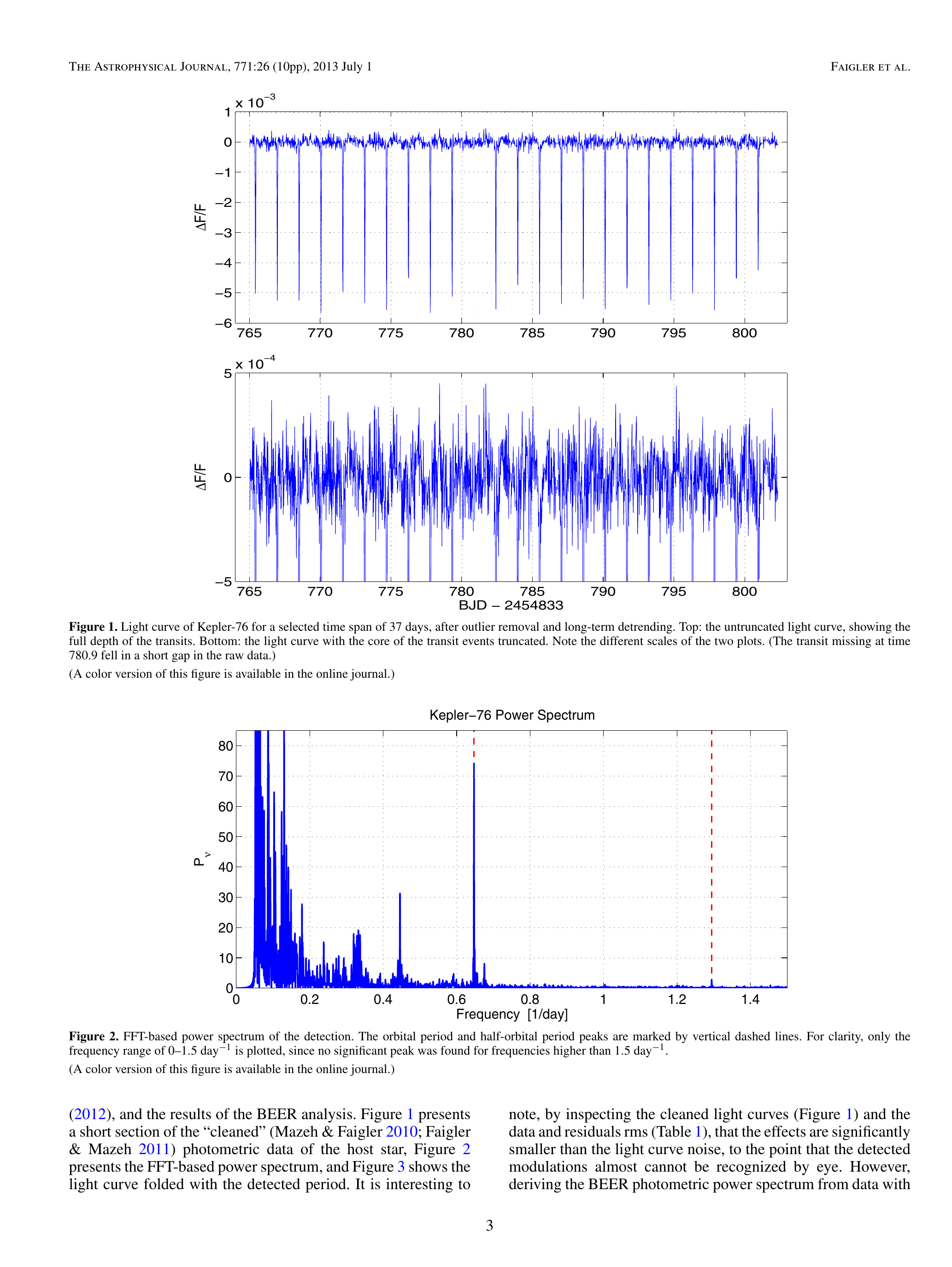,height=10in}}
\centerline{\psfig{figure=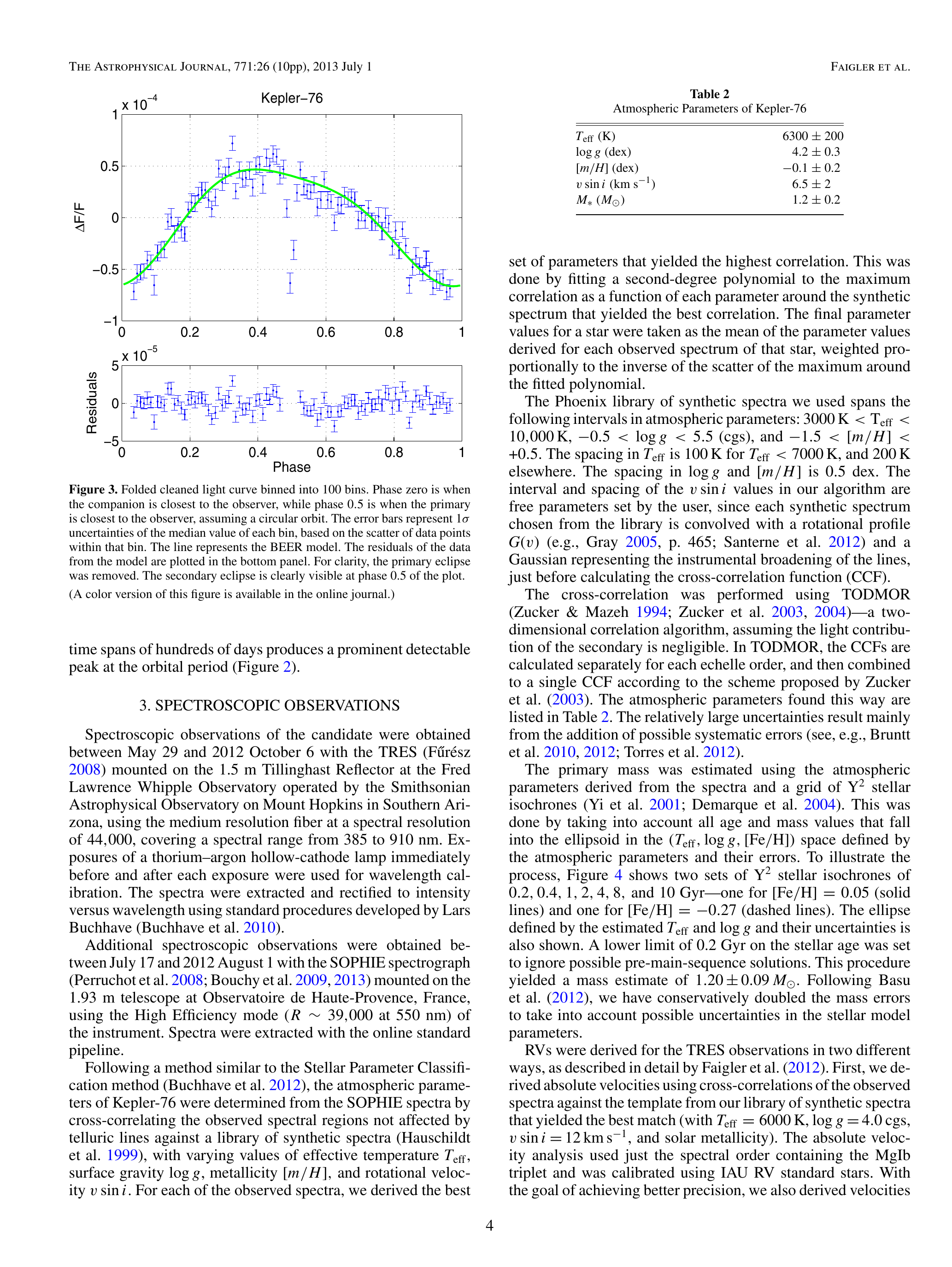,height=10in}}
\centerline{\psfig{figure=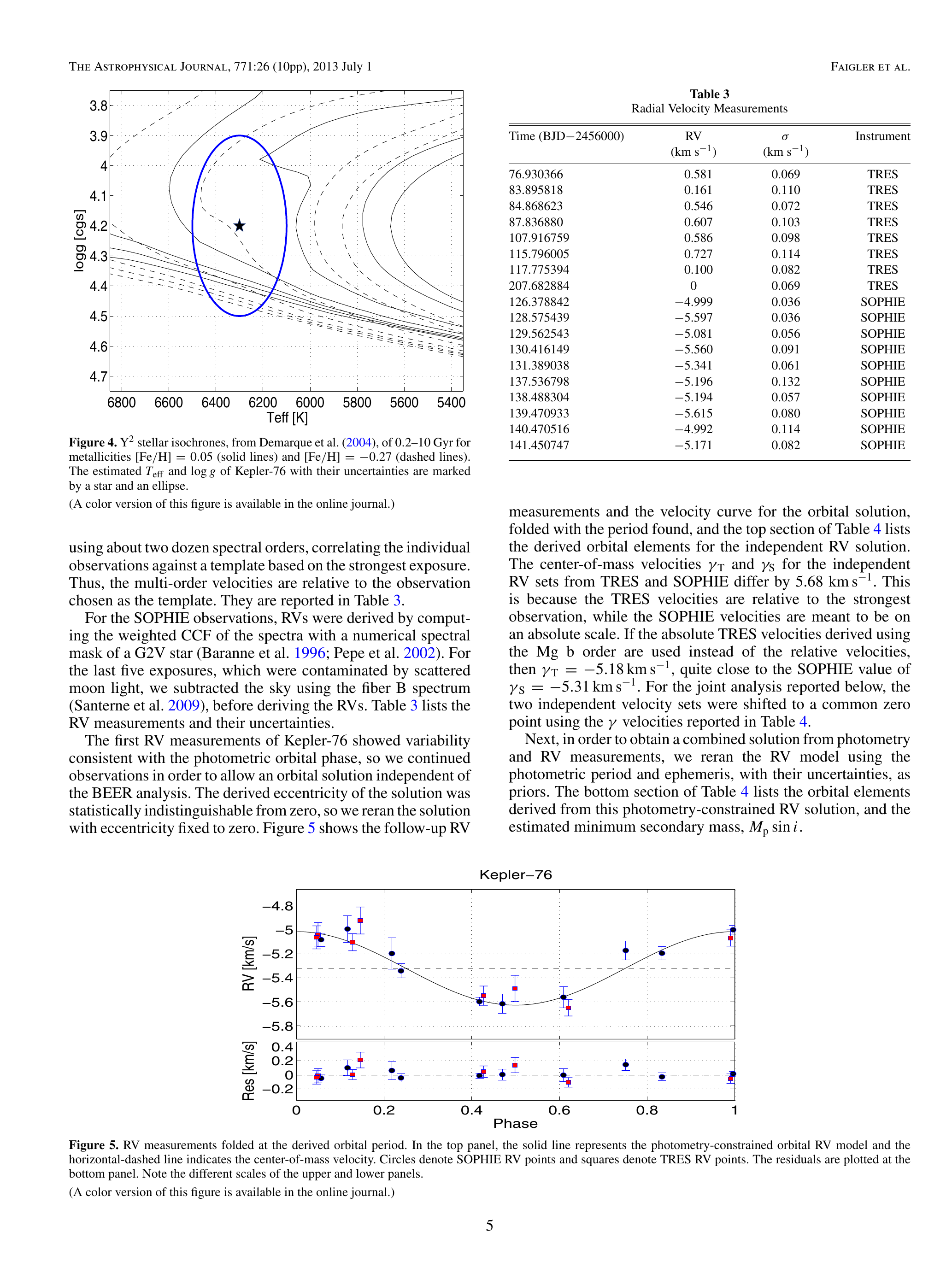,height=10in}}
\centerline{\psfig{figure=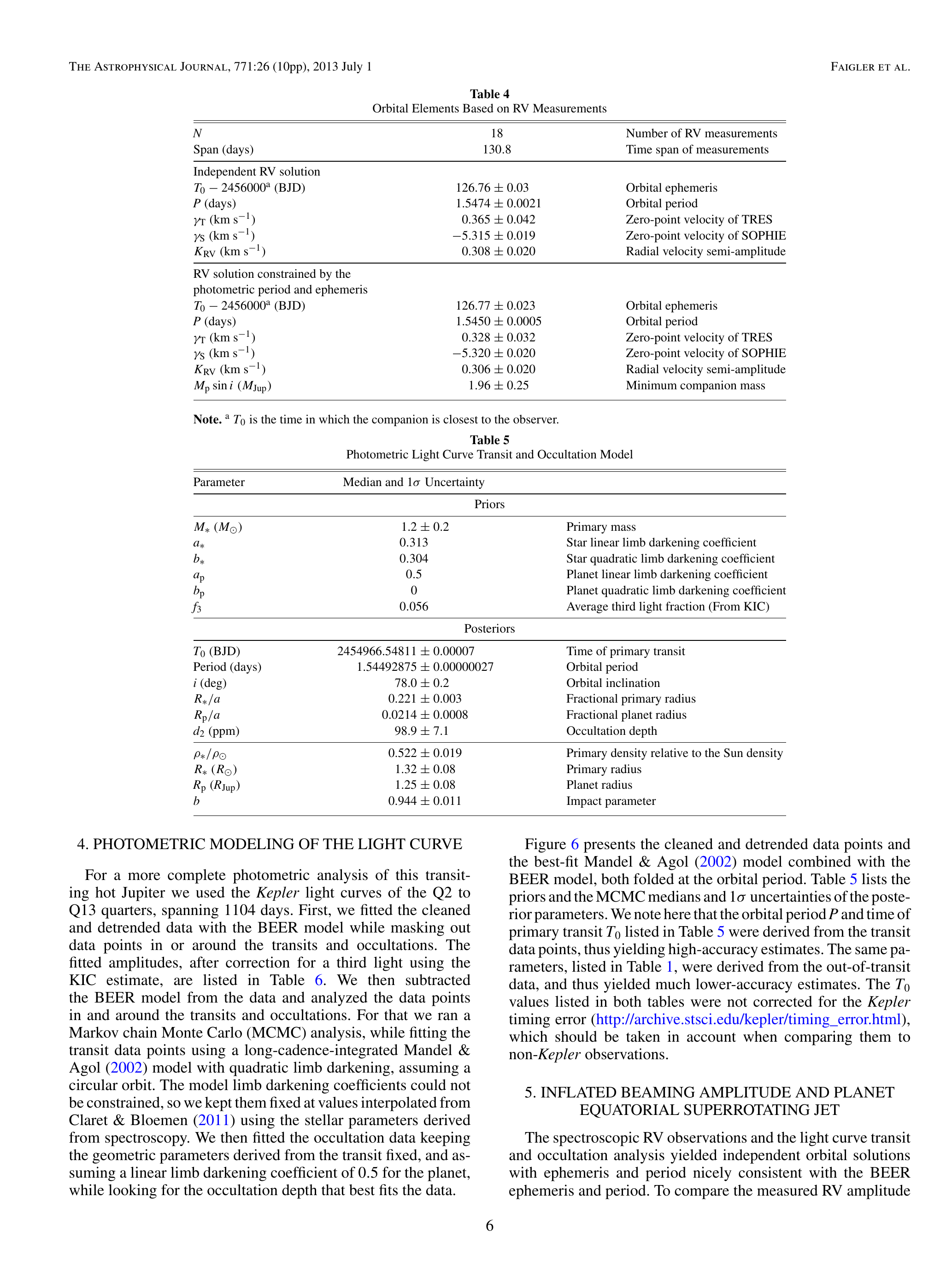,height=10in}}
\centerline{\psfig{figure=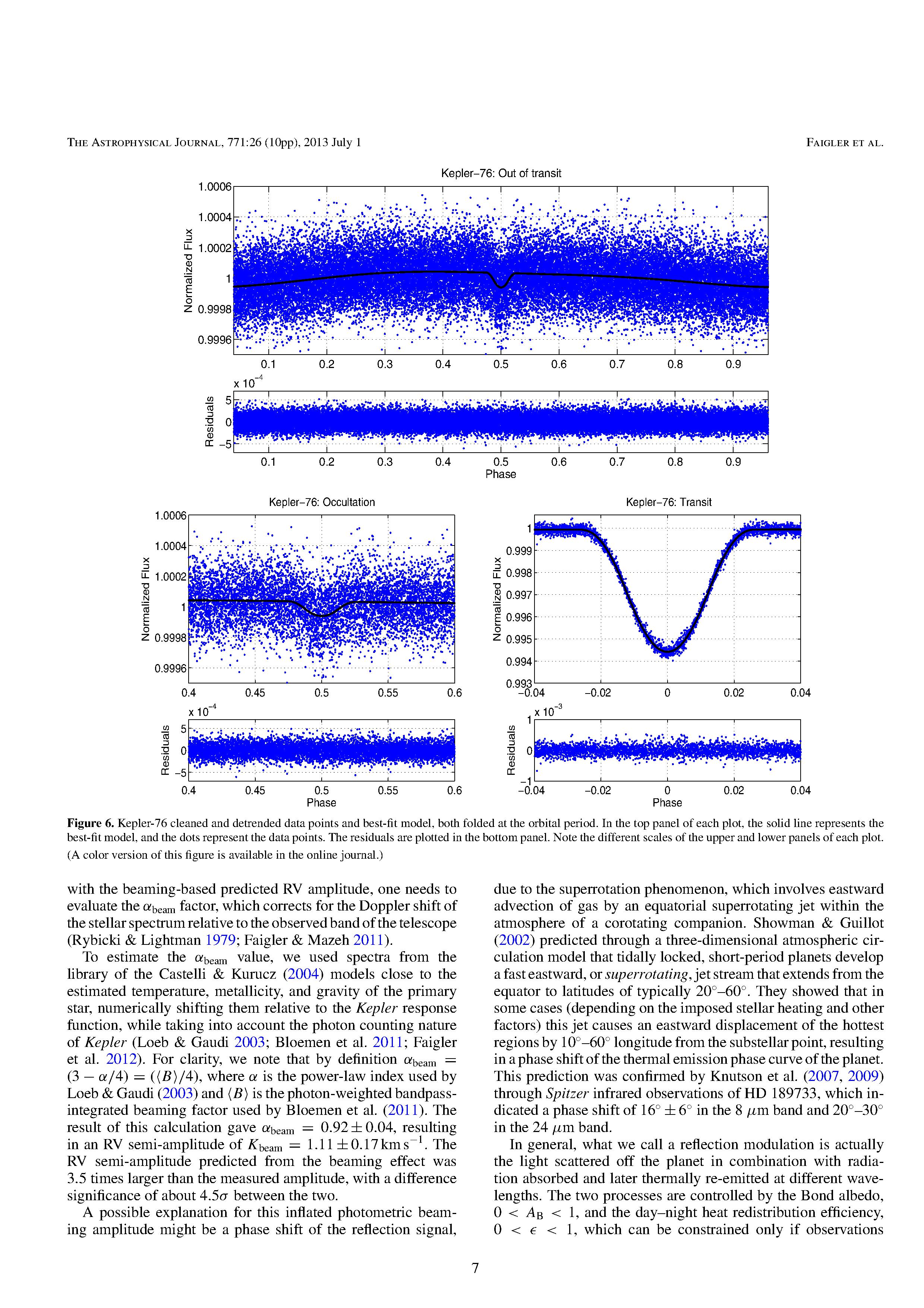,height=10in}}
\centerline{\psfig{figure=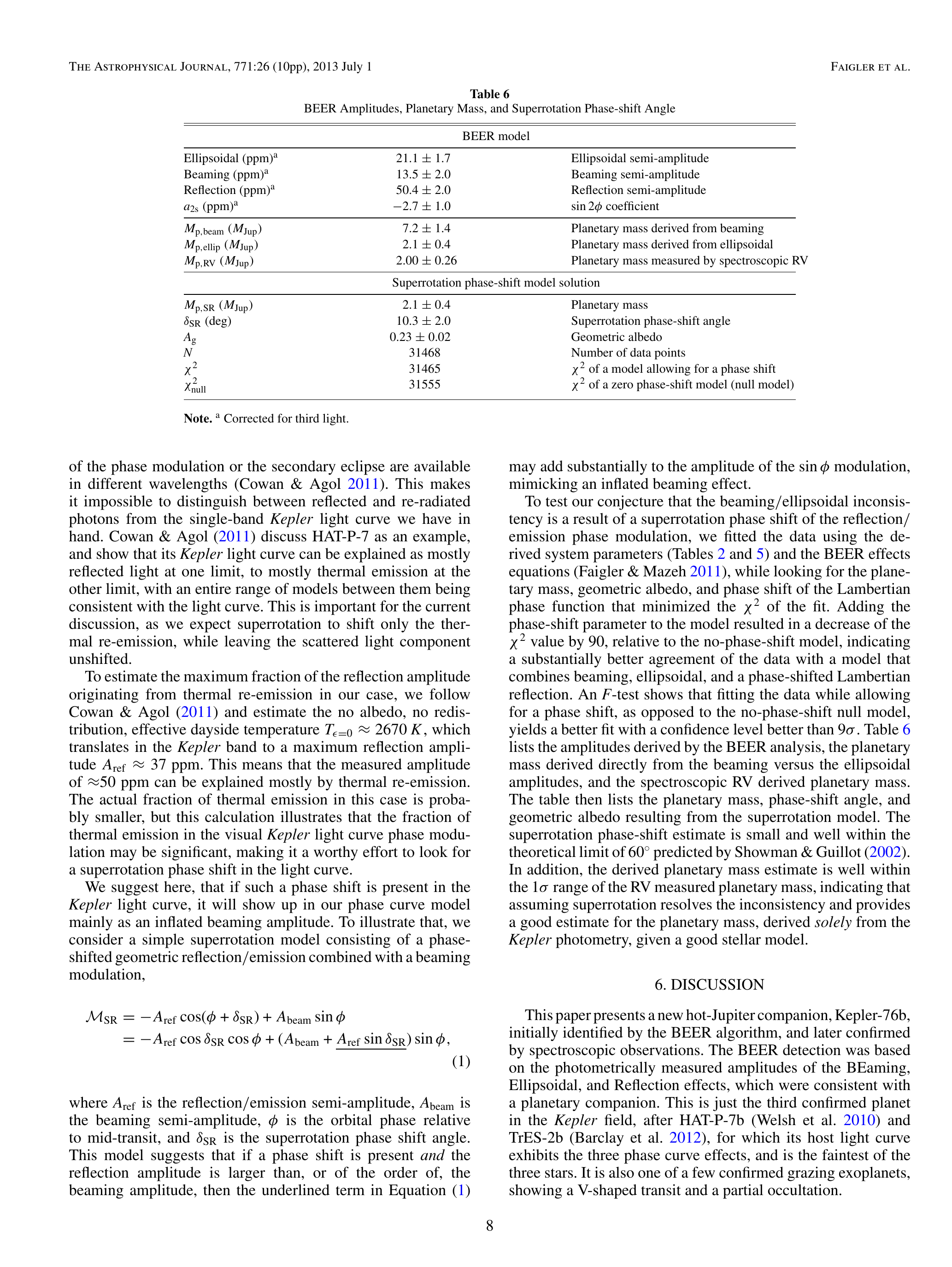,height=10in}}
\centerline{\psfig{figure=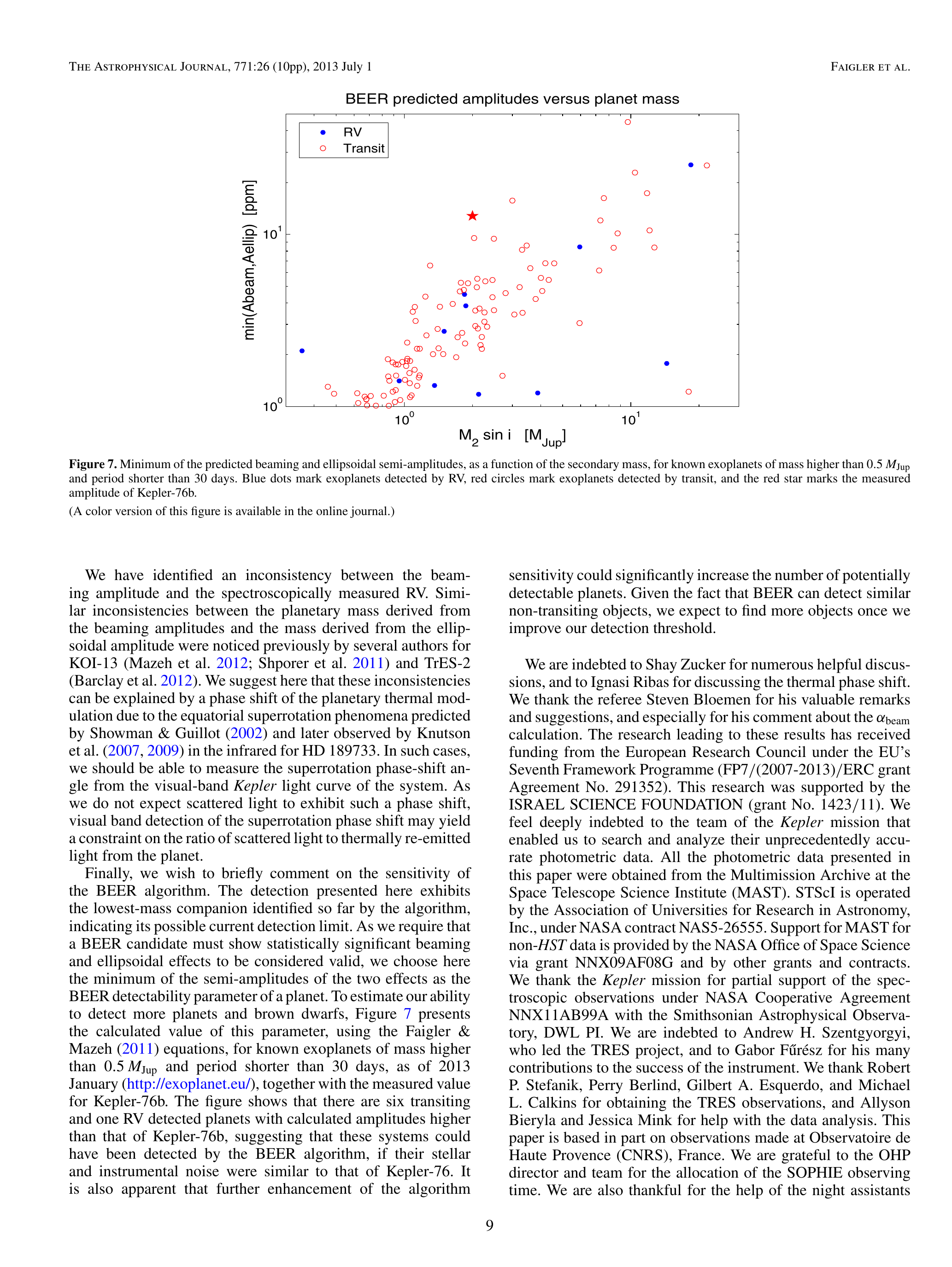,height=10in}}
\centerline{\psfig{figure=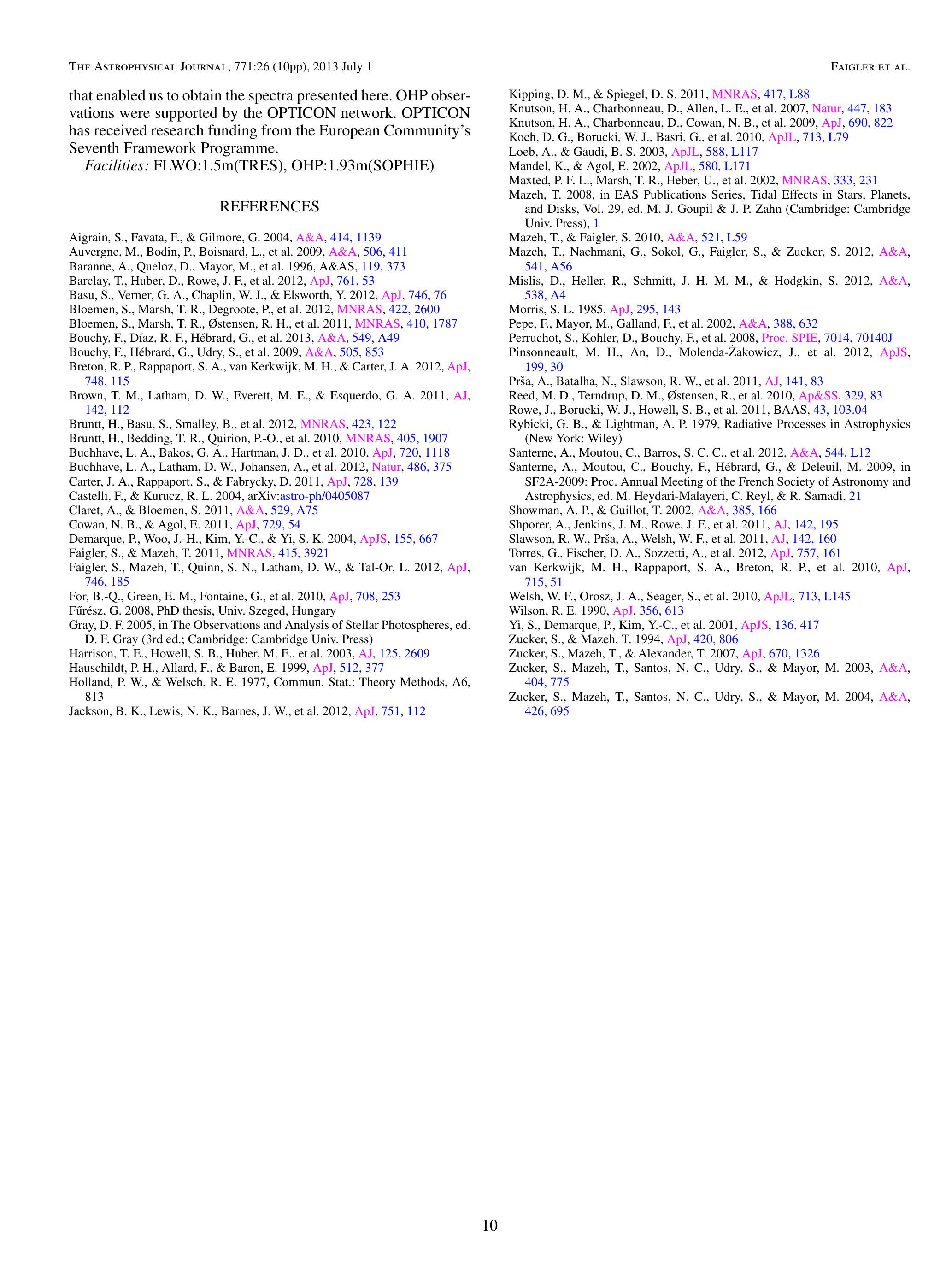,height=10in}}

\addcontentsline{toc}{section}{Paper V}
\centerline{\psfig{figure=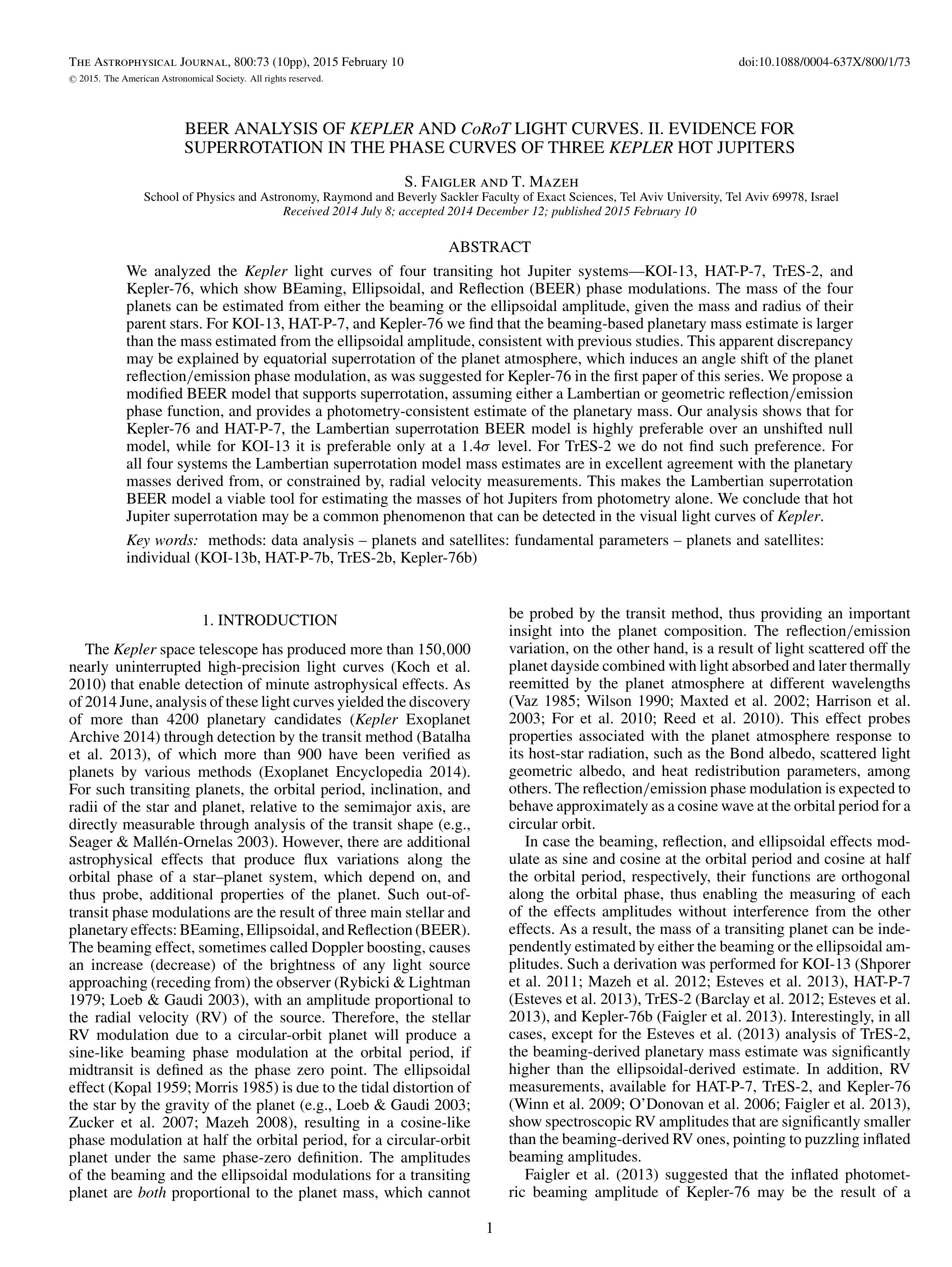,height=10in}}
\chead{Paper V -- Hot-Jupiters Superrotation}
\addtolength{\headsep}{0.0cm}
\centerline{\psfig{figure=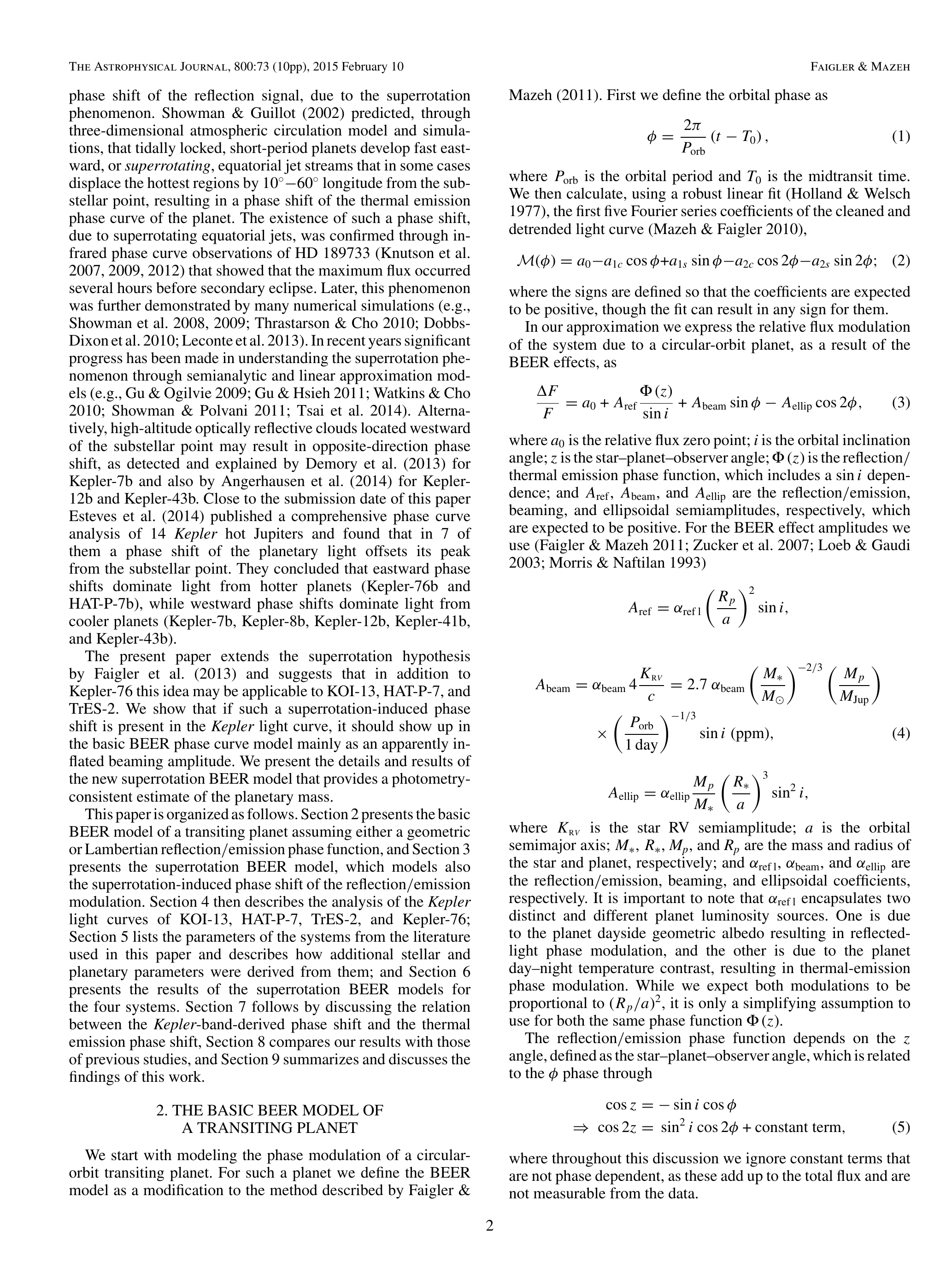,height=10in}}
\centerline{\psfig{figure=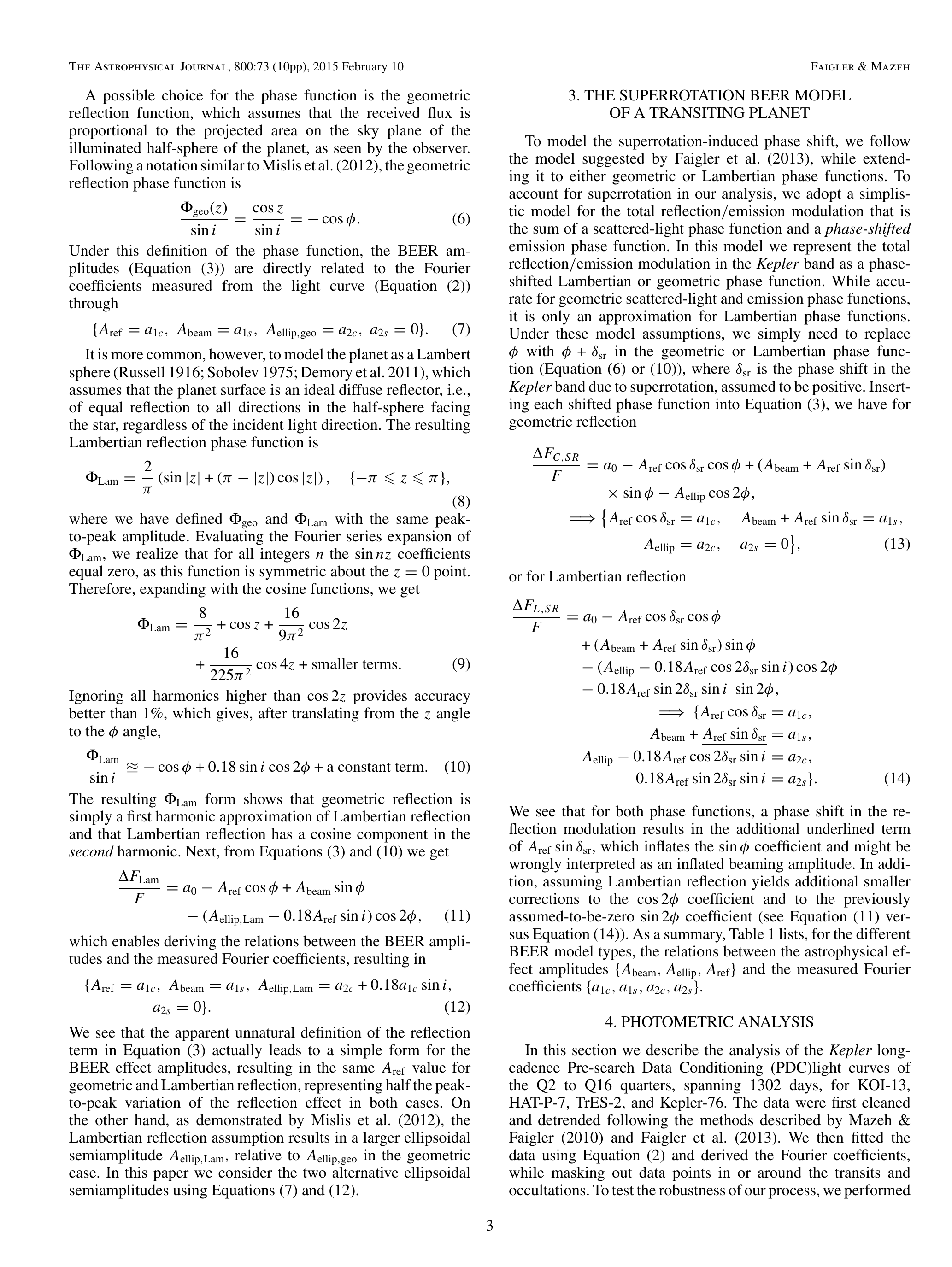,height=10in}}
\centerline{\psfig{figure=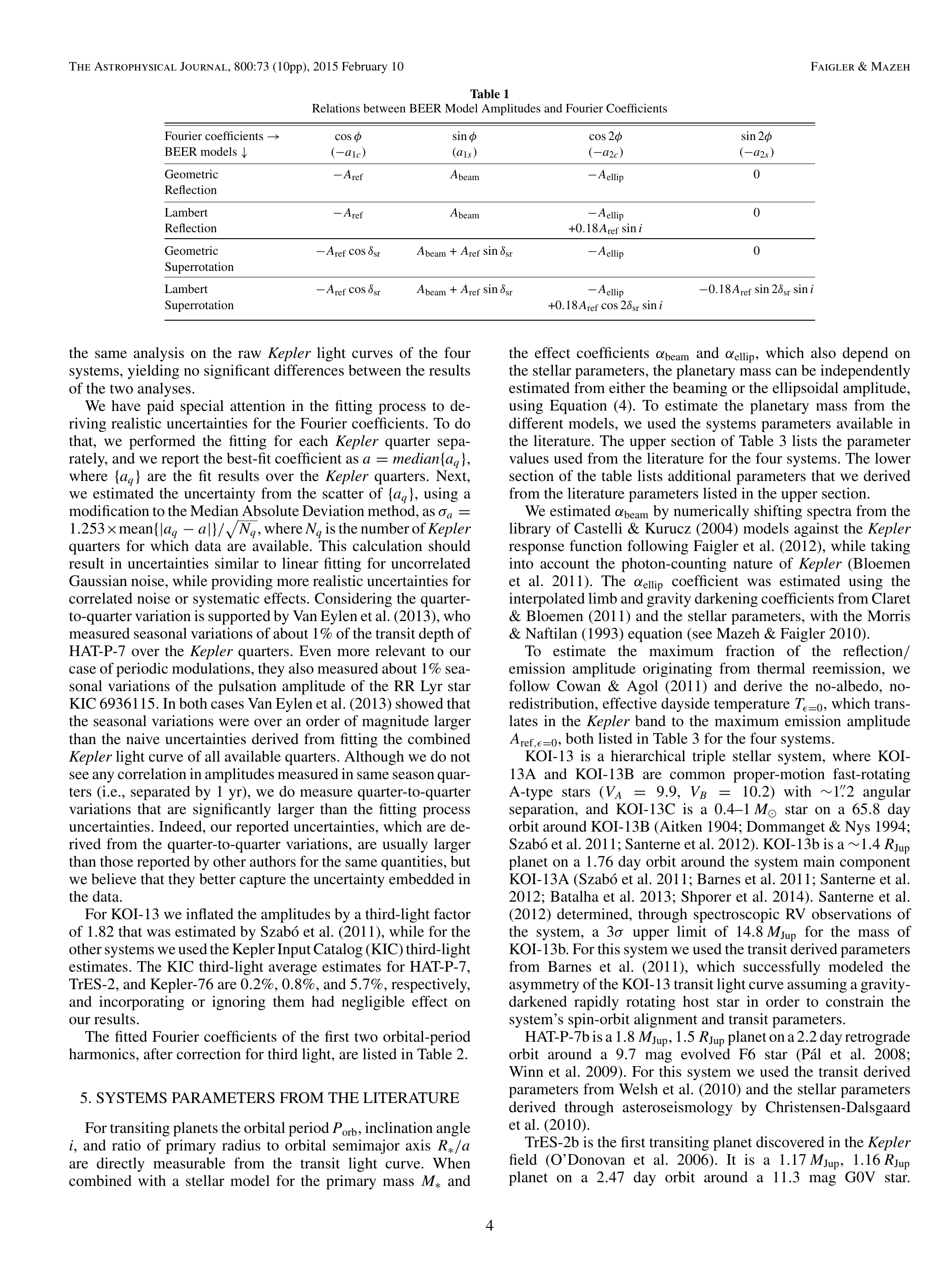,height=10in}}
\centerline{\psfig{figure=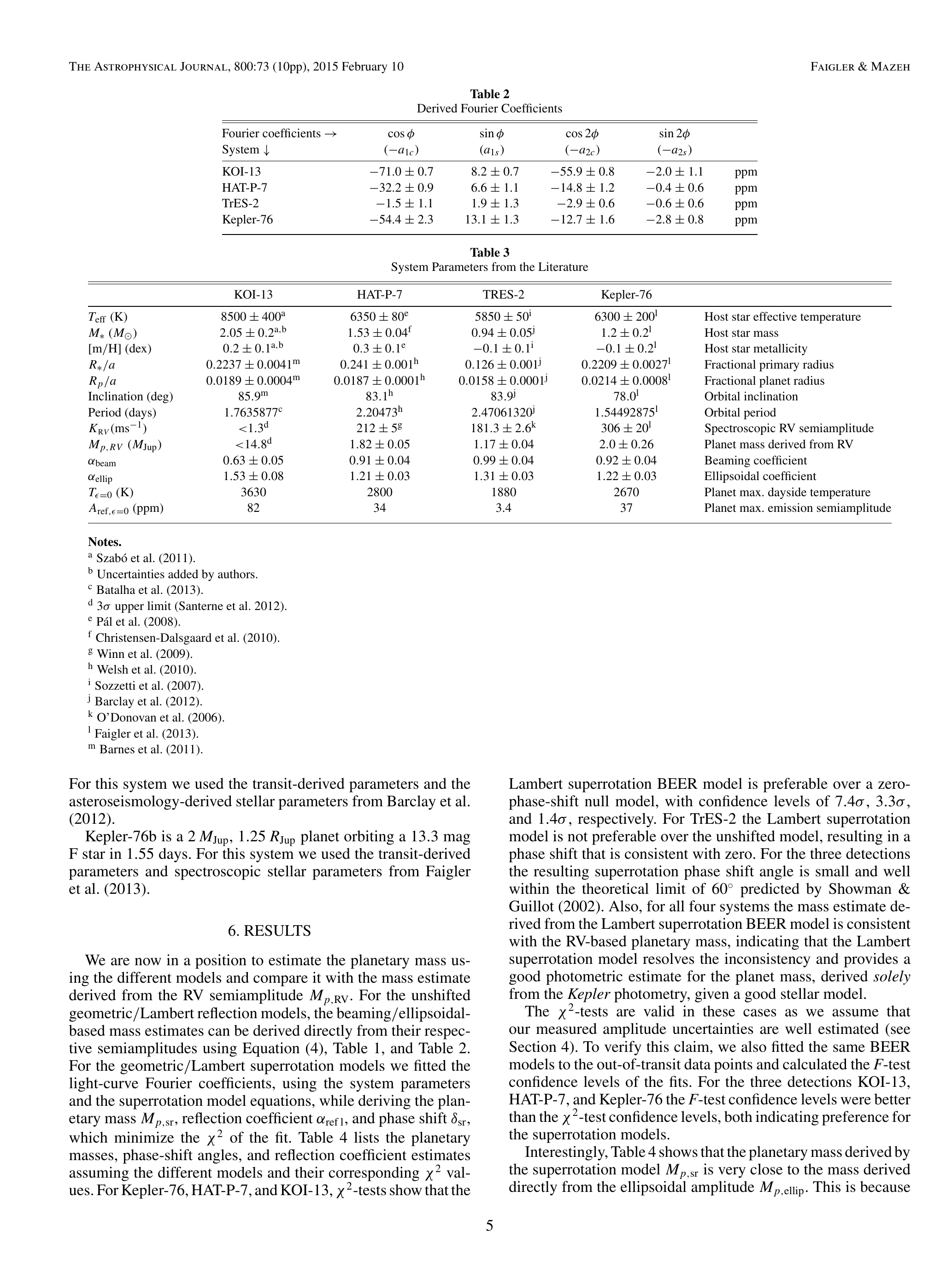,height=10in}}
\centerline{\psfig{figure=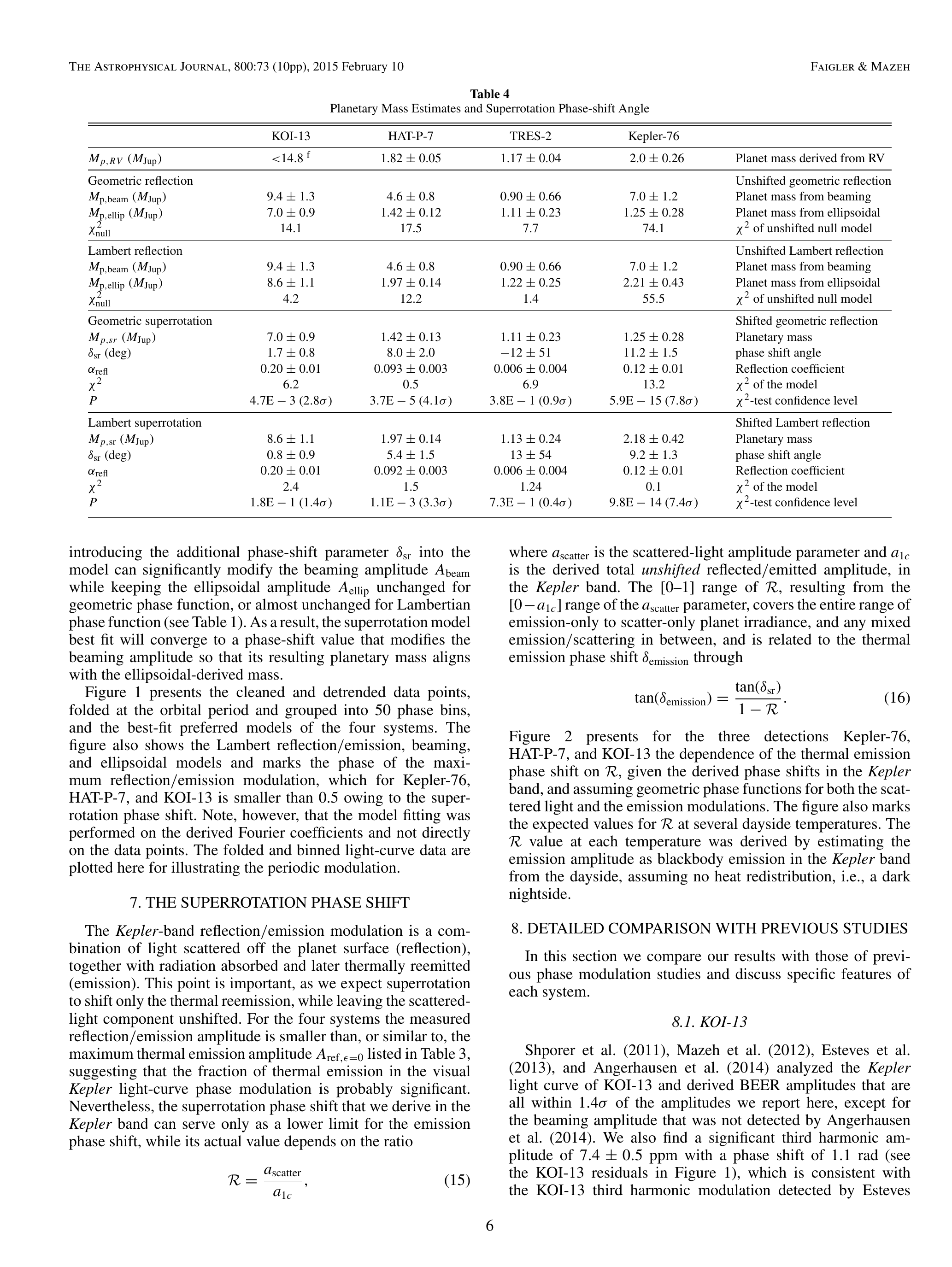,height=10in}}
\centerline{\psfig{figure=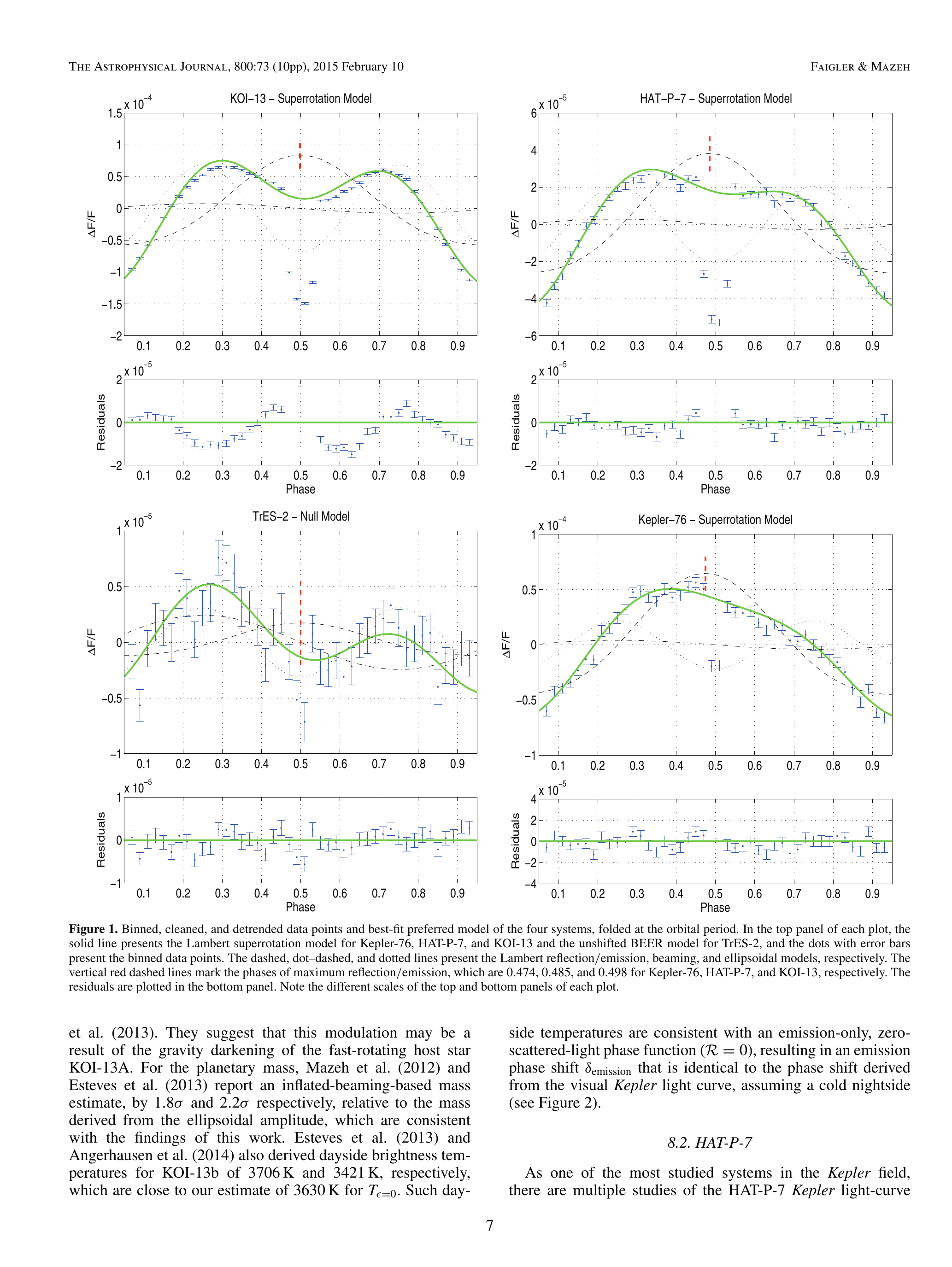,height=10in}}
\centerline{\psfig{figure=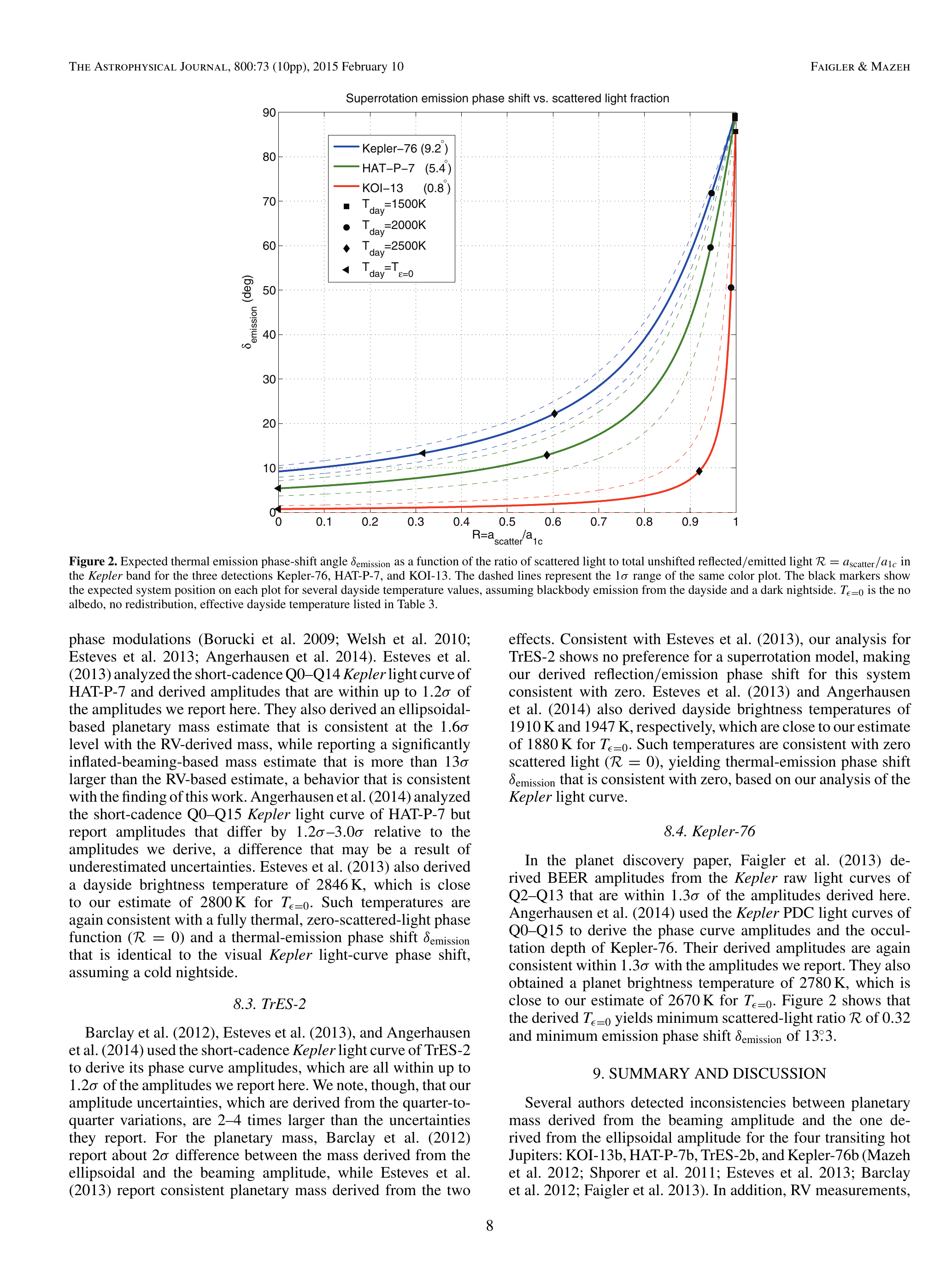,height=10in}}
\centerline{\psfig{figure=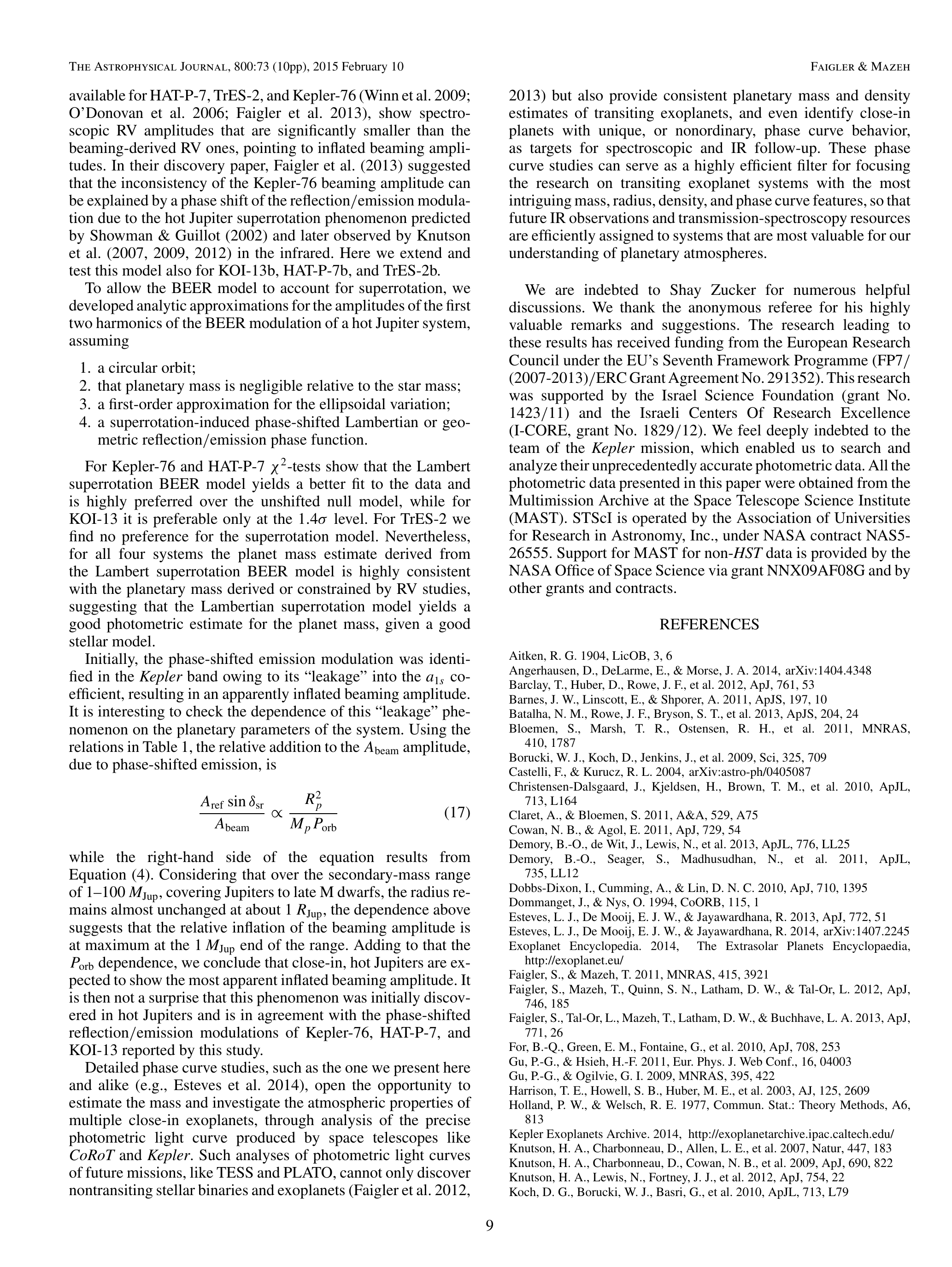,height=10in}}
\centerline{\psfig{figure=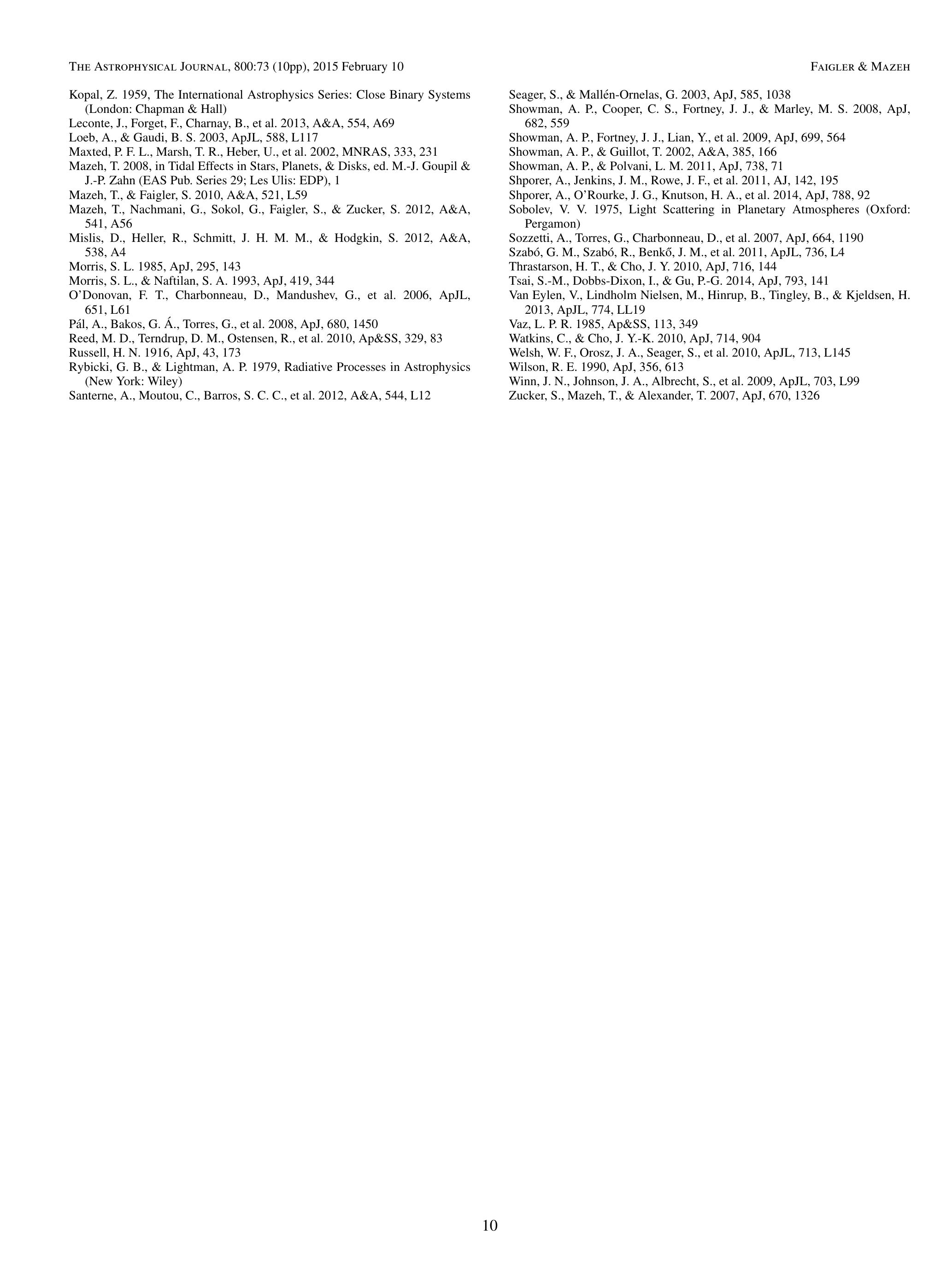,height=10in}}

\addcontentsline{toc}{section}{Paper VI}
\centerline{\psfig{figure=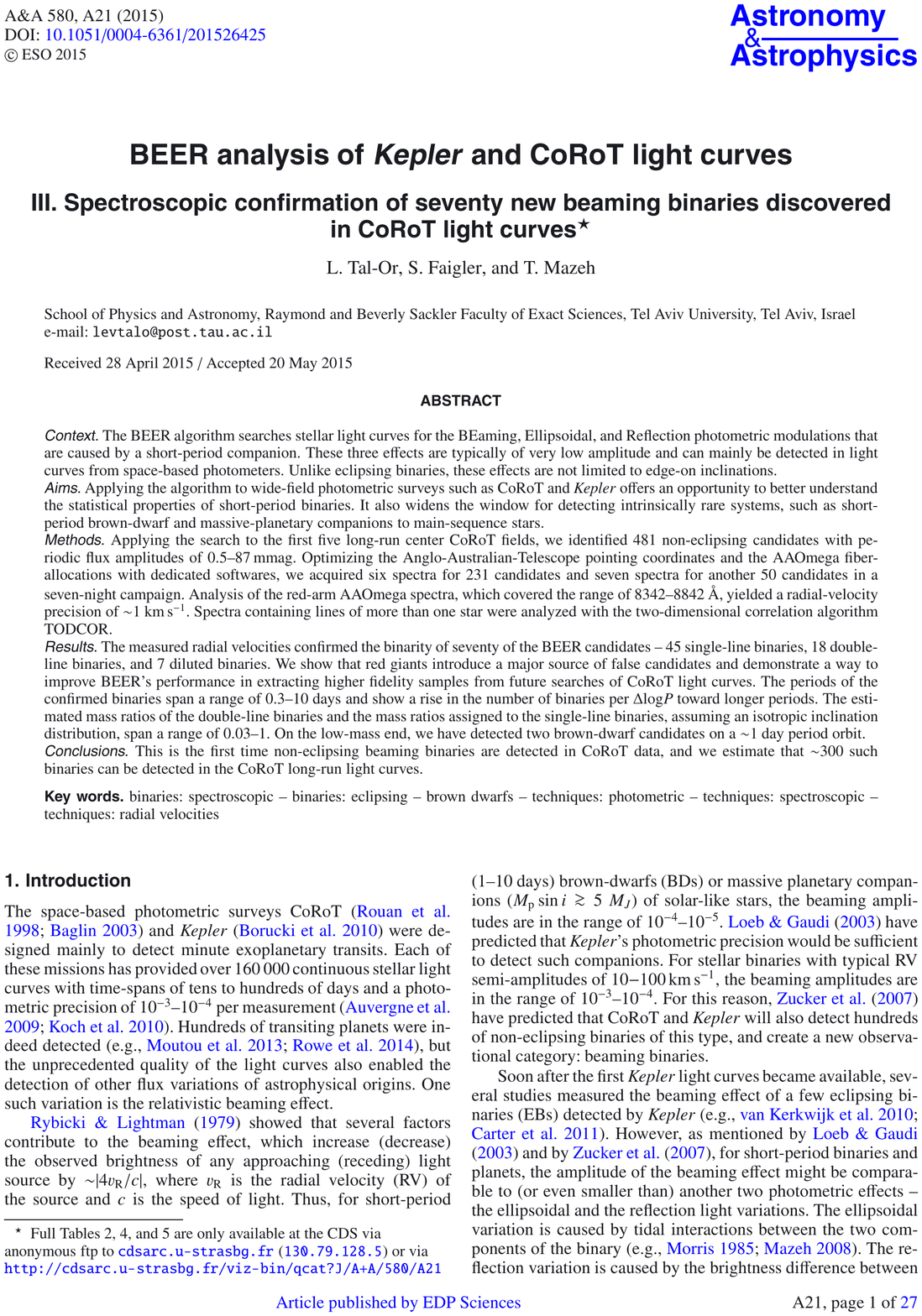,height=10.5in}}
\chead{Paper VI -- Seventy {\it CoRoT} BEER binaries}
\addtolength{\headsep}{-1.0cm}
\centerline{\psfig{figure=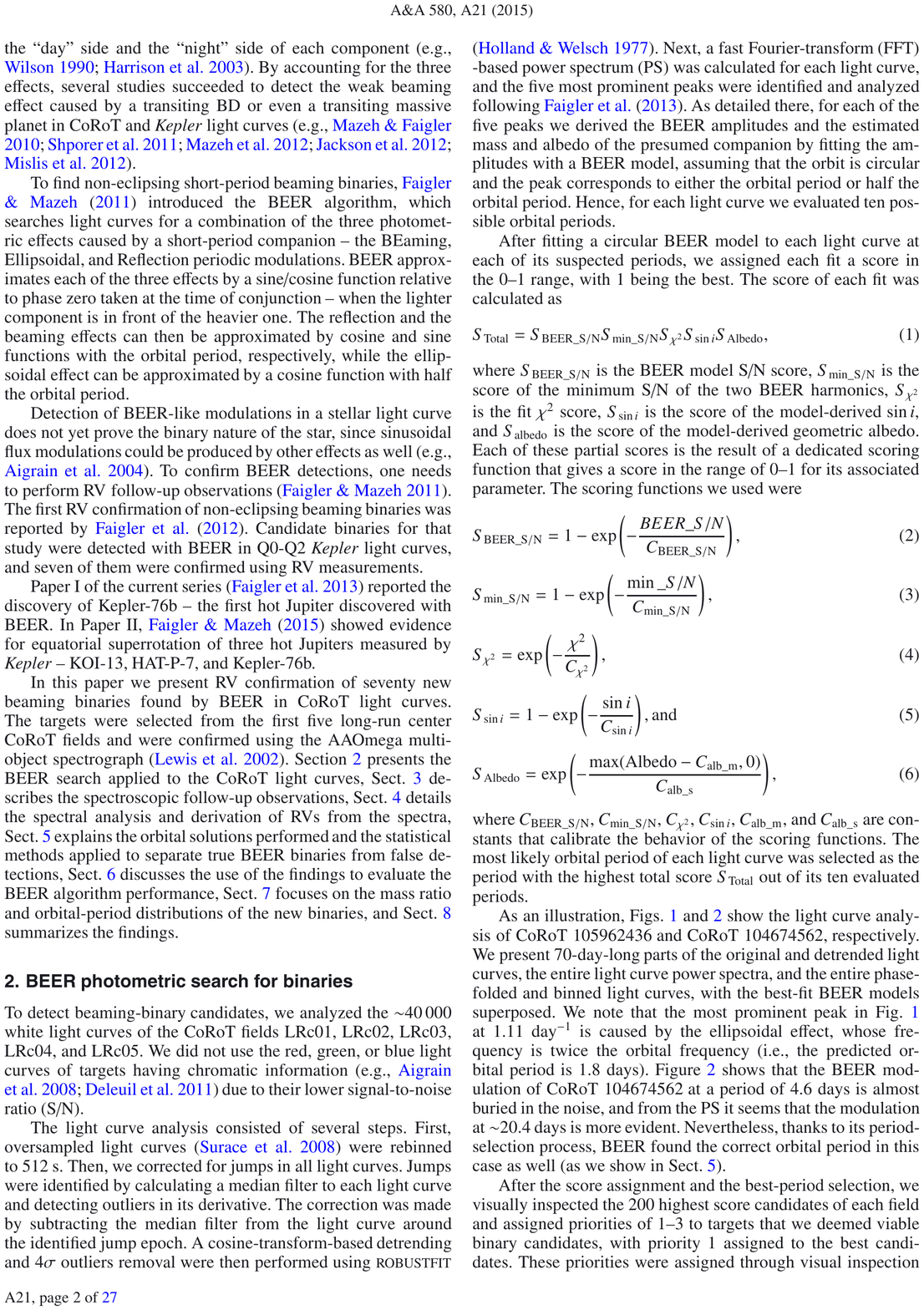,height=10.5in}}
\centerline{\psfig{figure=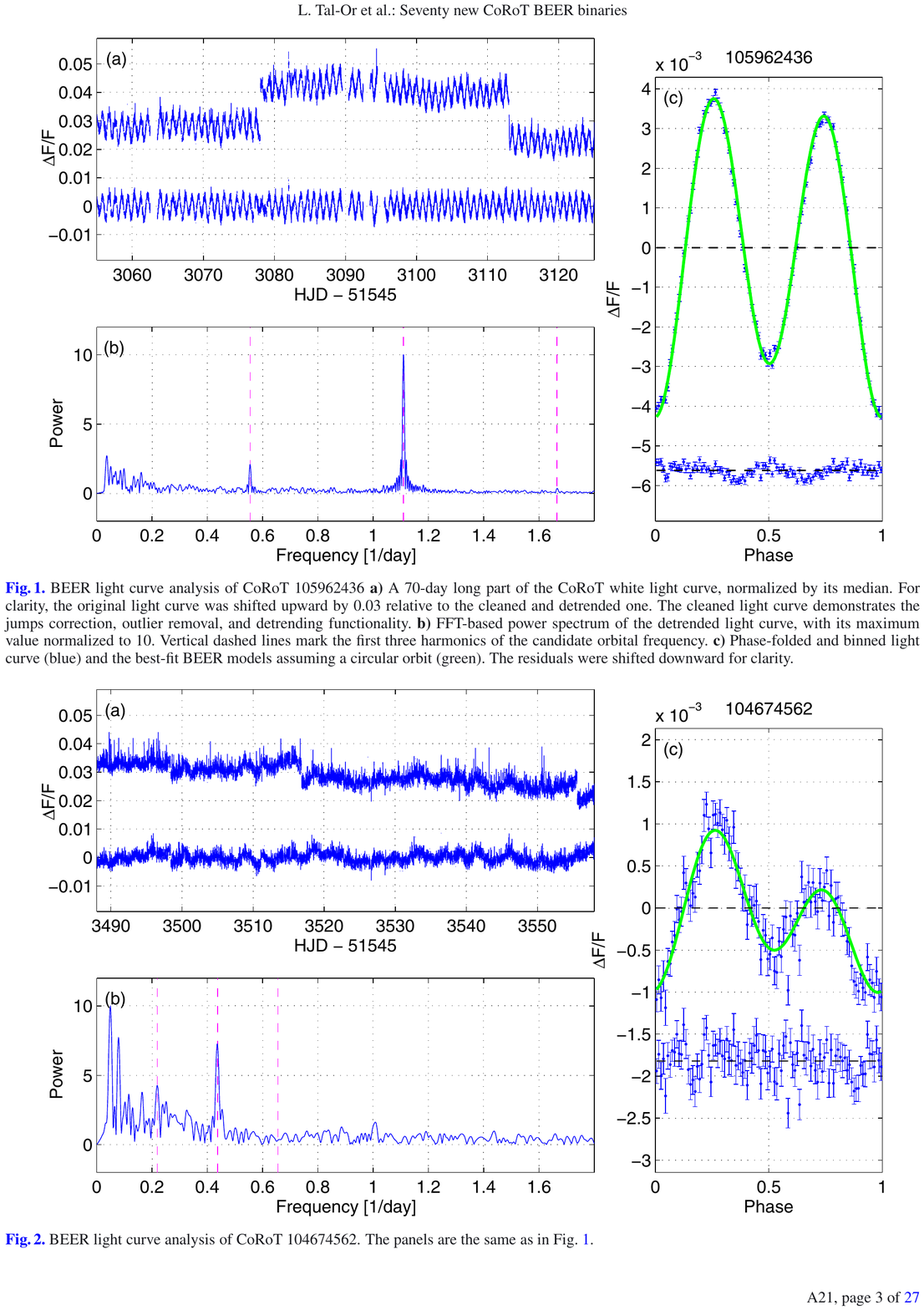,height=10.5in}}
\centerline{\psfig{figure=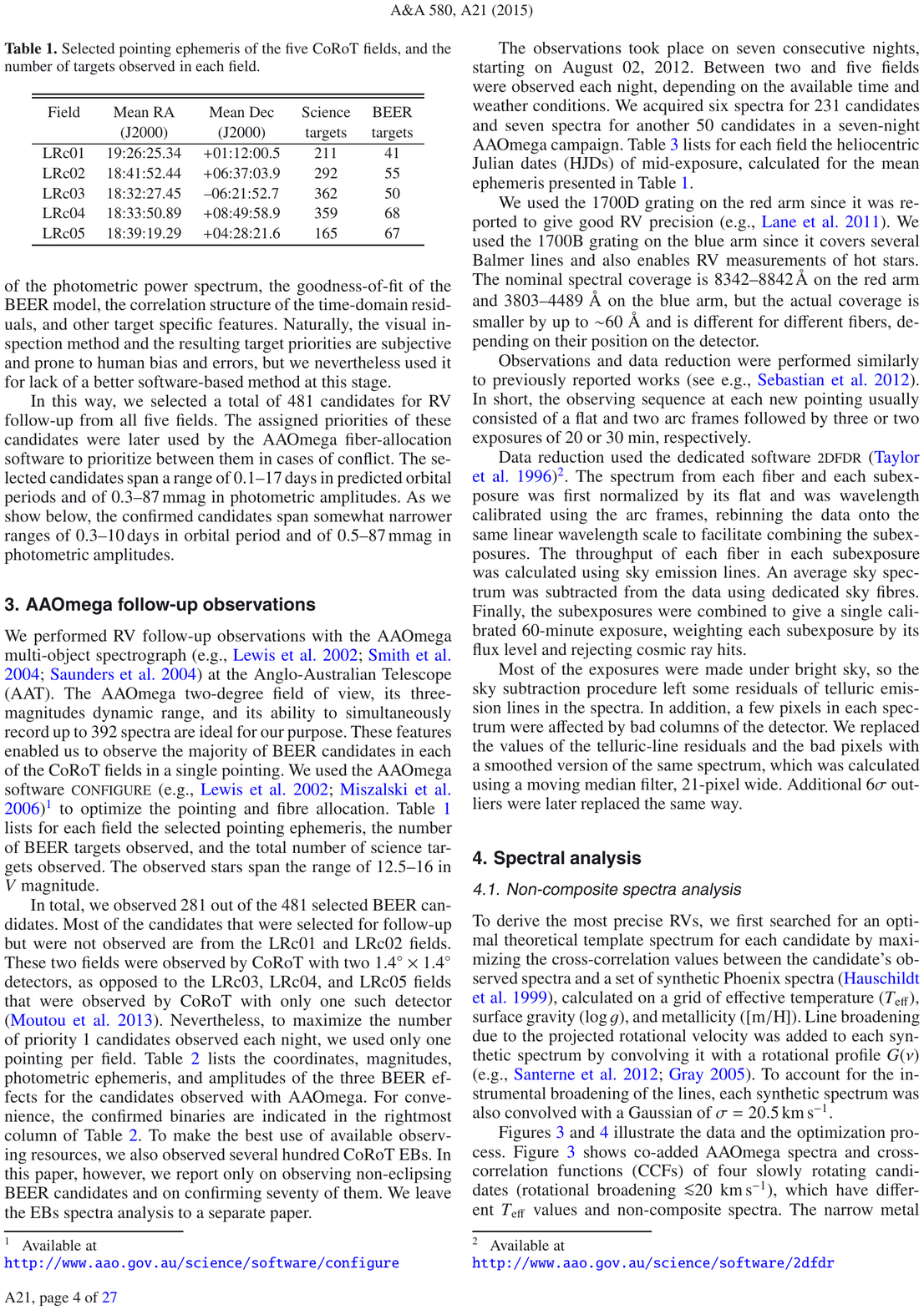,height=10.5in}}
\centerline{\psfig{figure=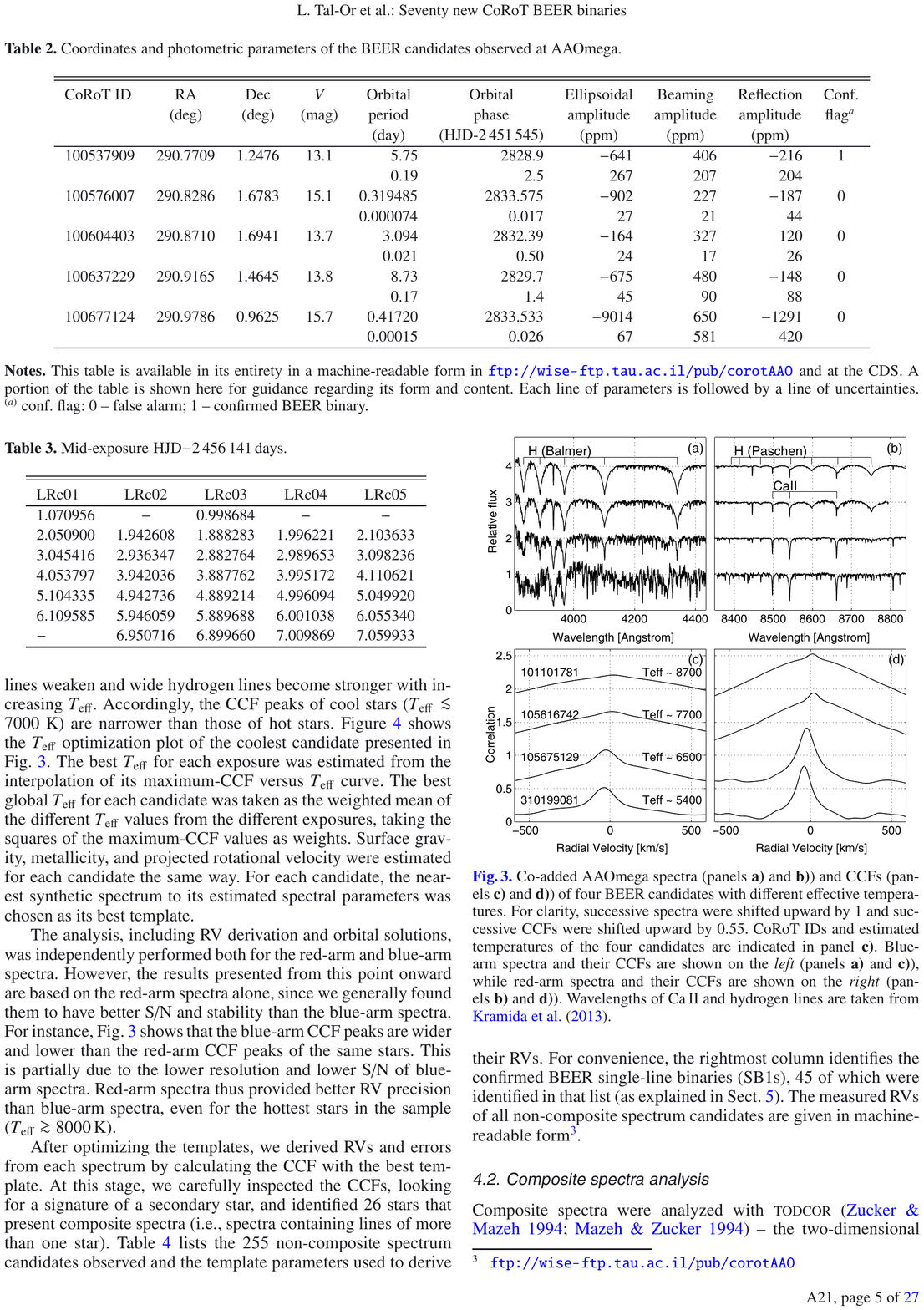,height=10.5in}}
\centerline{\psfig{figure=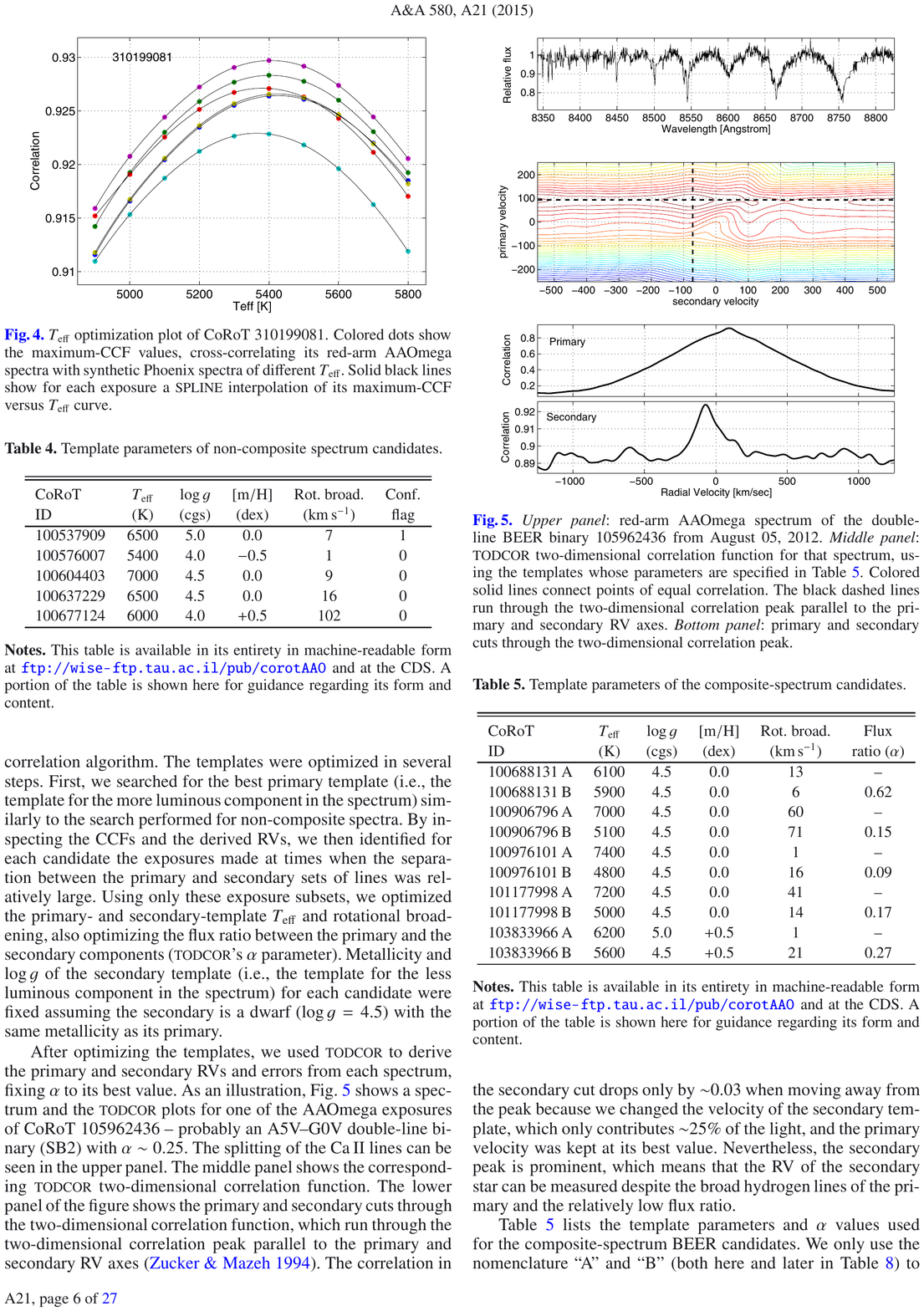,height=10.5in}}
\centerline{\psfig{figure=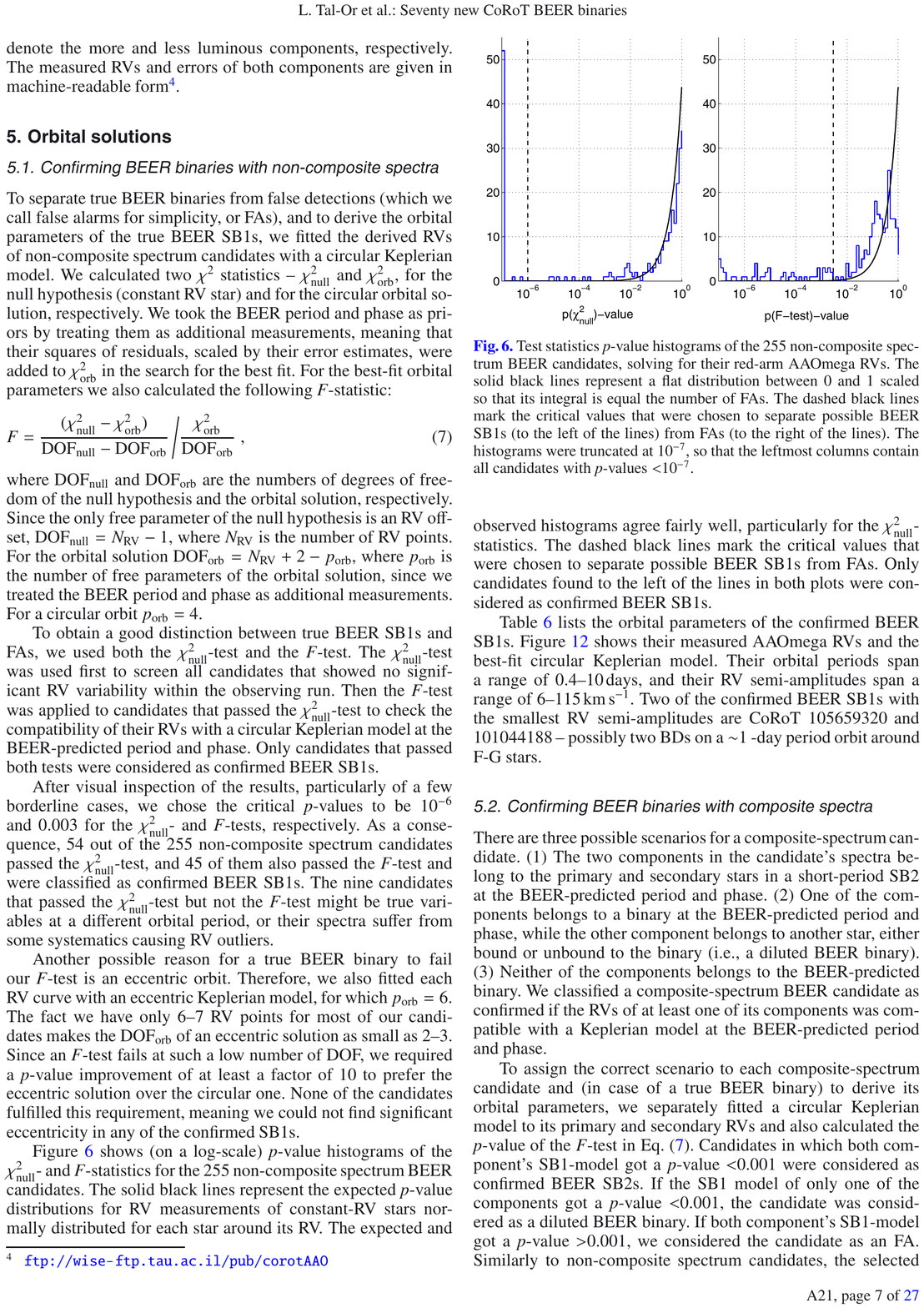,height=10.5in}}
\centerline{\psfig{figure=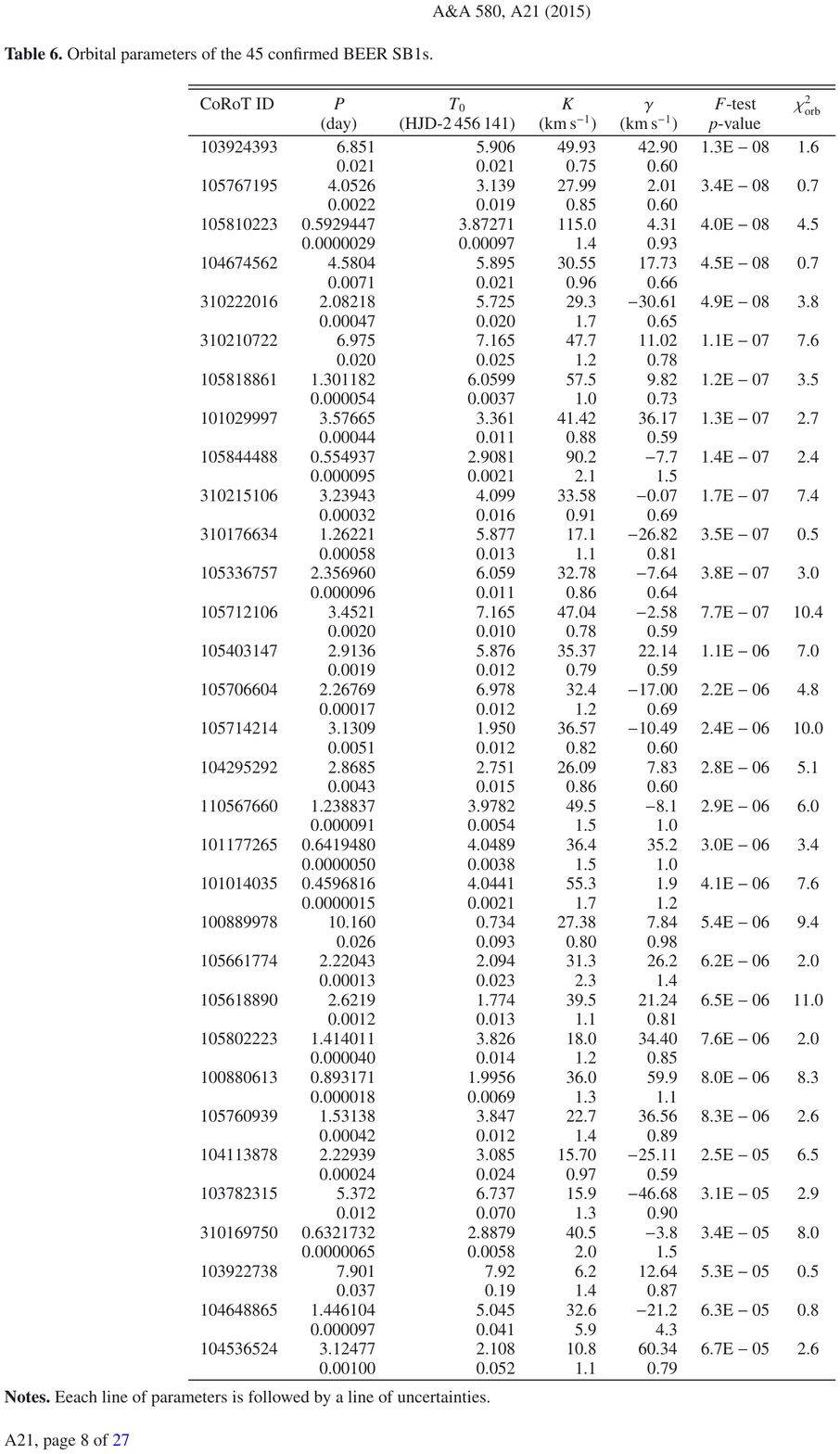,height=10.5in}}
\centerline{\psfig{figure=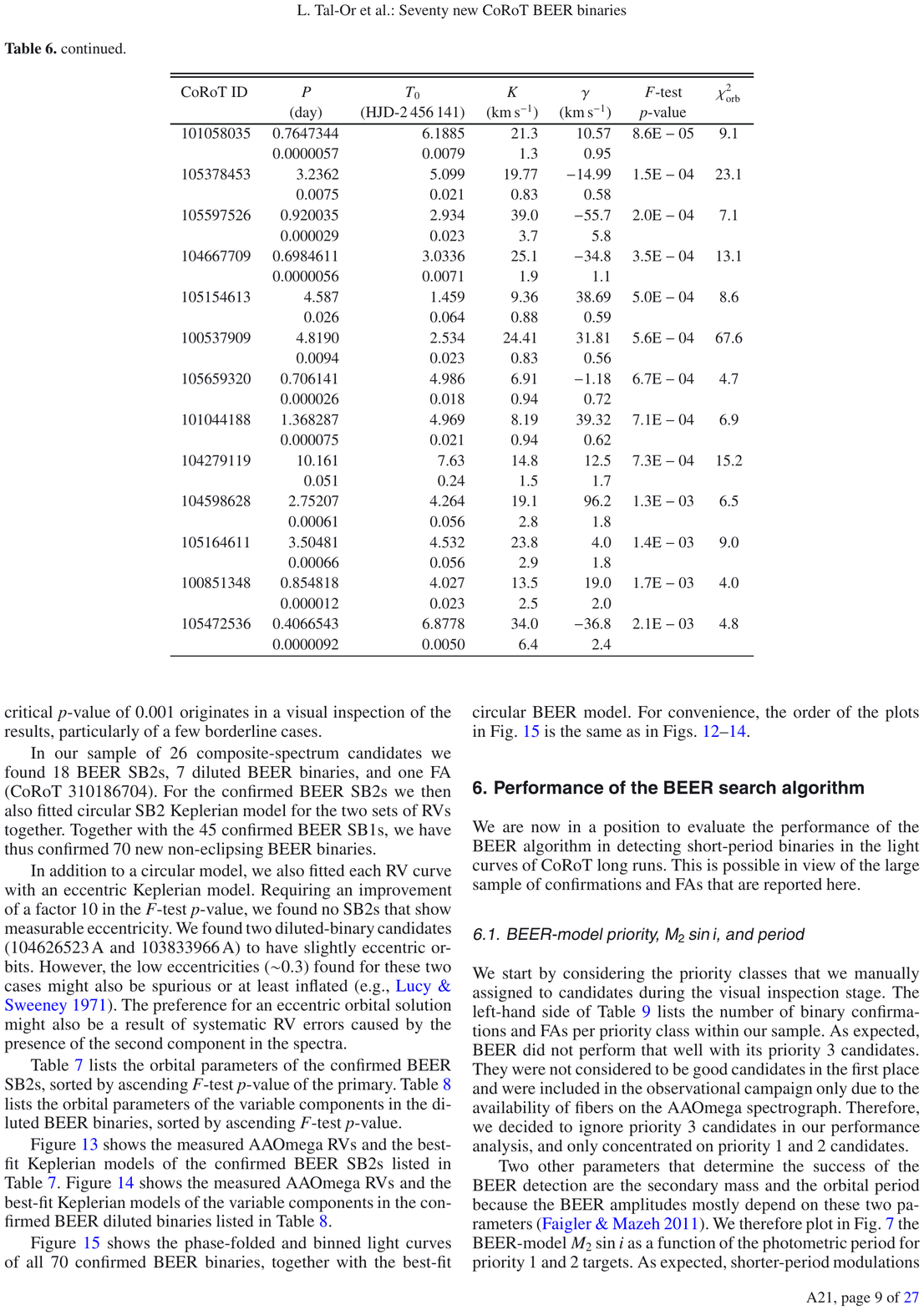,height=10.5in}}
\centerline{\psfig{figure=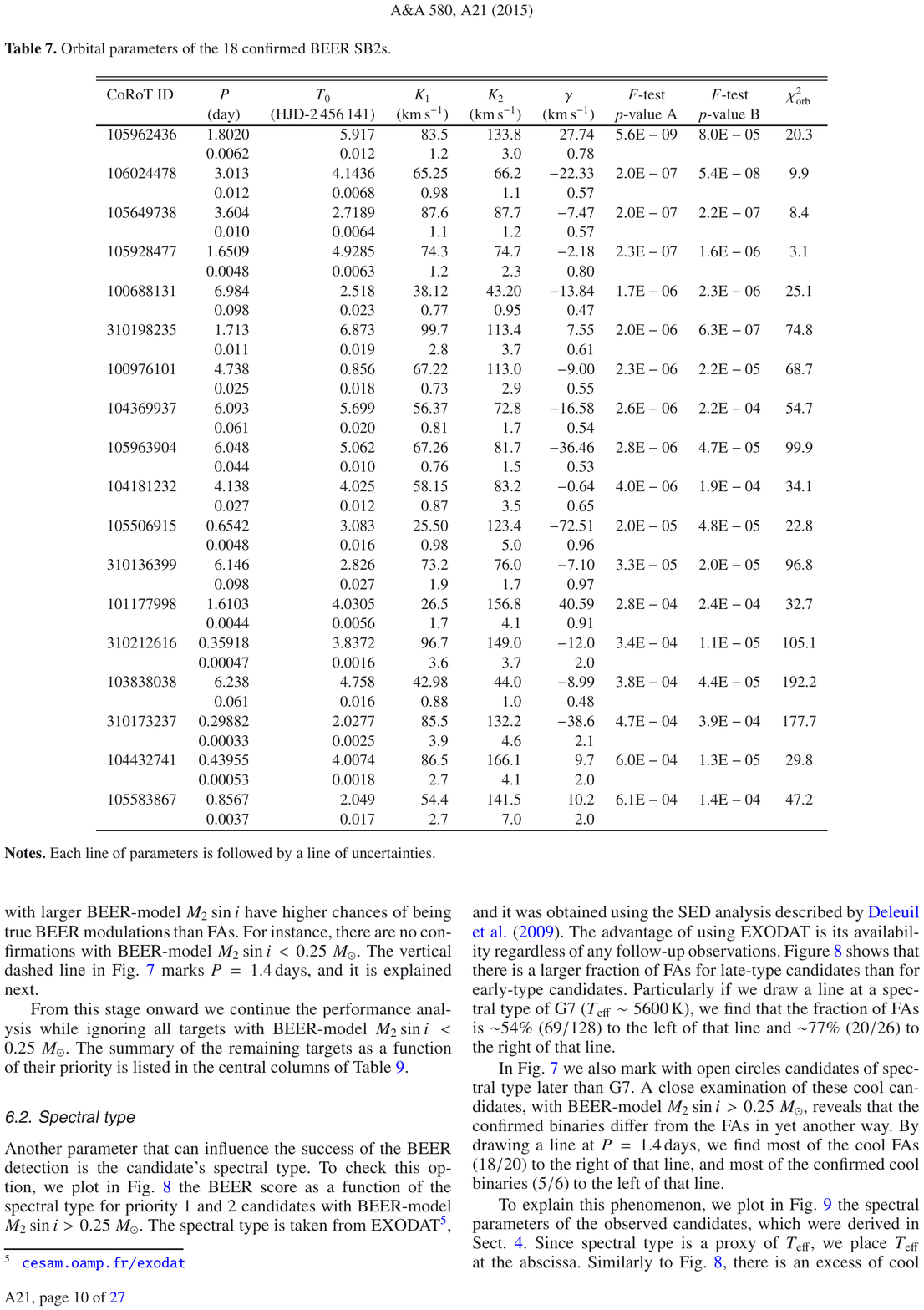,height=10.5in}}
\centerline{\psfig{figure=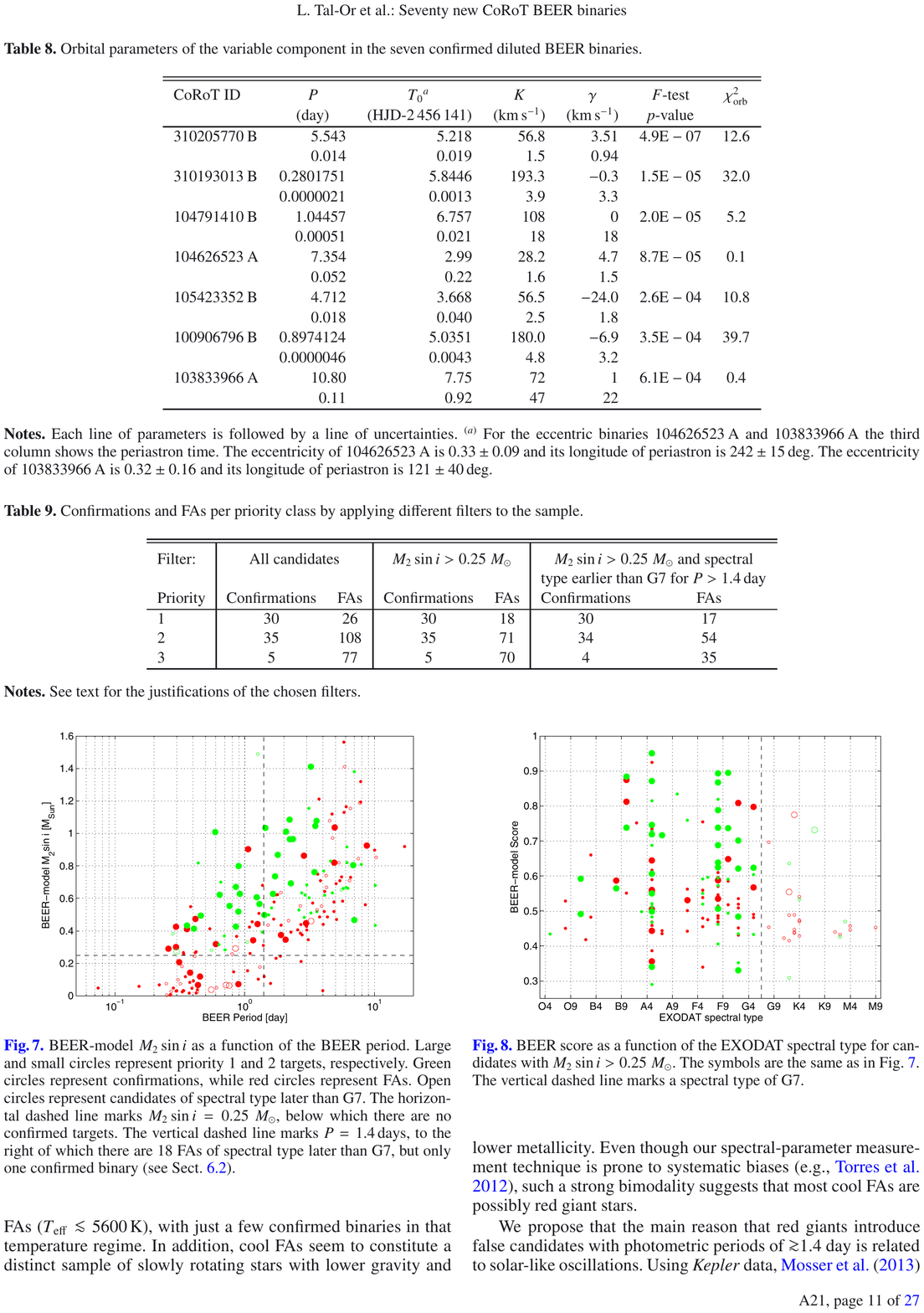,height=10.5in}}
\centerline{\psfig{figure=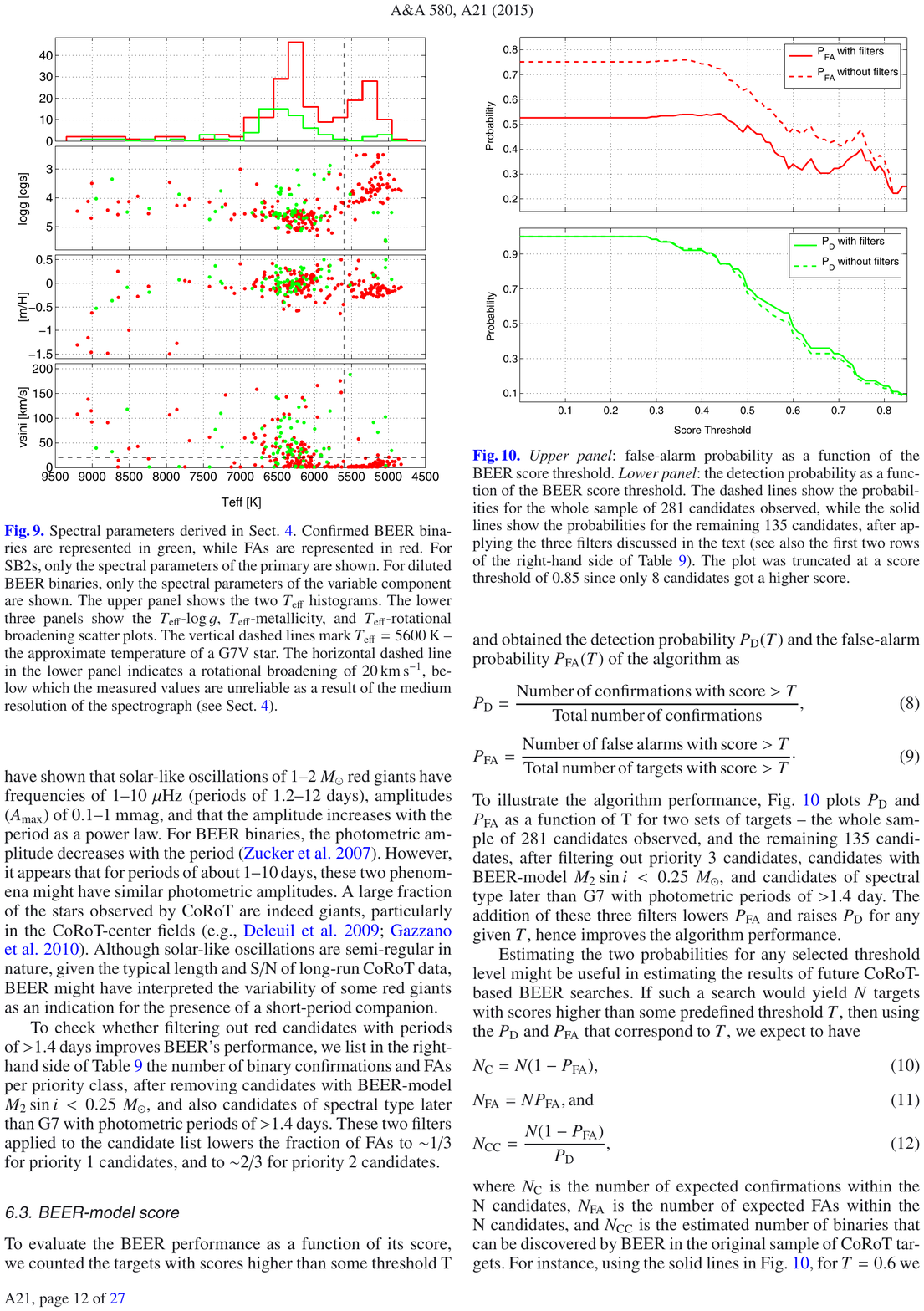,height=10.5in}}
\centerline{\psfig{figure=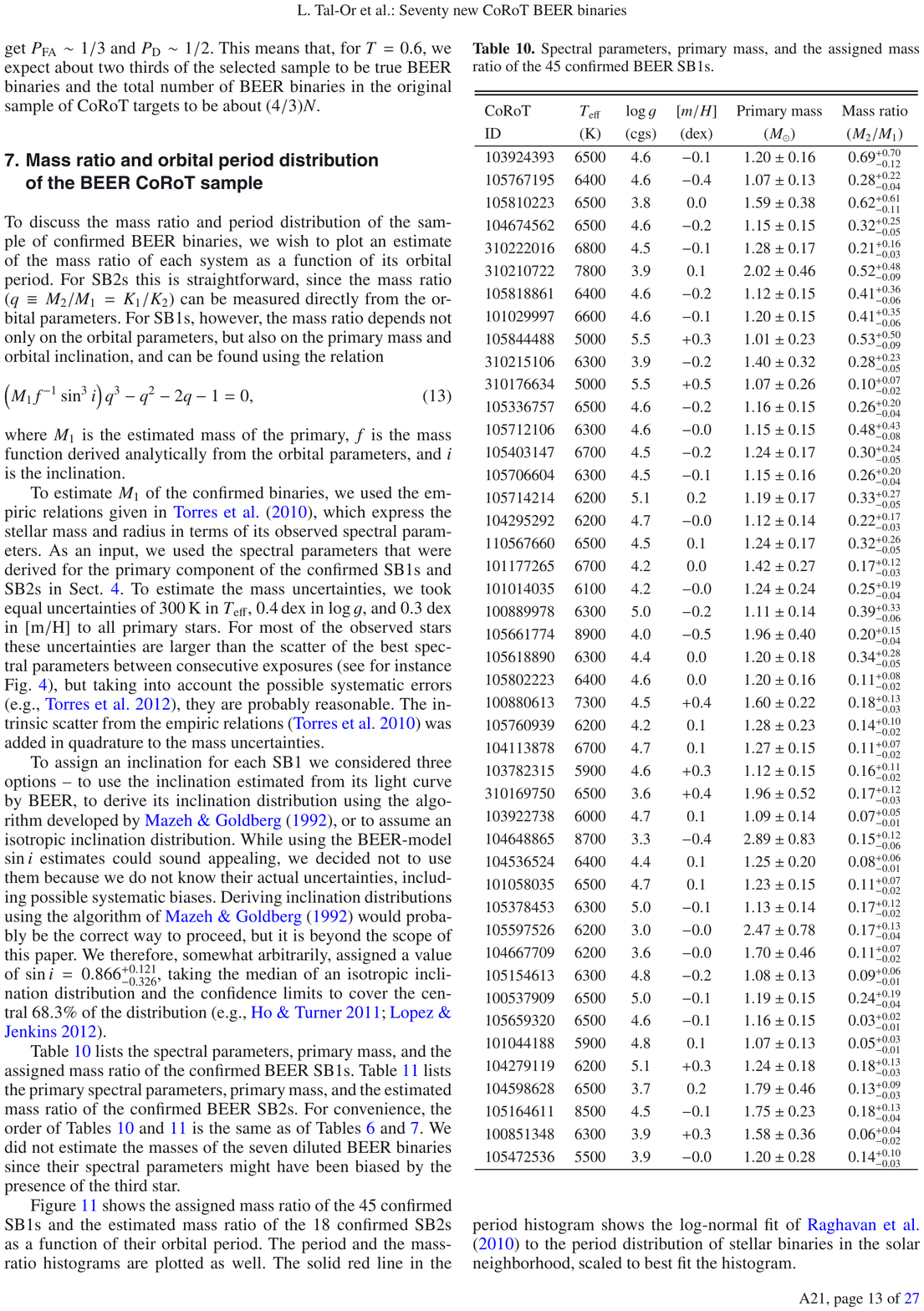,height=10.5in}}
\centerline{\psfig{figure=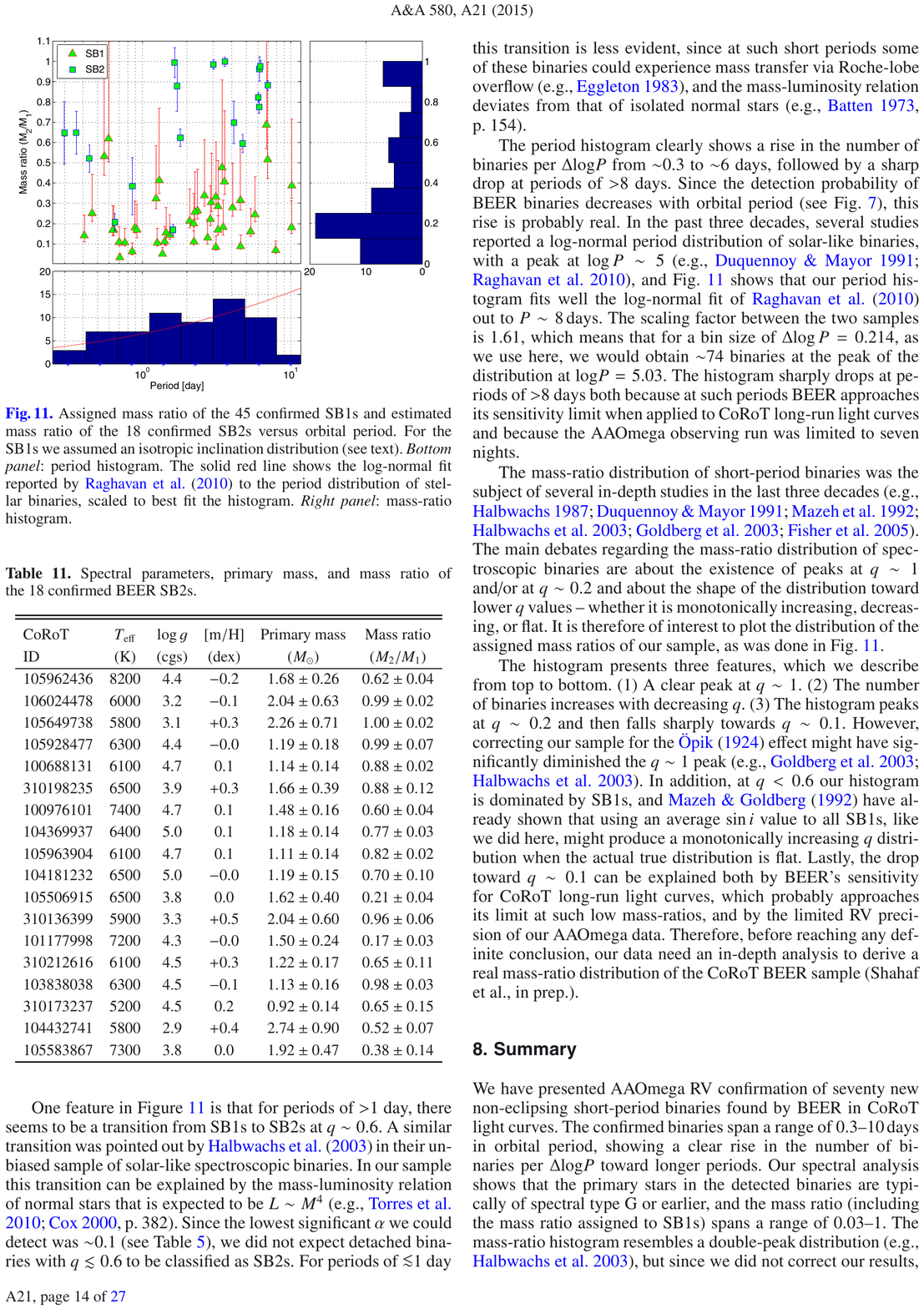,height=10.5in}}
\centerline{\psfig{figure=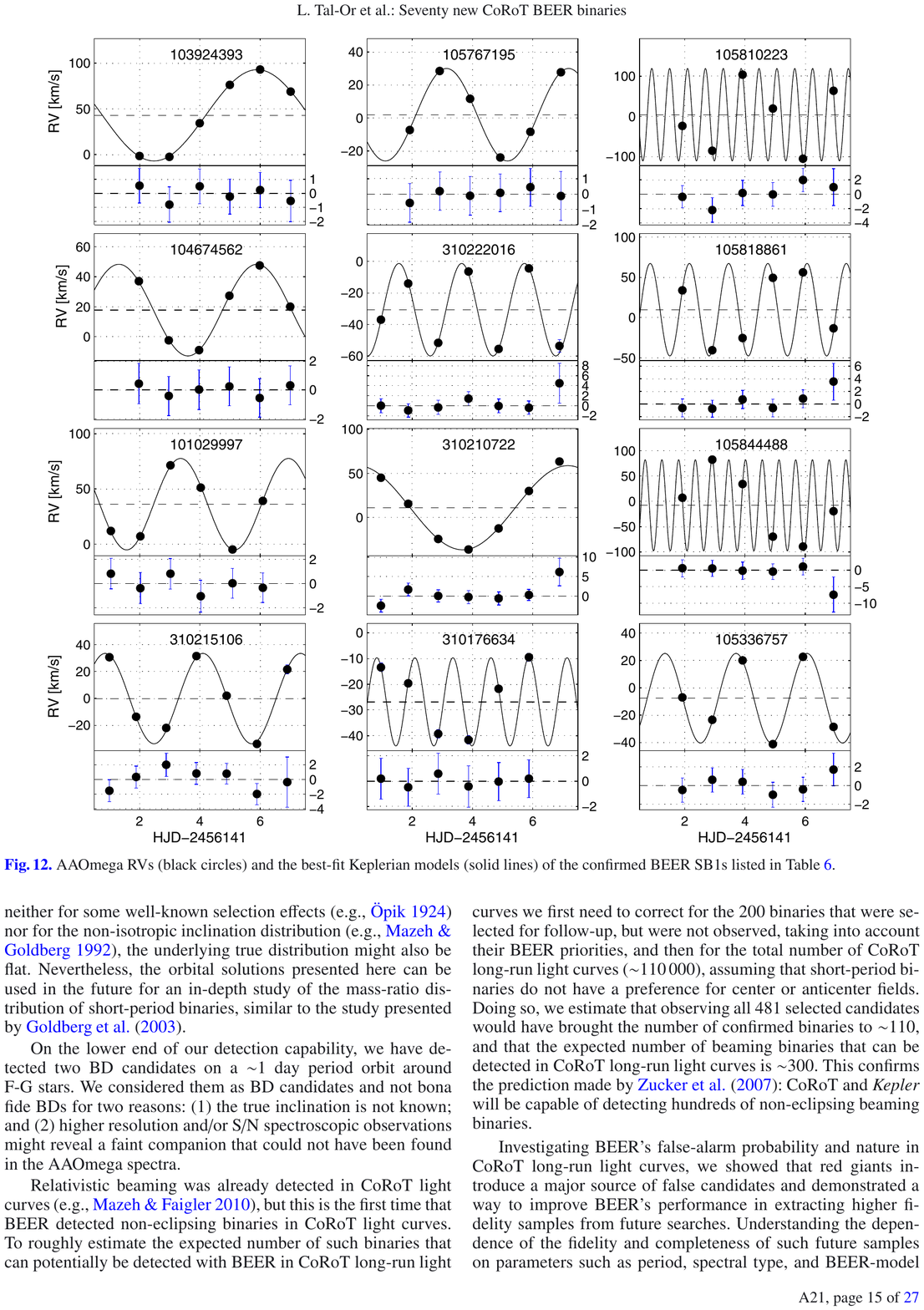,height=10.5in}}
\centerline{\psfig{figure=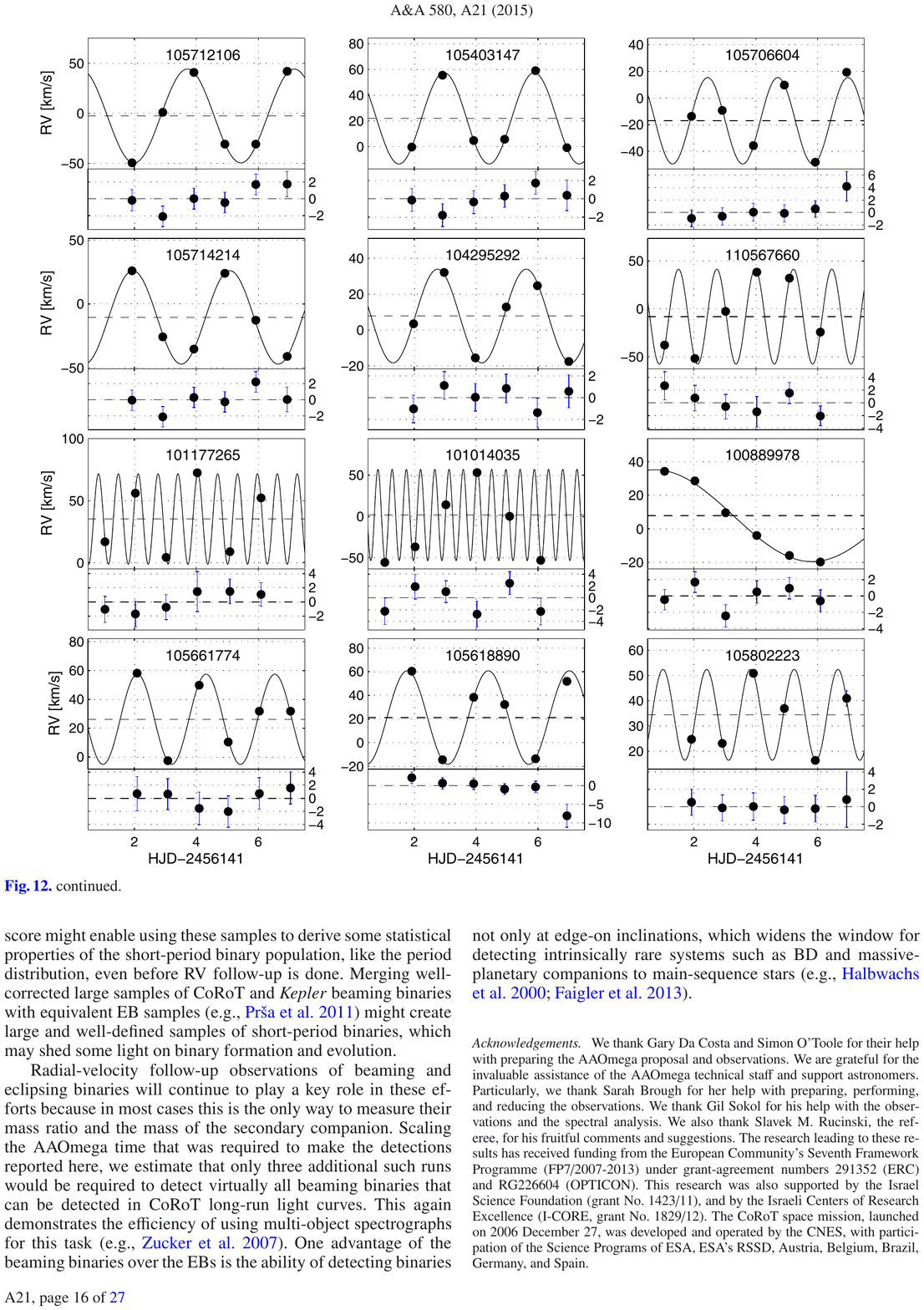,height=10.5in}}
\centerline{\psfig{figure=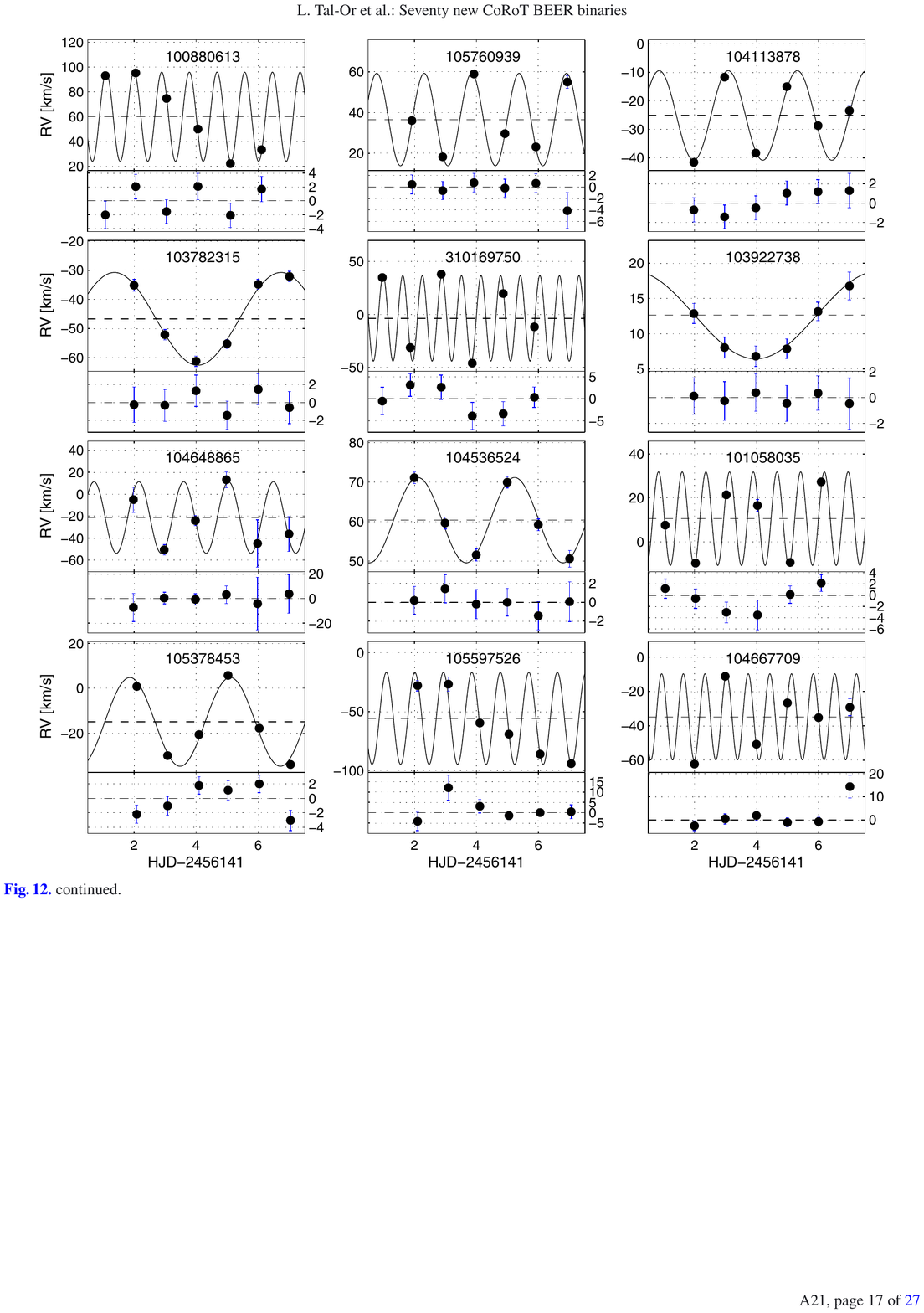,height=10.5in}}
\centerline{\psfig{figure=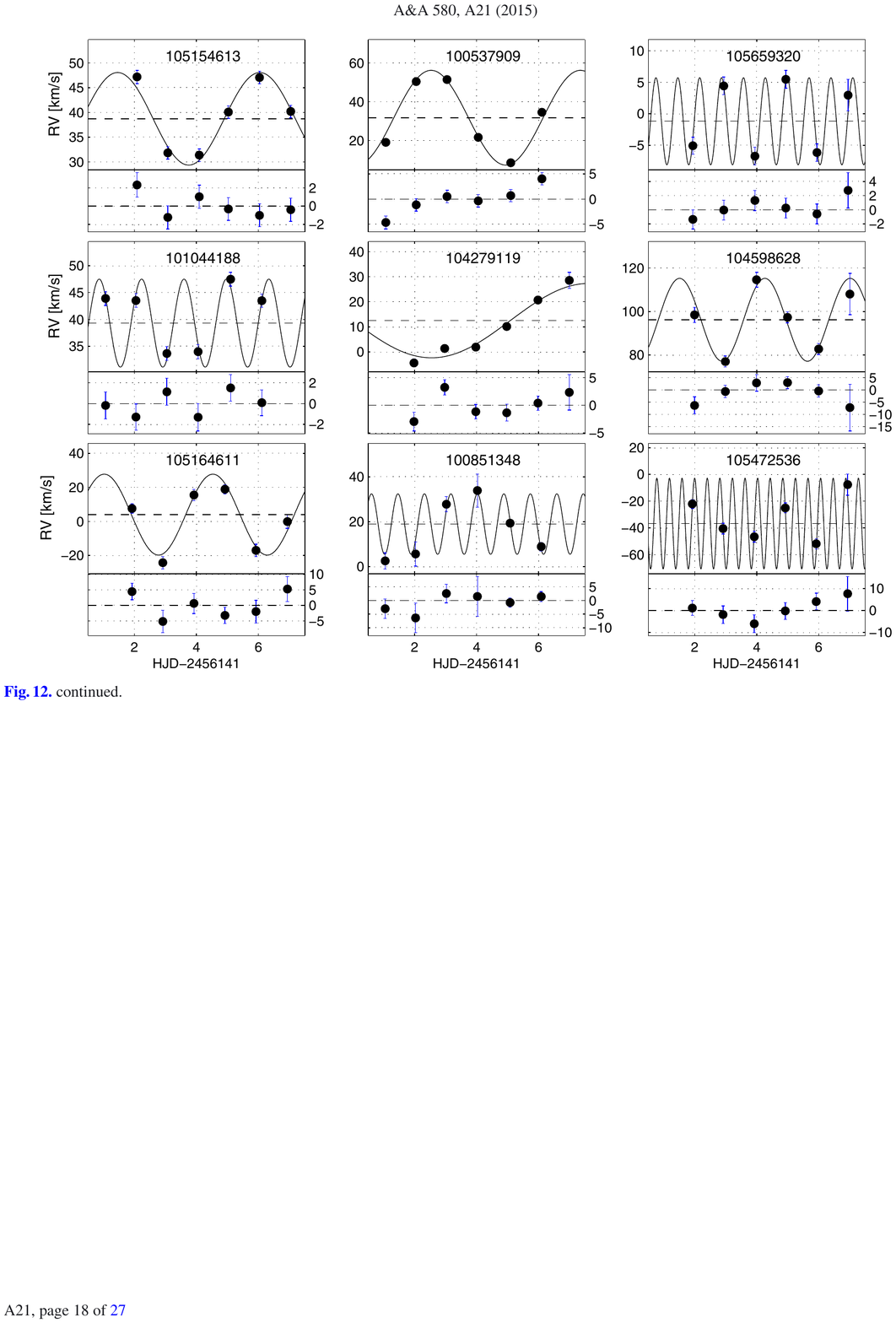,height=10.5in}}
\centerline{\psfig{figure=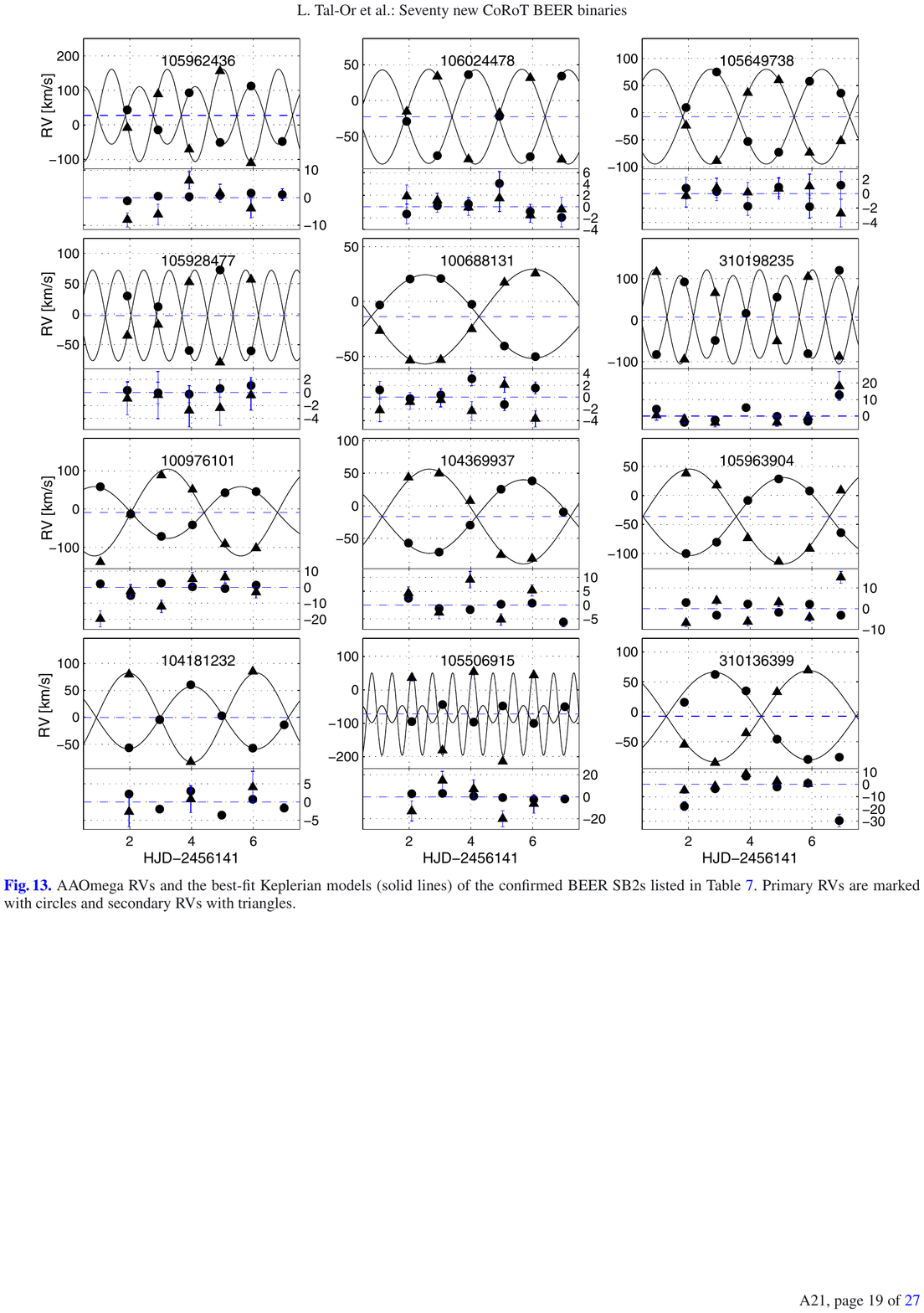,height=10.5in}}
\centerline{\psfig{figure=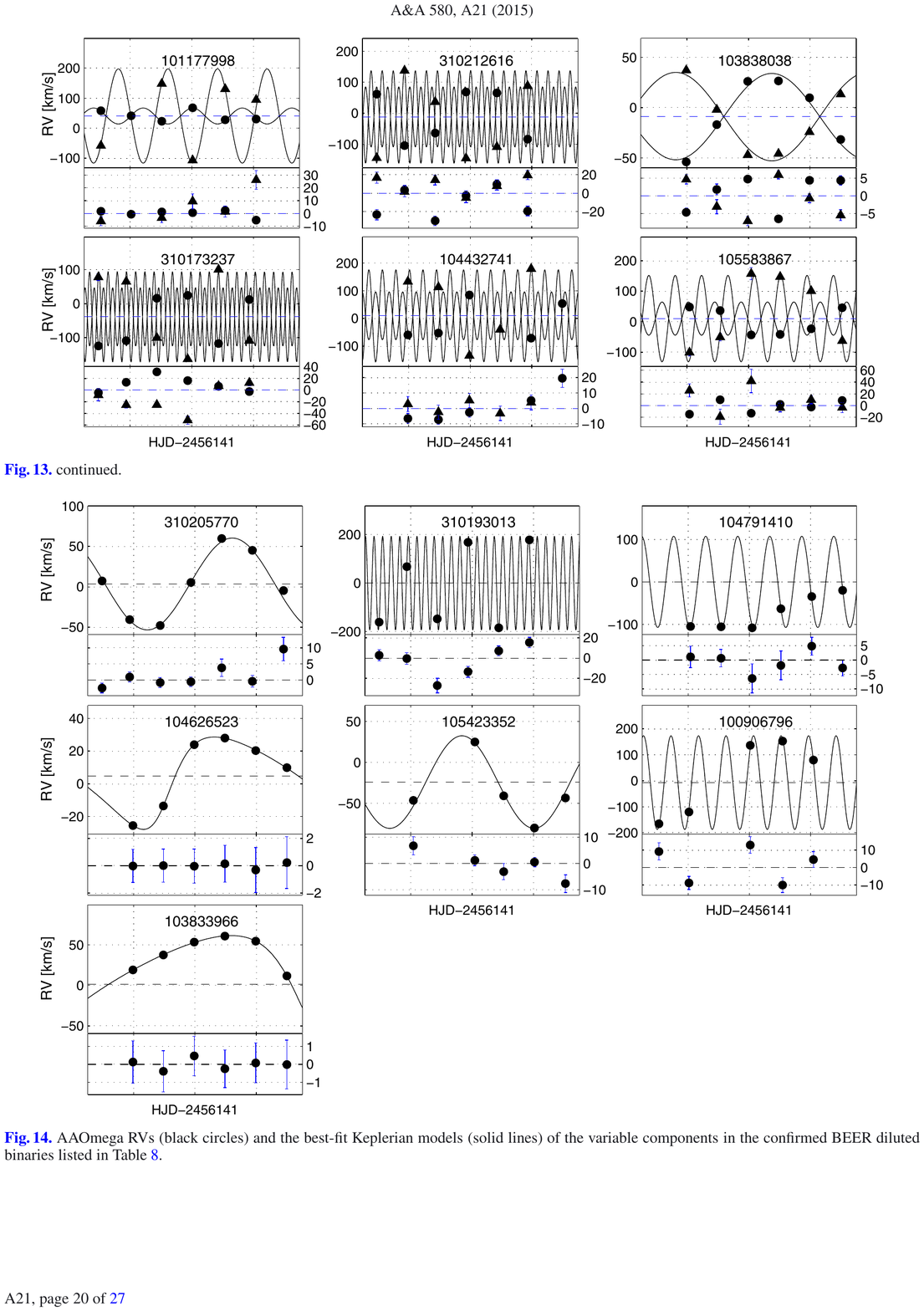,height=10.5in}}
\centerline{\psfig{figure=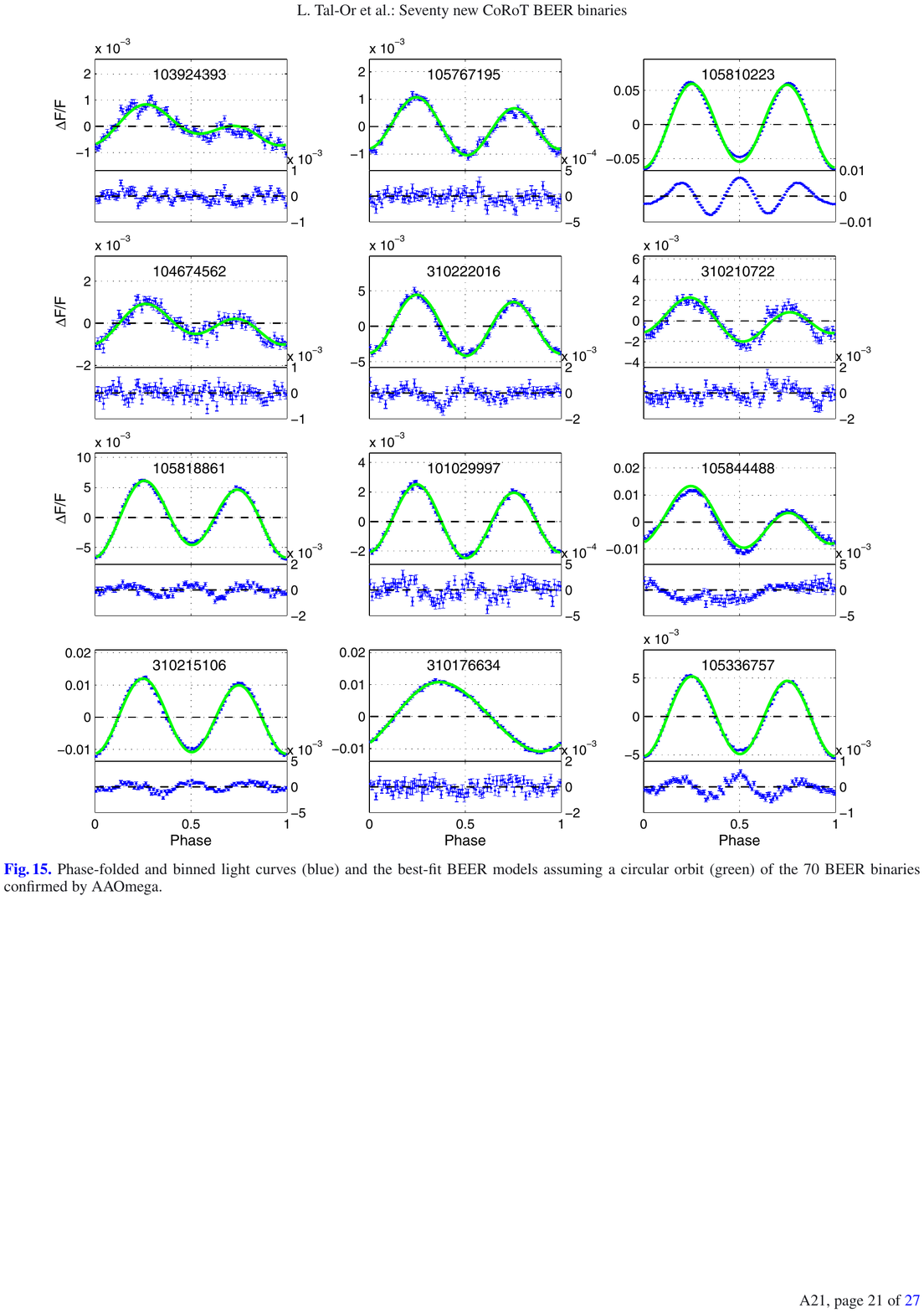,height=10.5in}}
\centerline{\psfig{figure=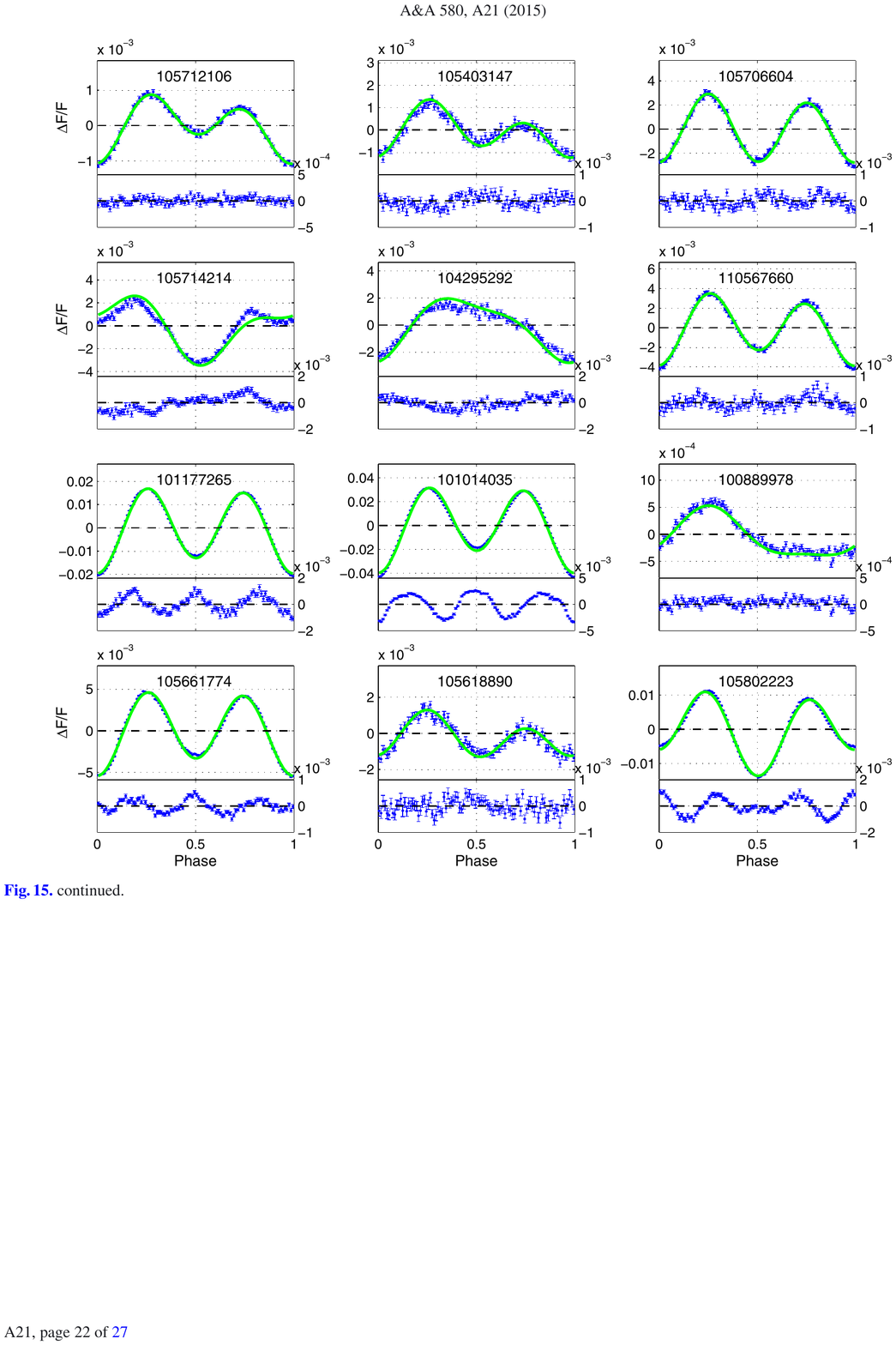,height=10.5in}}
\centerline{\psfig{figure=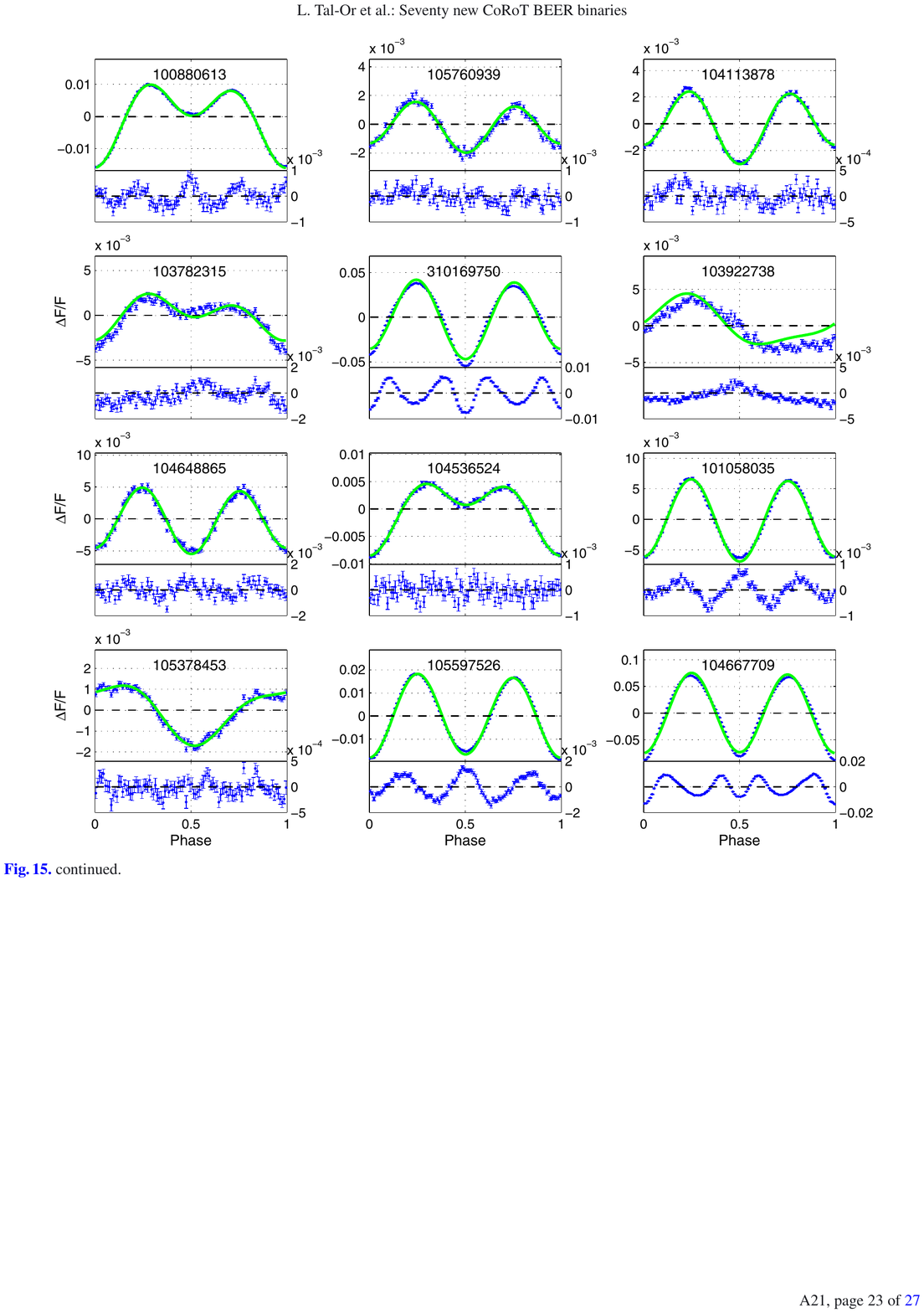,height=10.5in}}
\centerline{\psfig{figure=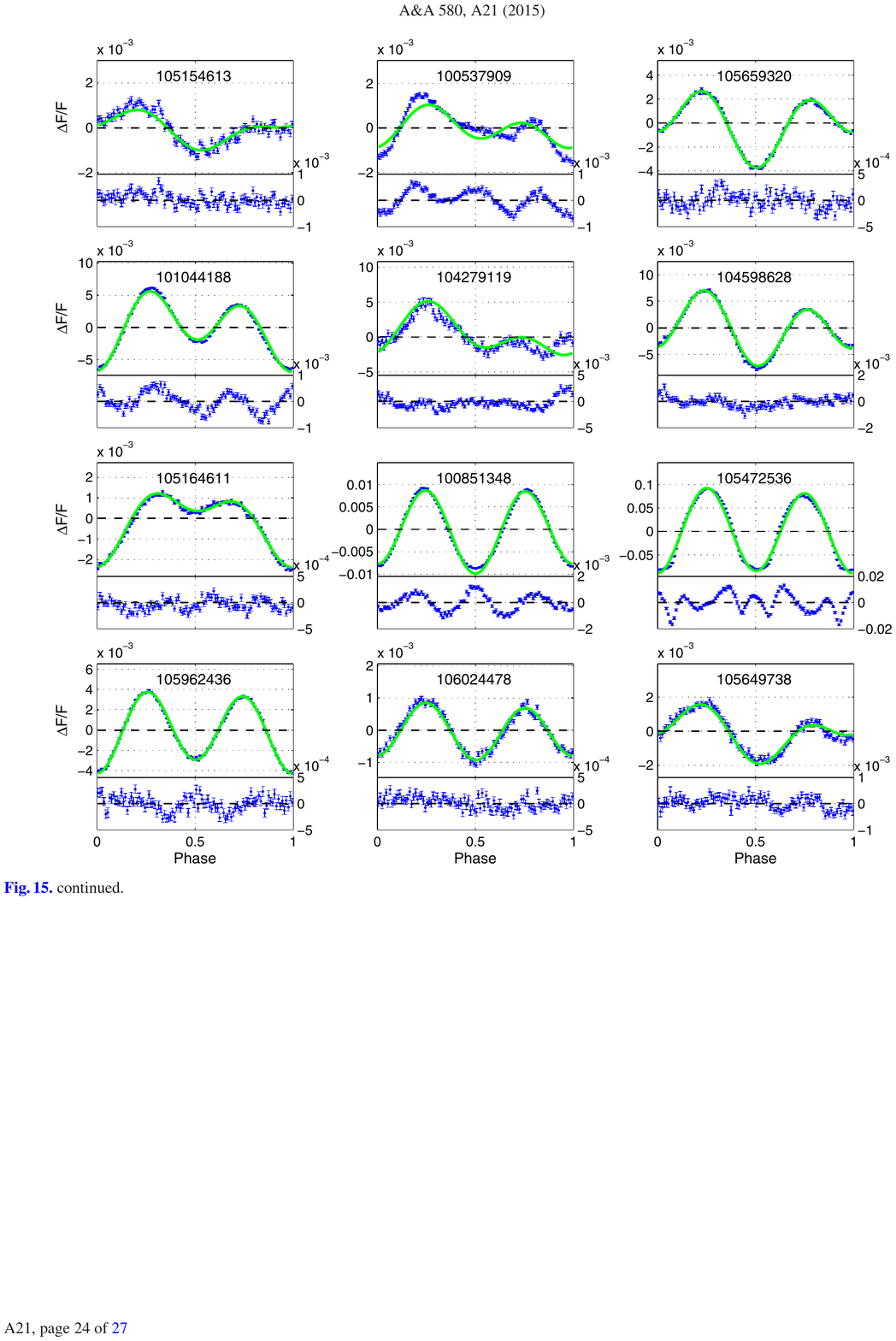,height=10.5in}}
\centerline{\psfig{figure=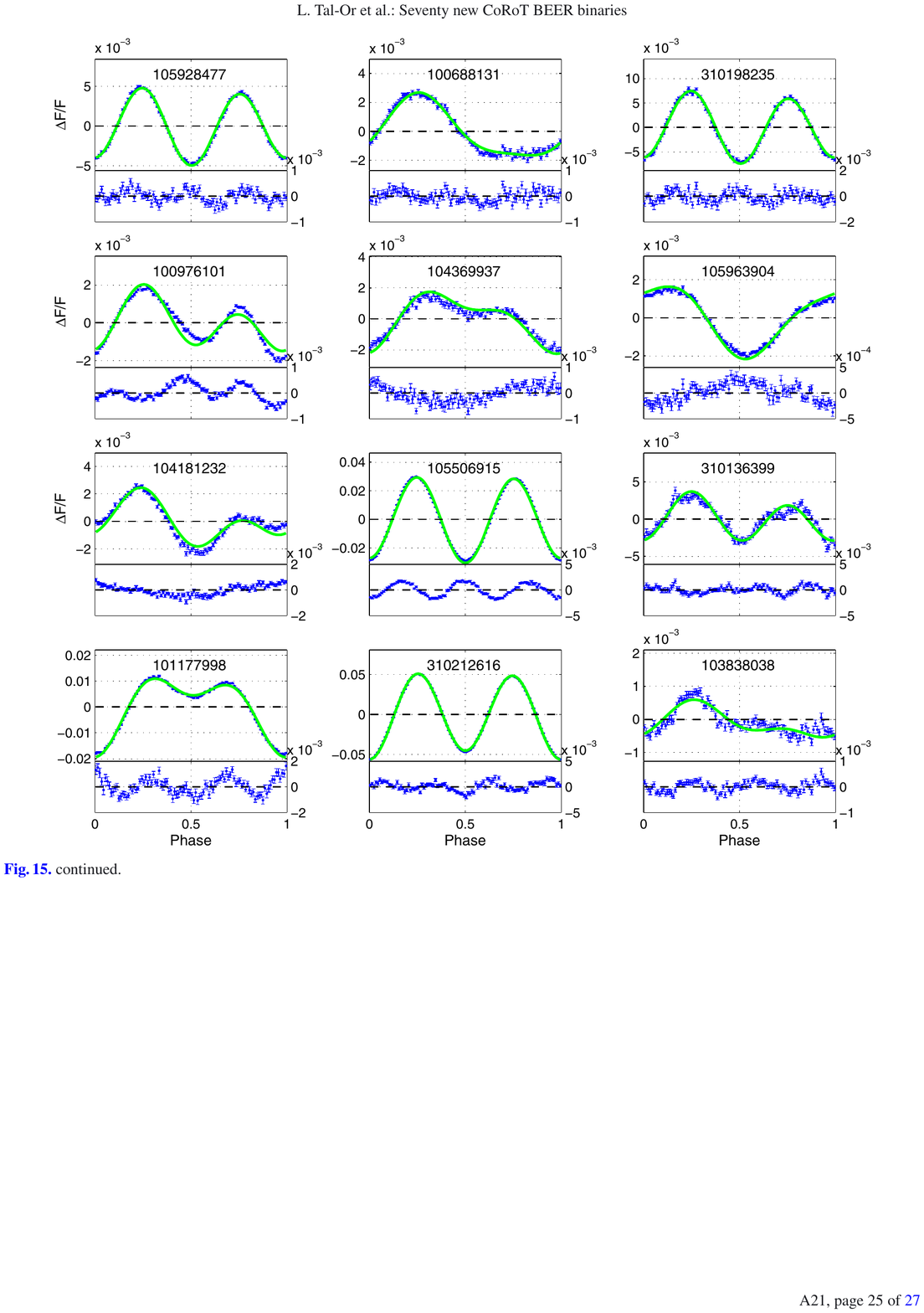,height=10.5in}}
\centerline{\psfig{figure=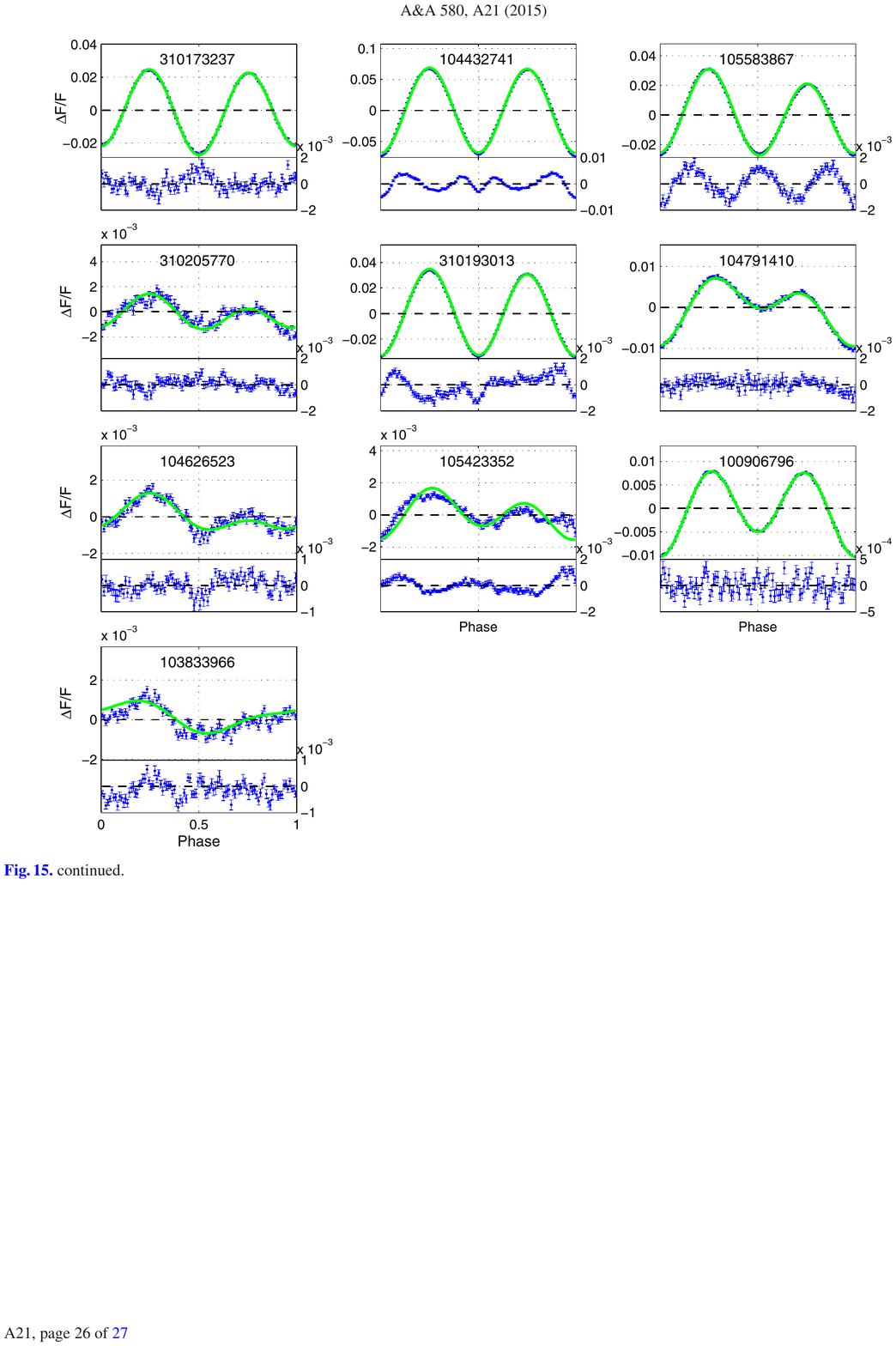,height=10.5in}}
\centerline{\psfig{figure=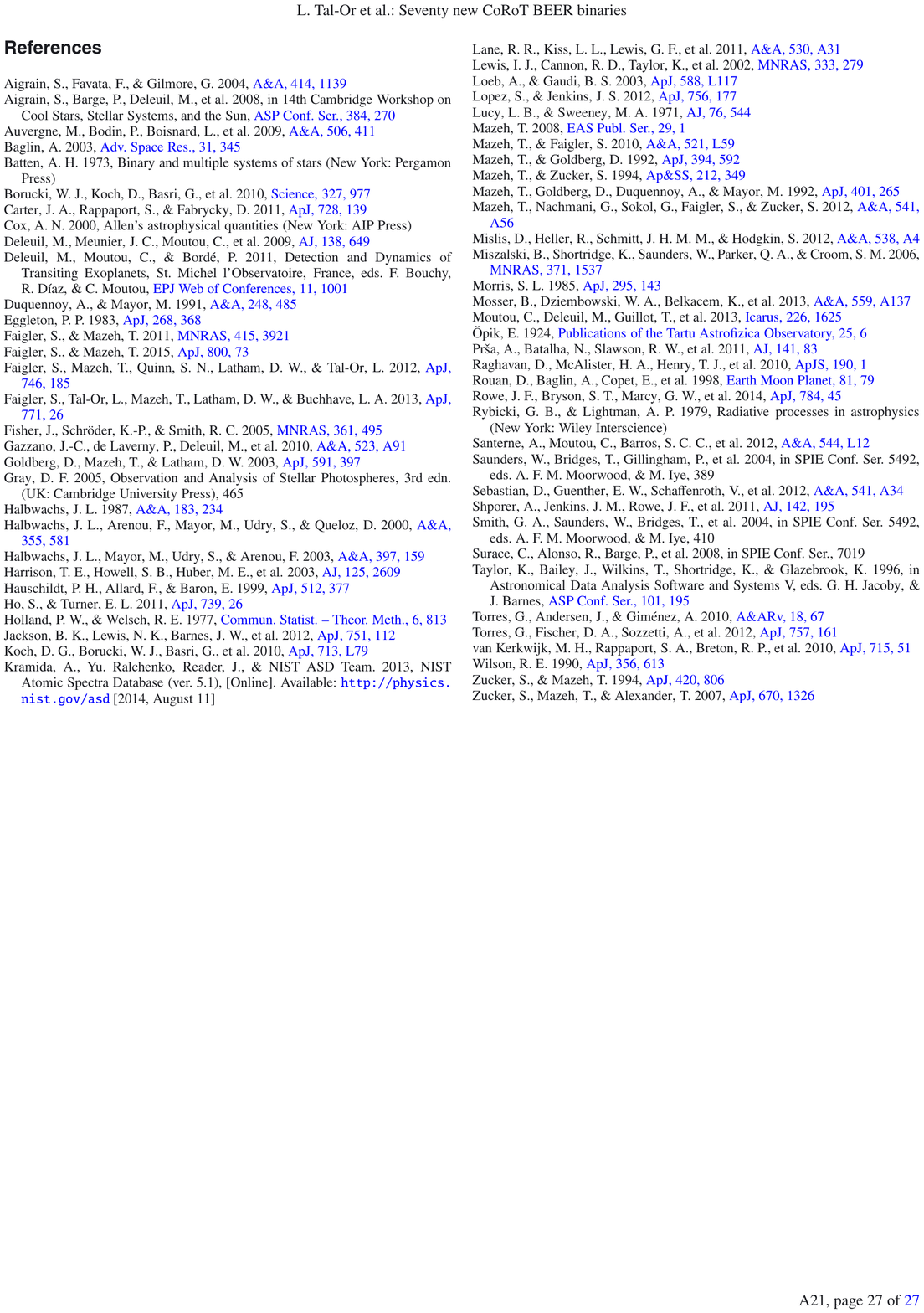,height=10.5in}}

\clearpage
\thispagestyle{empty}

\addcontentsline{toc}{section}{Paper VII}

\centerline{\psfig{figure=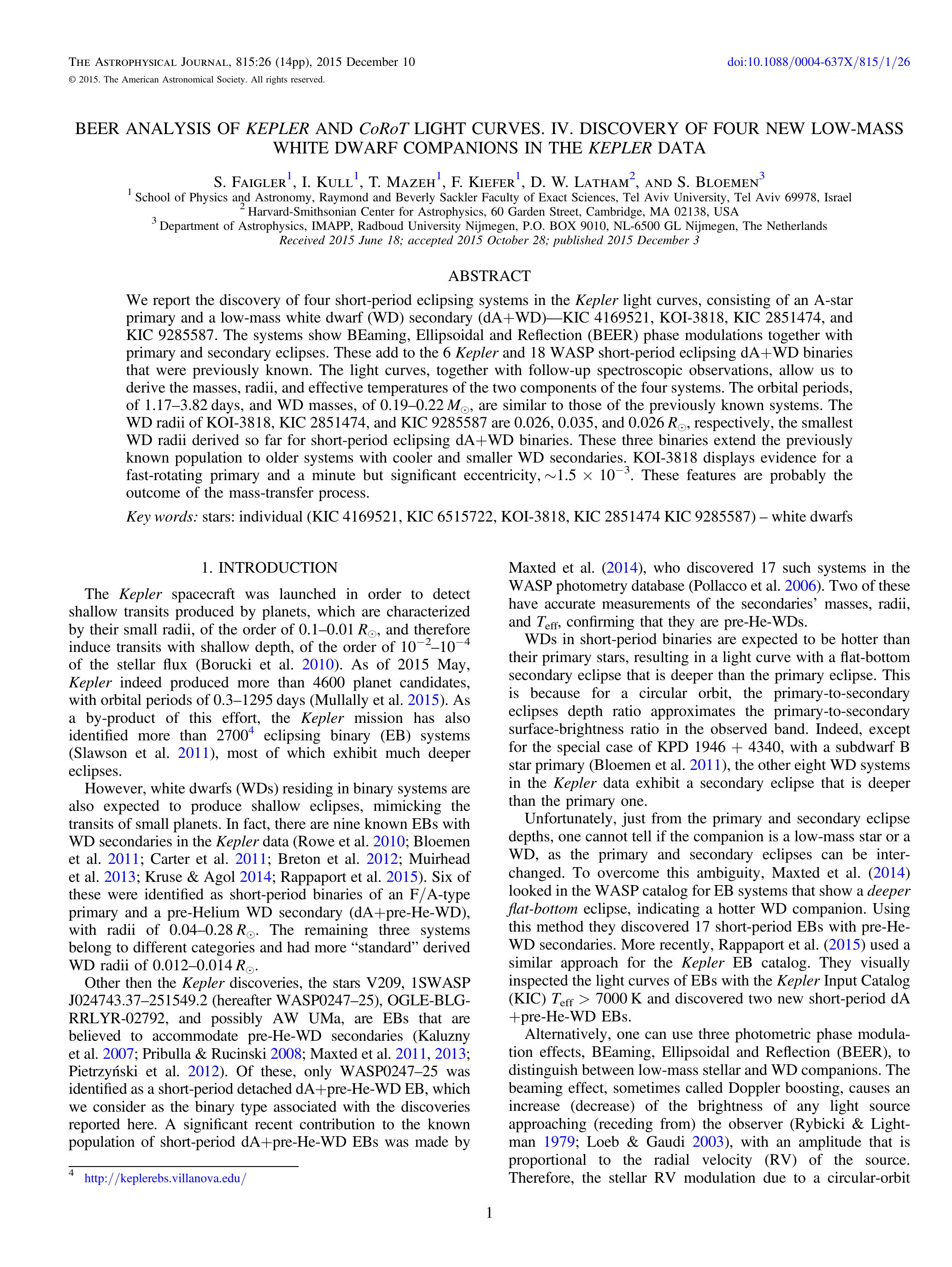,height=10in}}
\chead{Paper VII -- Four {\it Kepler} WDs}
\addtolength{\headsep}{1.0cm}
\centerline{\psfig{figure=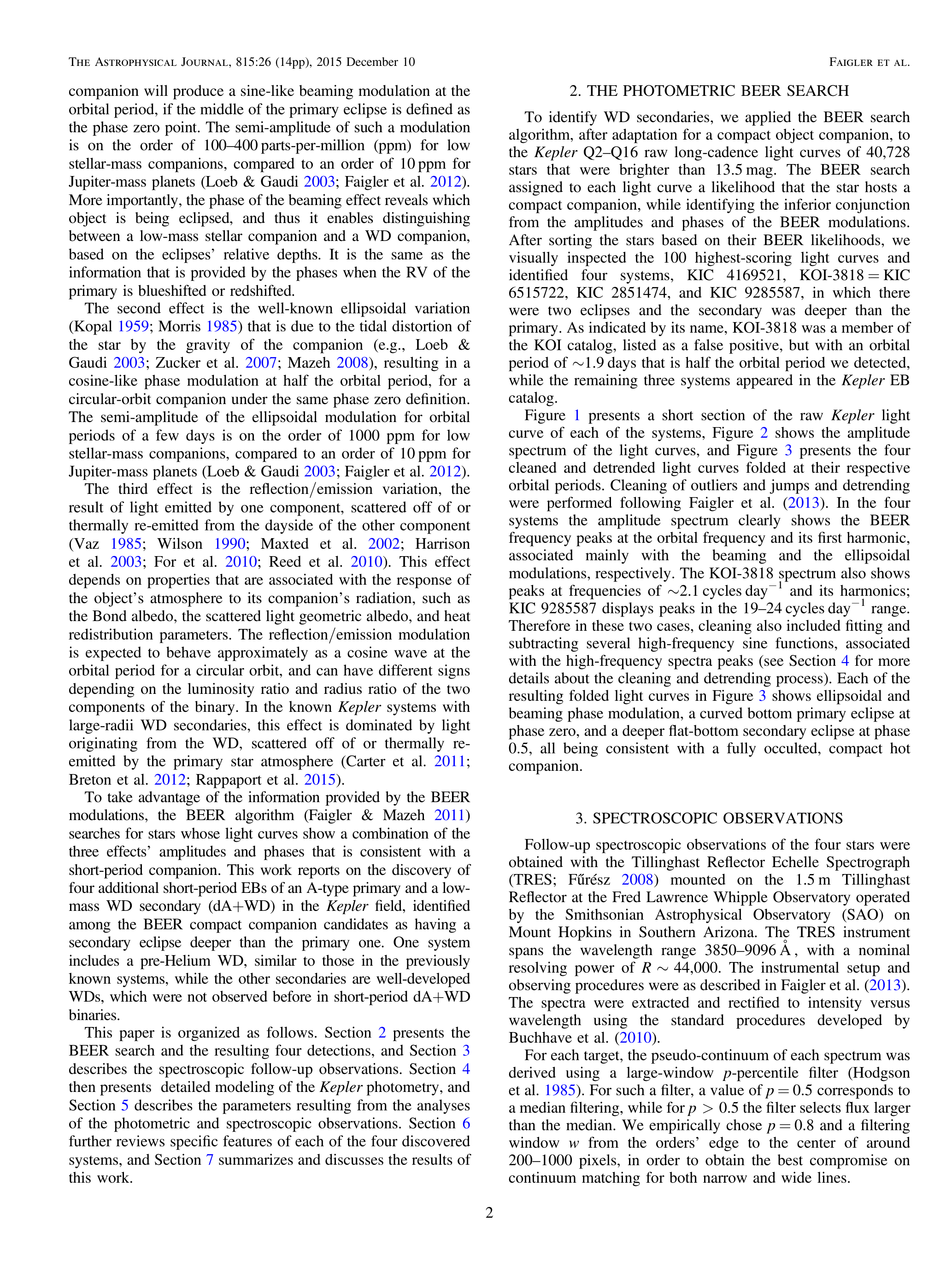,height=10.0in}}
\centerline{\psfig{figure=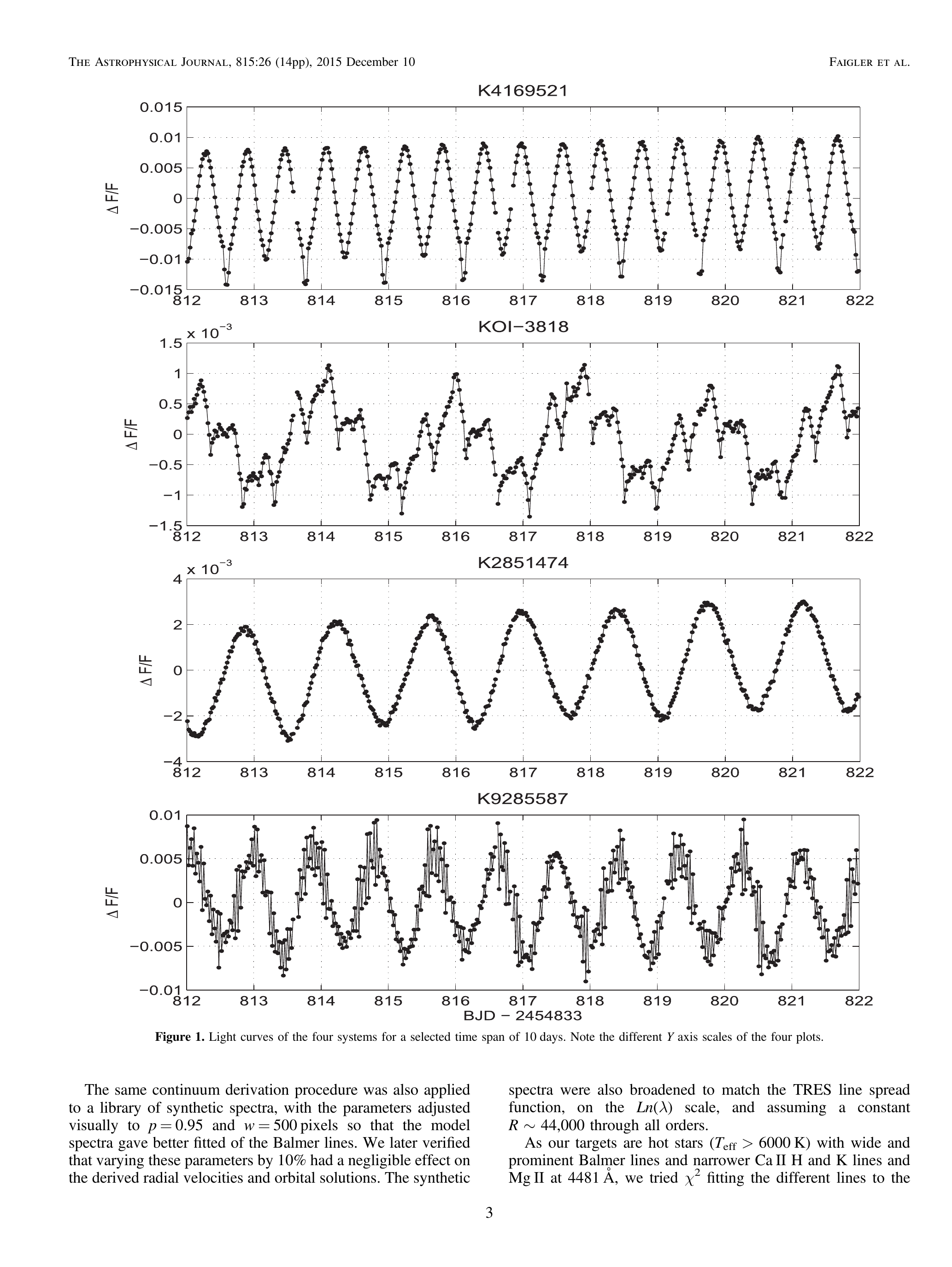,height=10.0in}}
\centerline{\psfig{figure=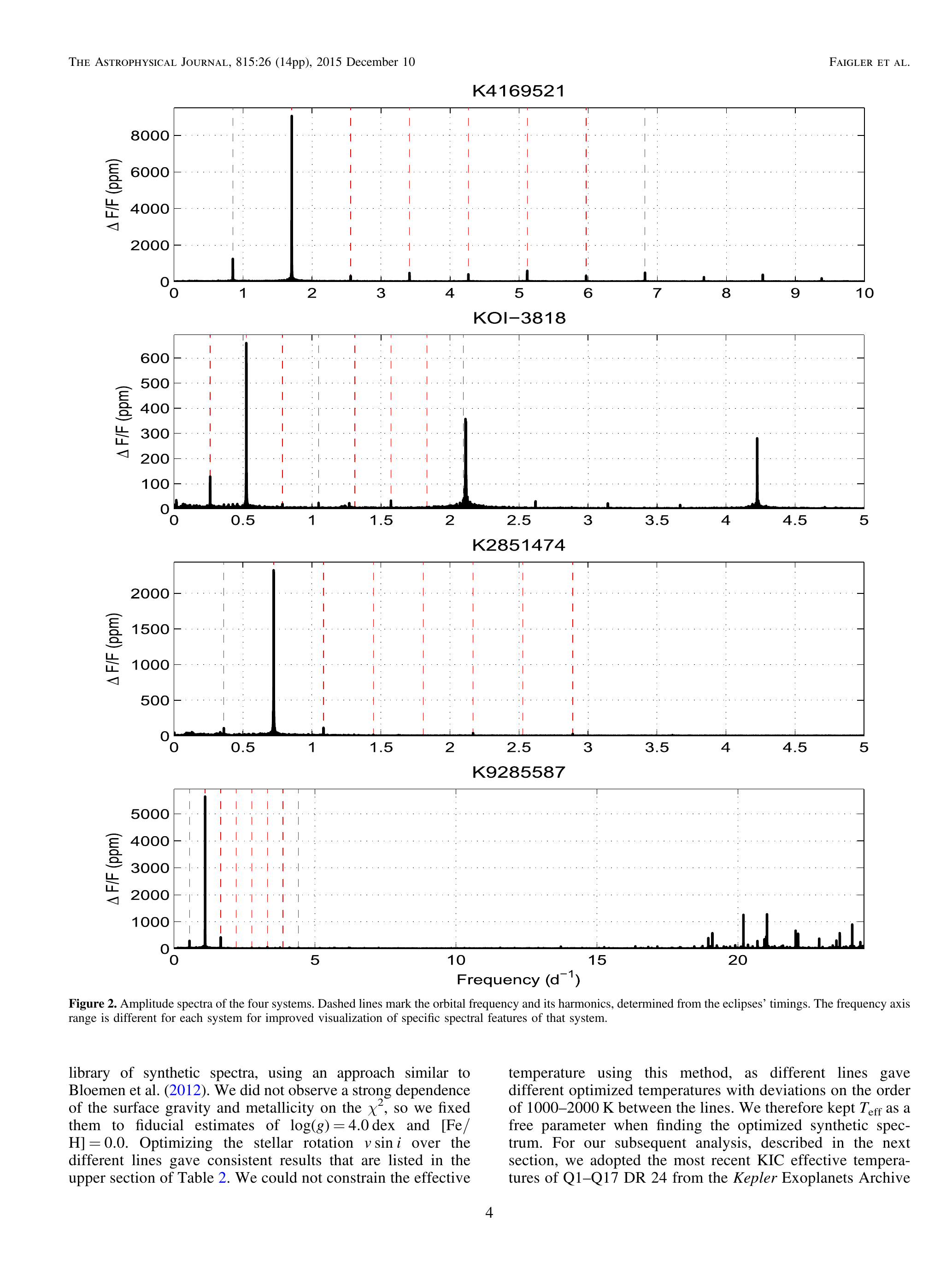,height=10.0in}}
\centerline{\psfig{figure=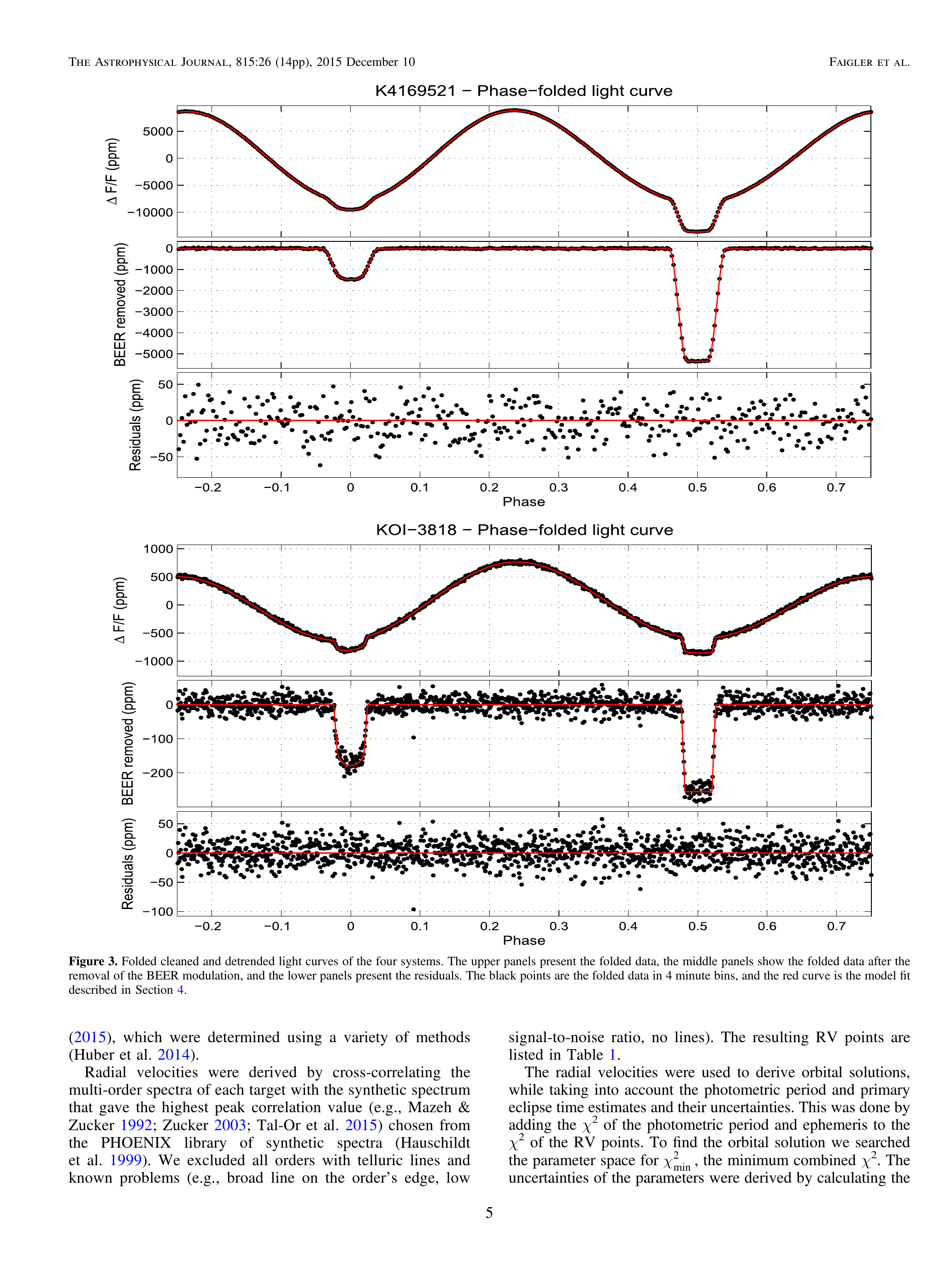,height=10.0in}}
\centerline{\psfig{figure=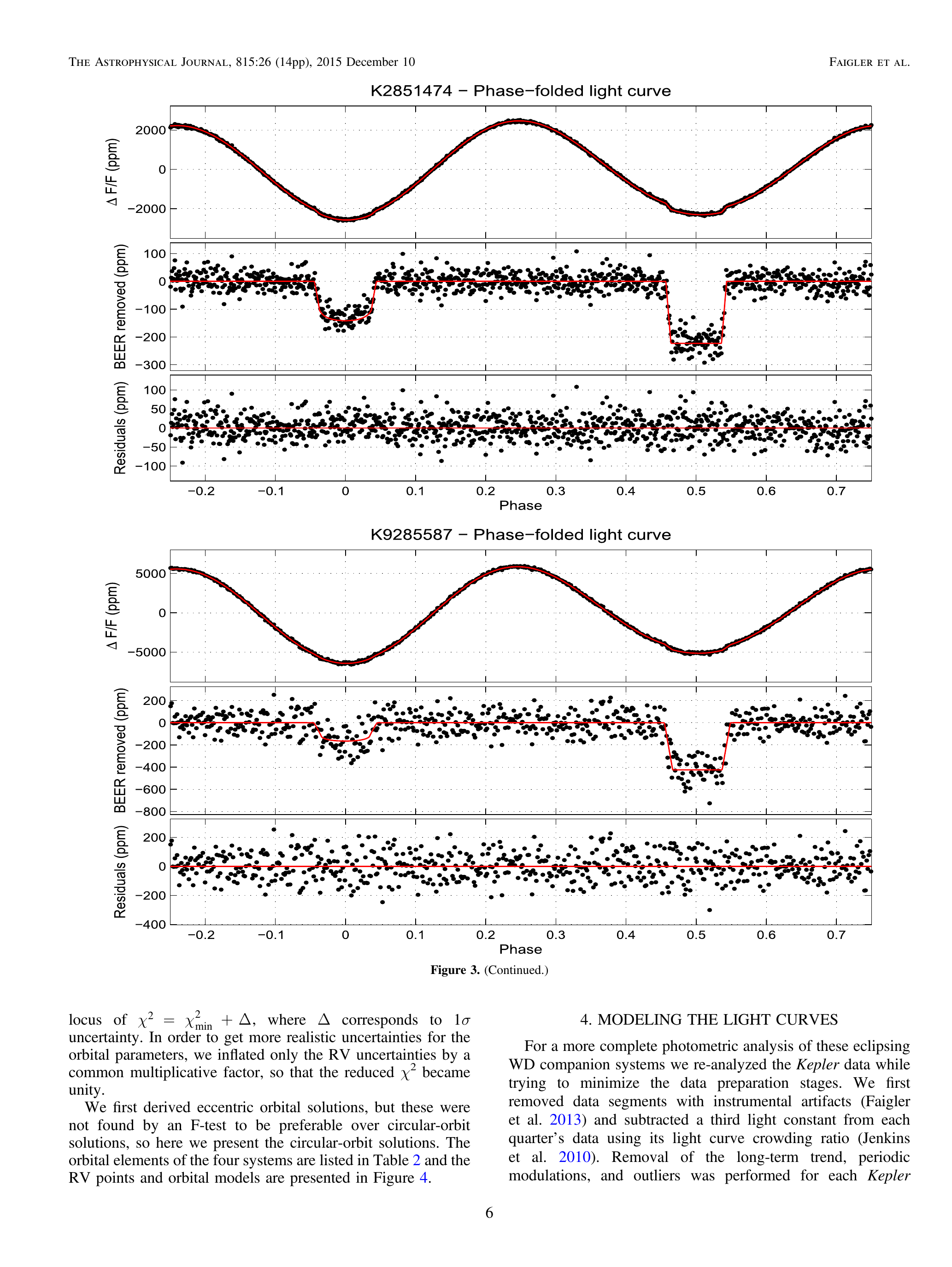,height=10.0in}}
\centerline{\psfig{figure=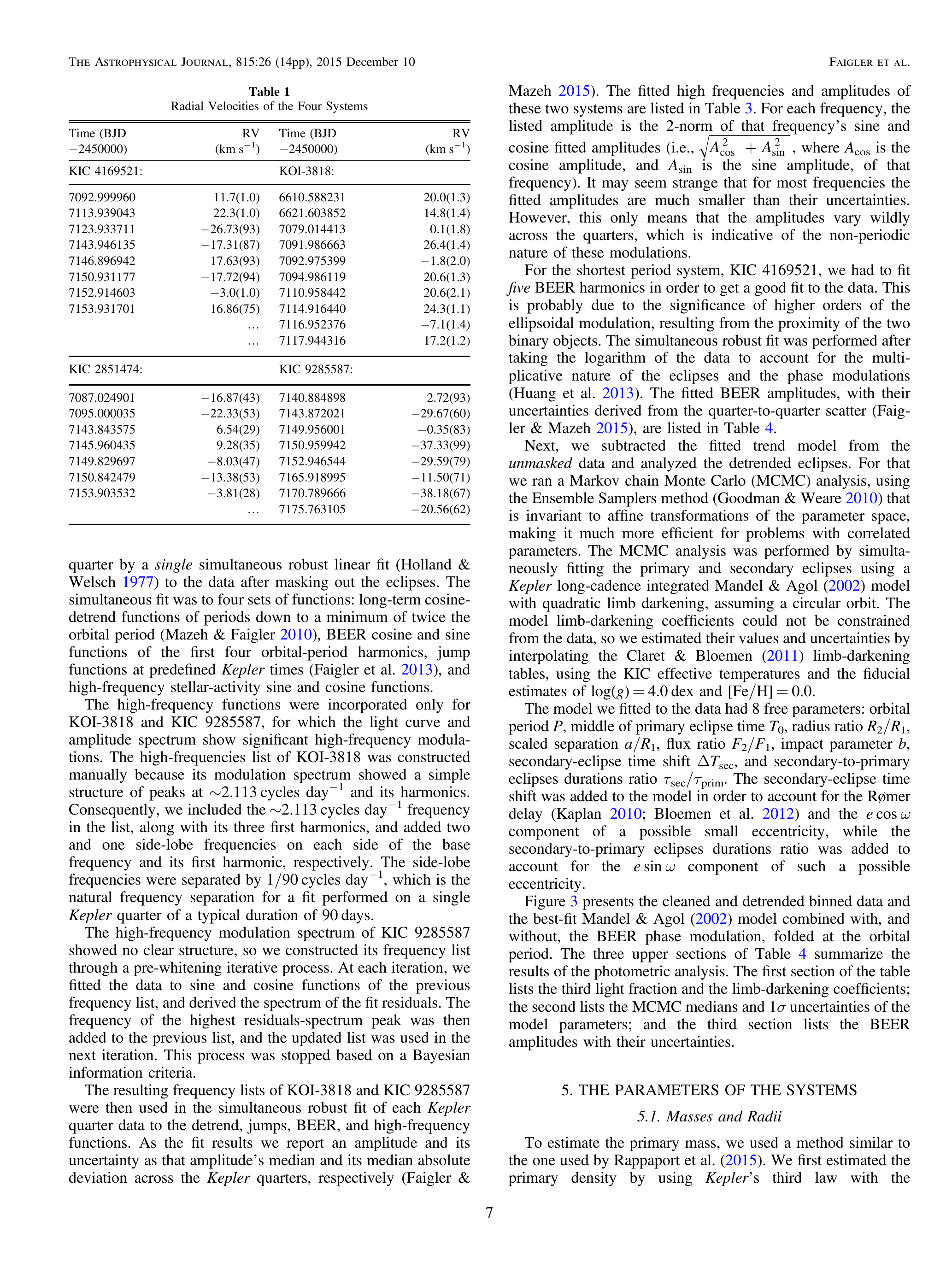,height=10.0in}}
\centerline{\psfig{figure=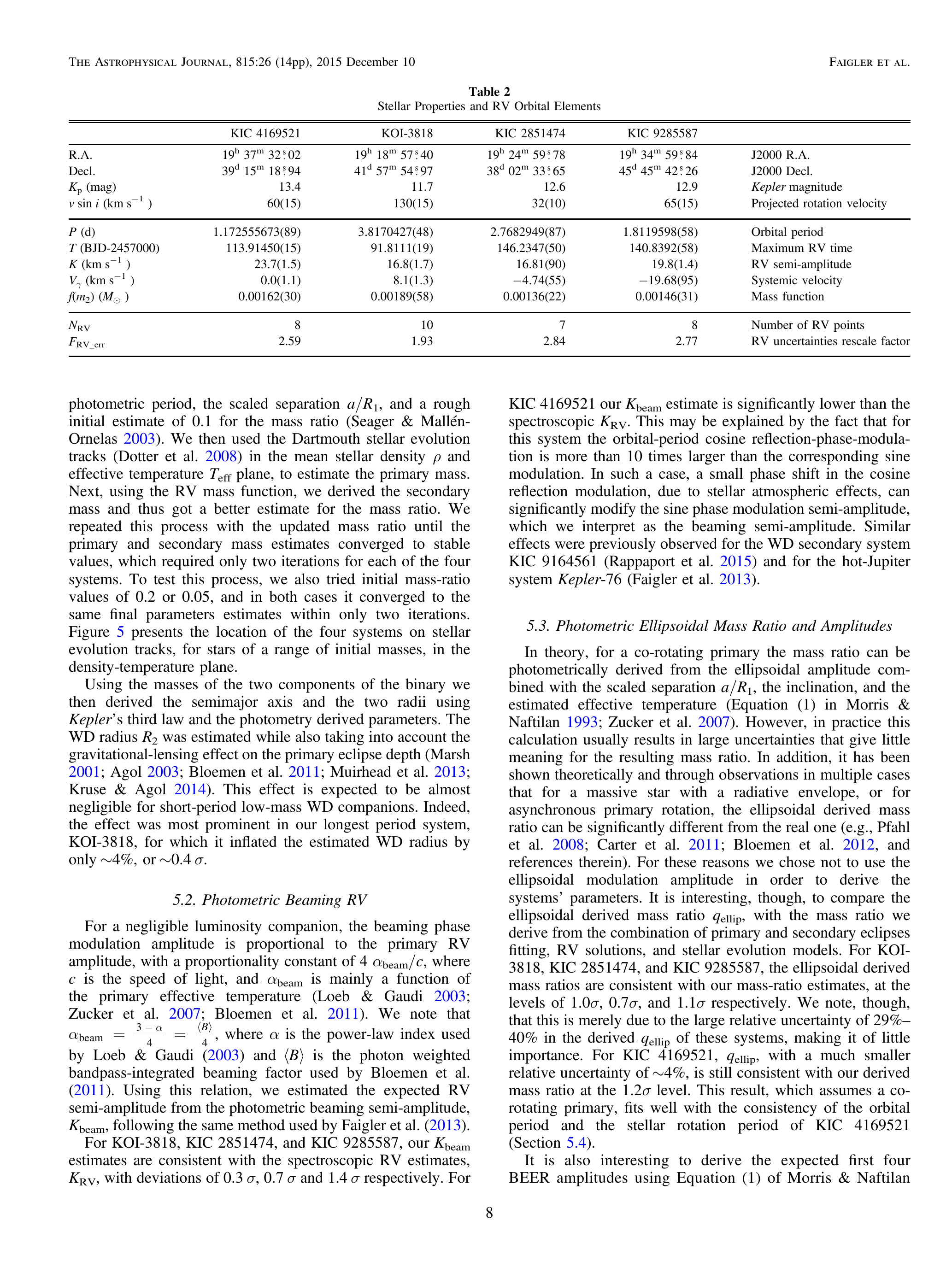,height=10.0in}}
\centerline{\psfig{figure=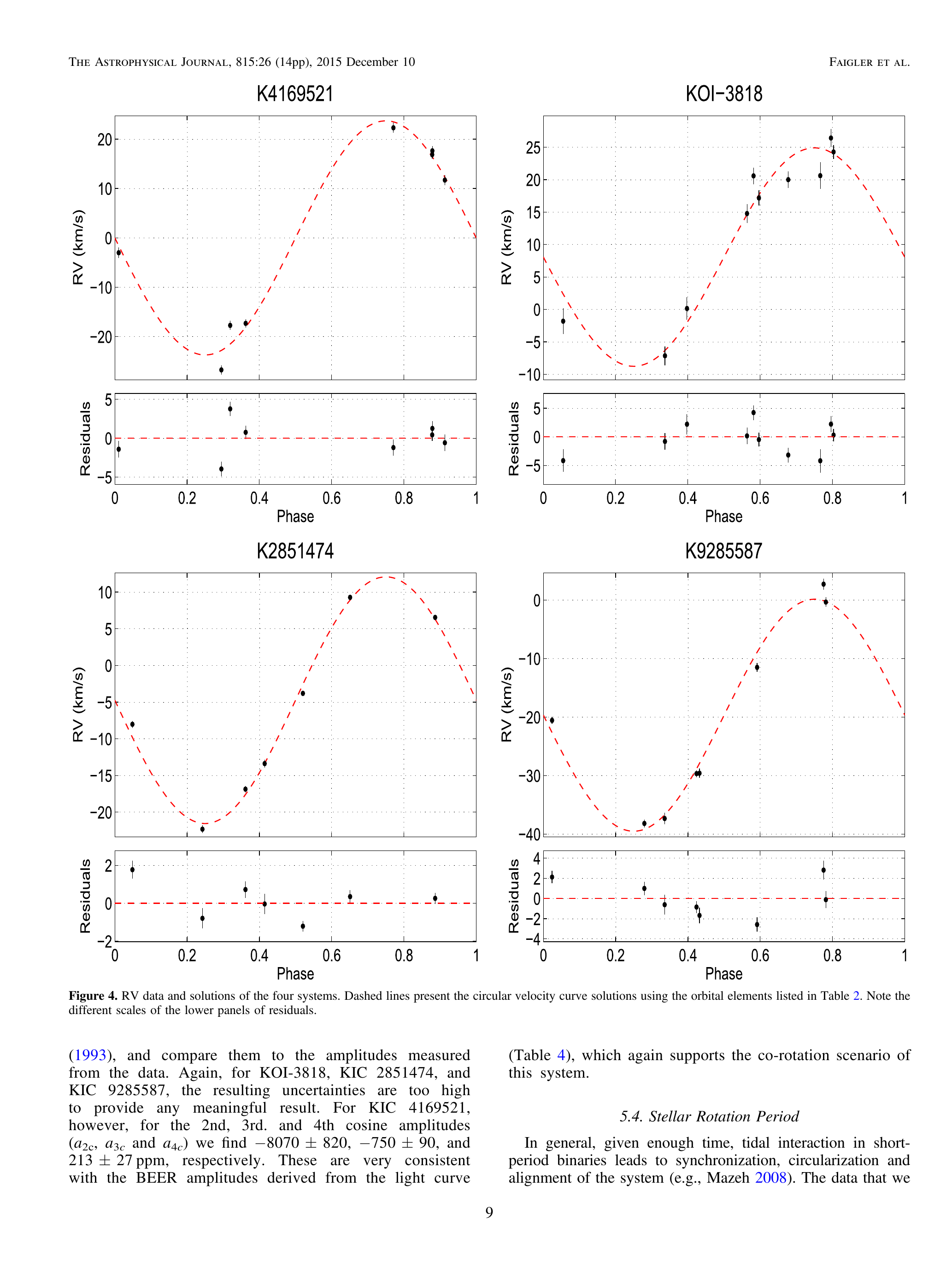,height=10.0in}}
\centerline{\psfig{figure=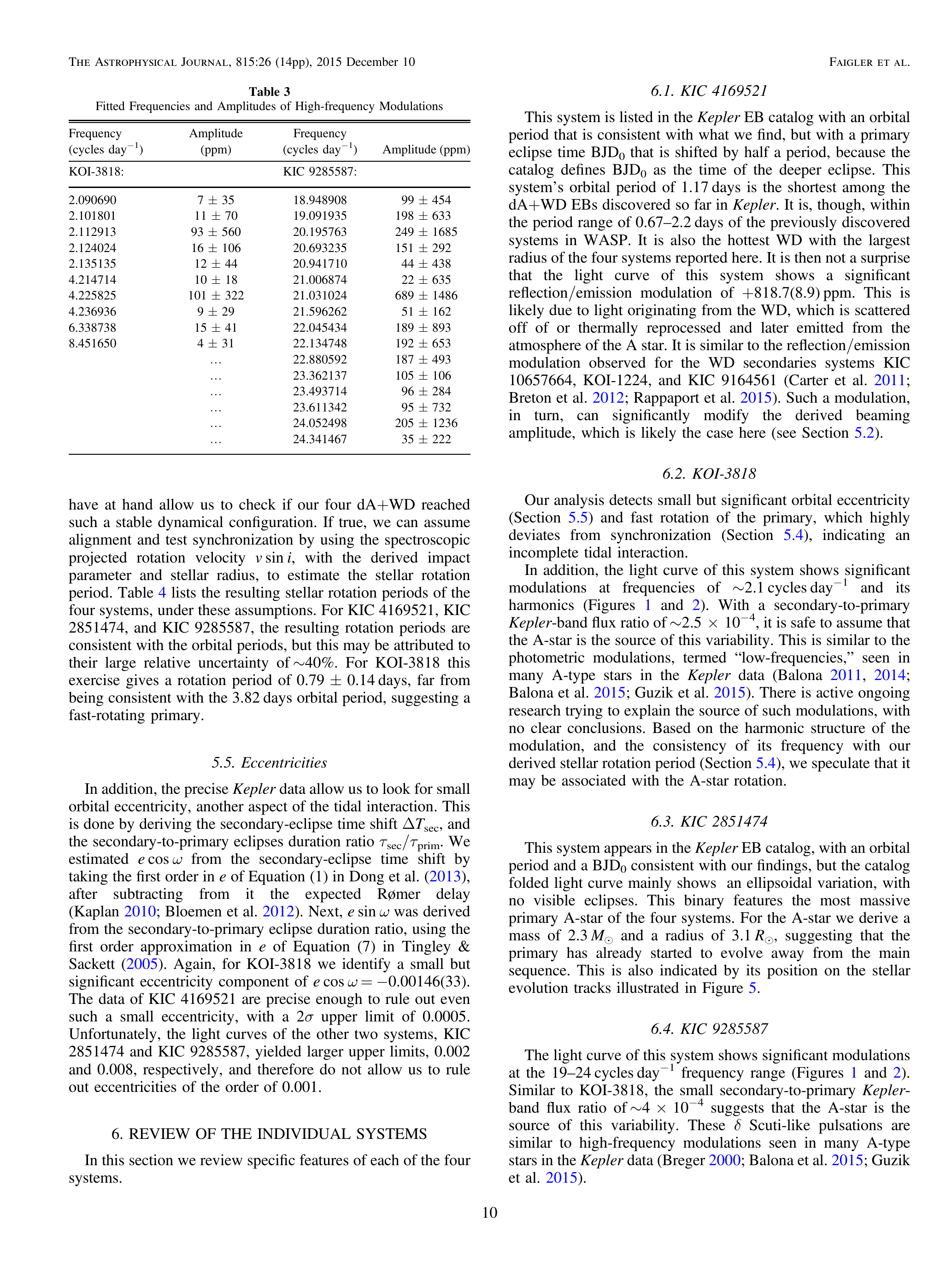,height=10.0in}}
\centerline{\psfig{figure=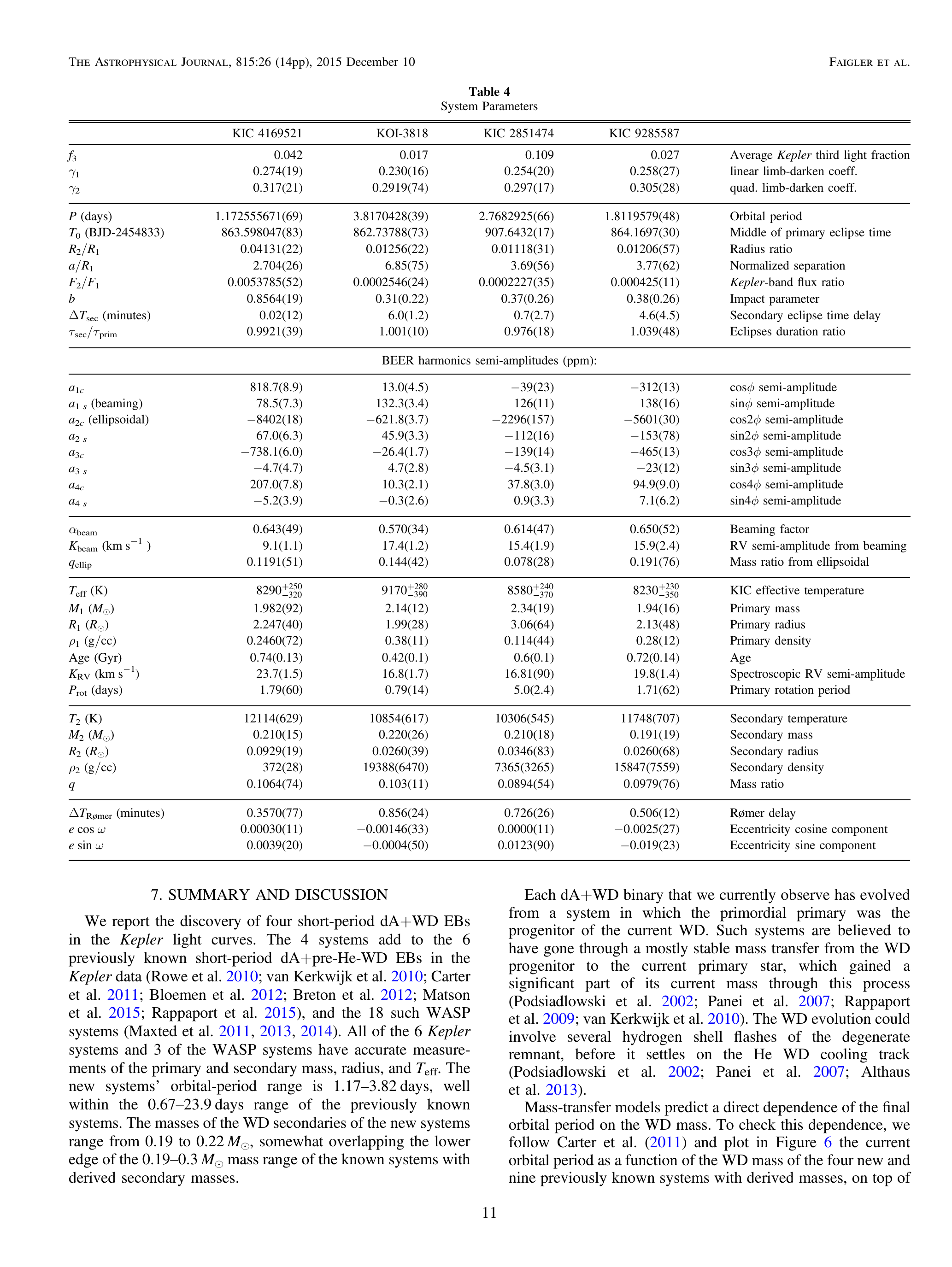,height=10.0in}}
\centerline{\psfig{figure=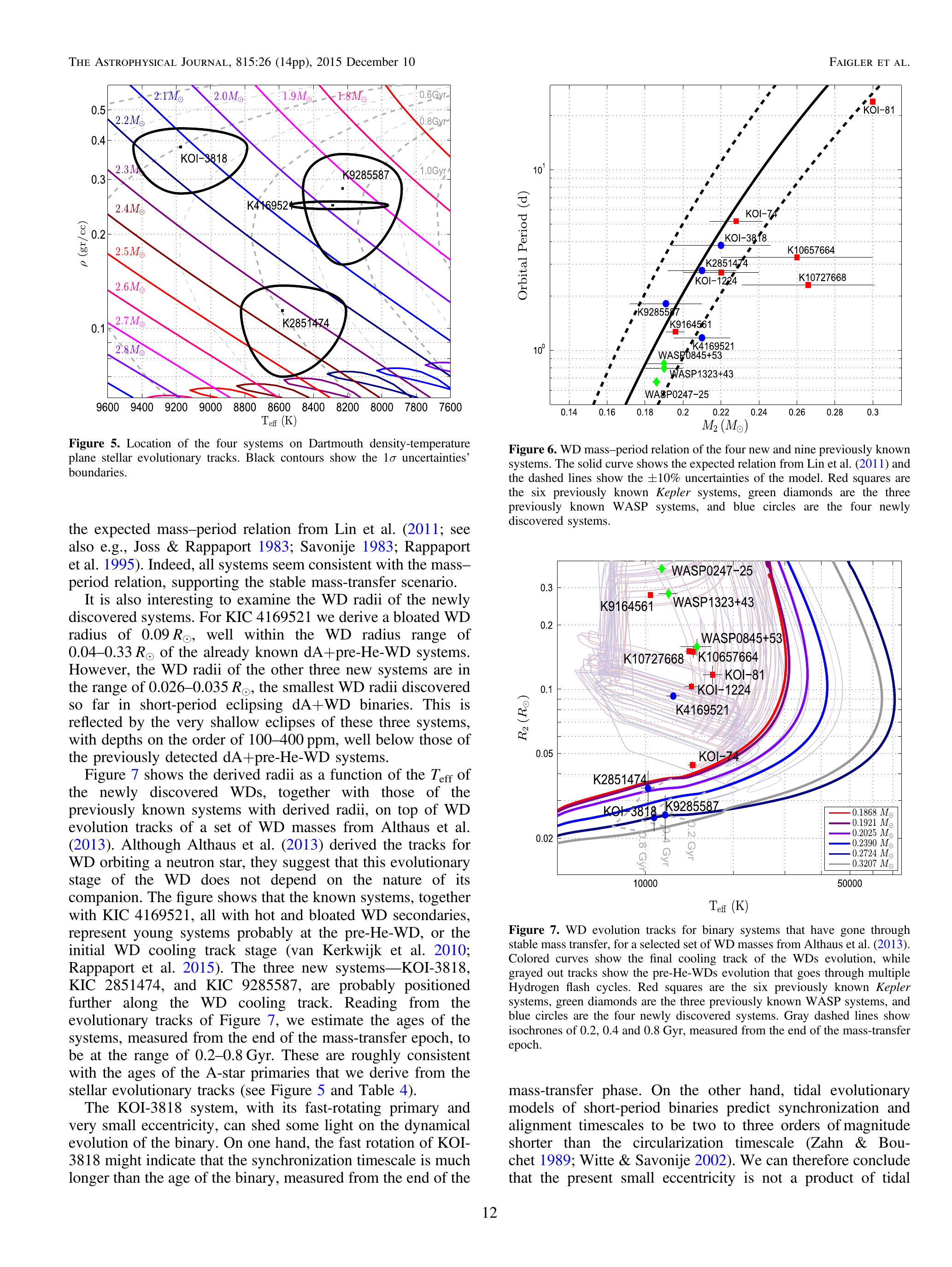,height=10.0in}}
\centerline{\psfig{figure=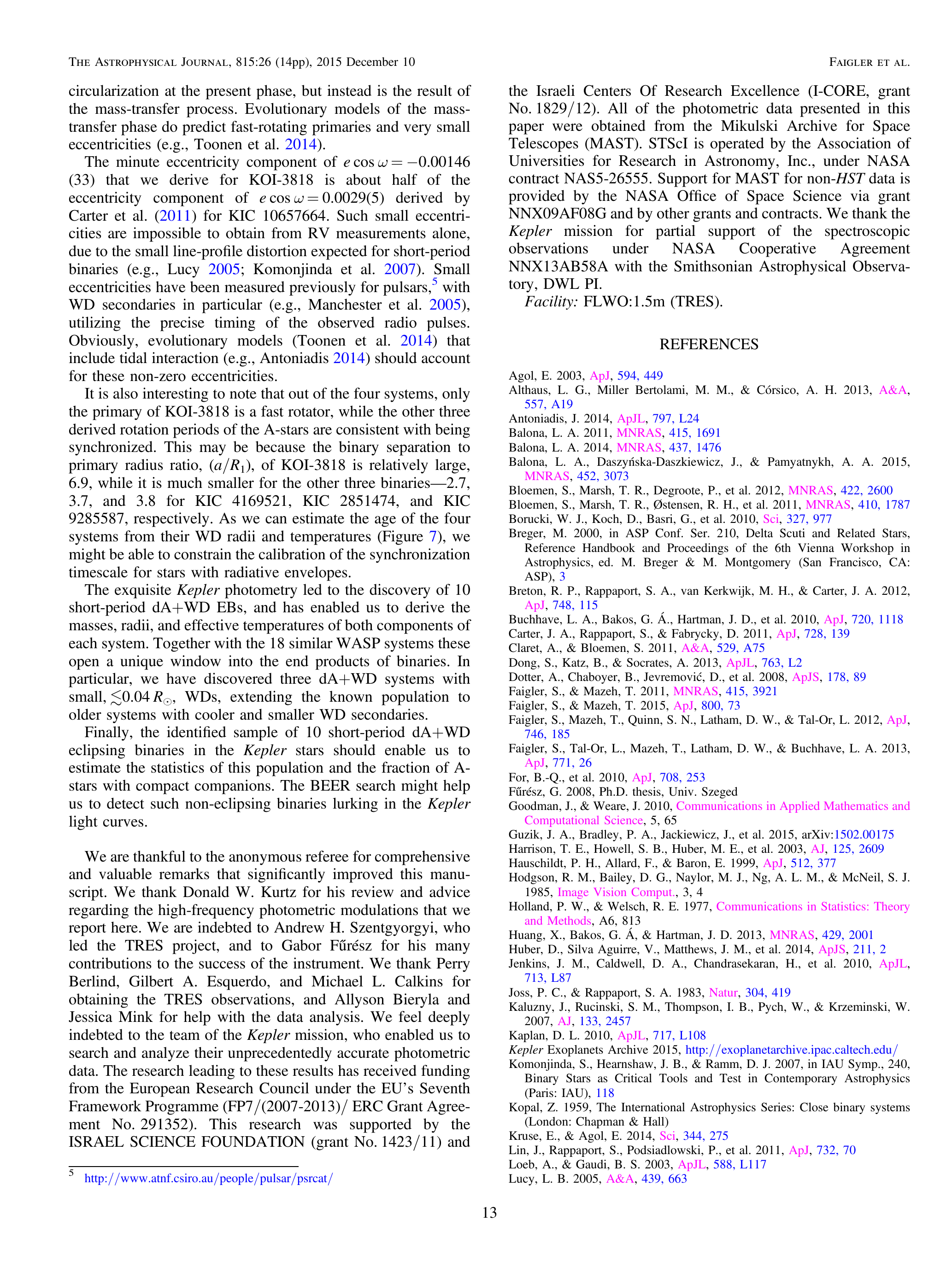,height=10.0in}}
\centerline{\psfig{figure=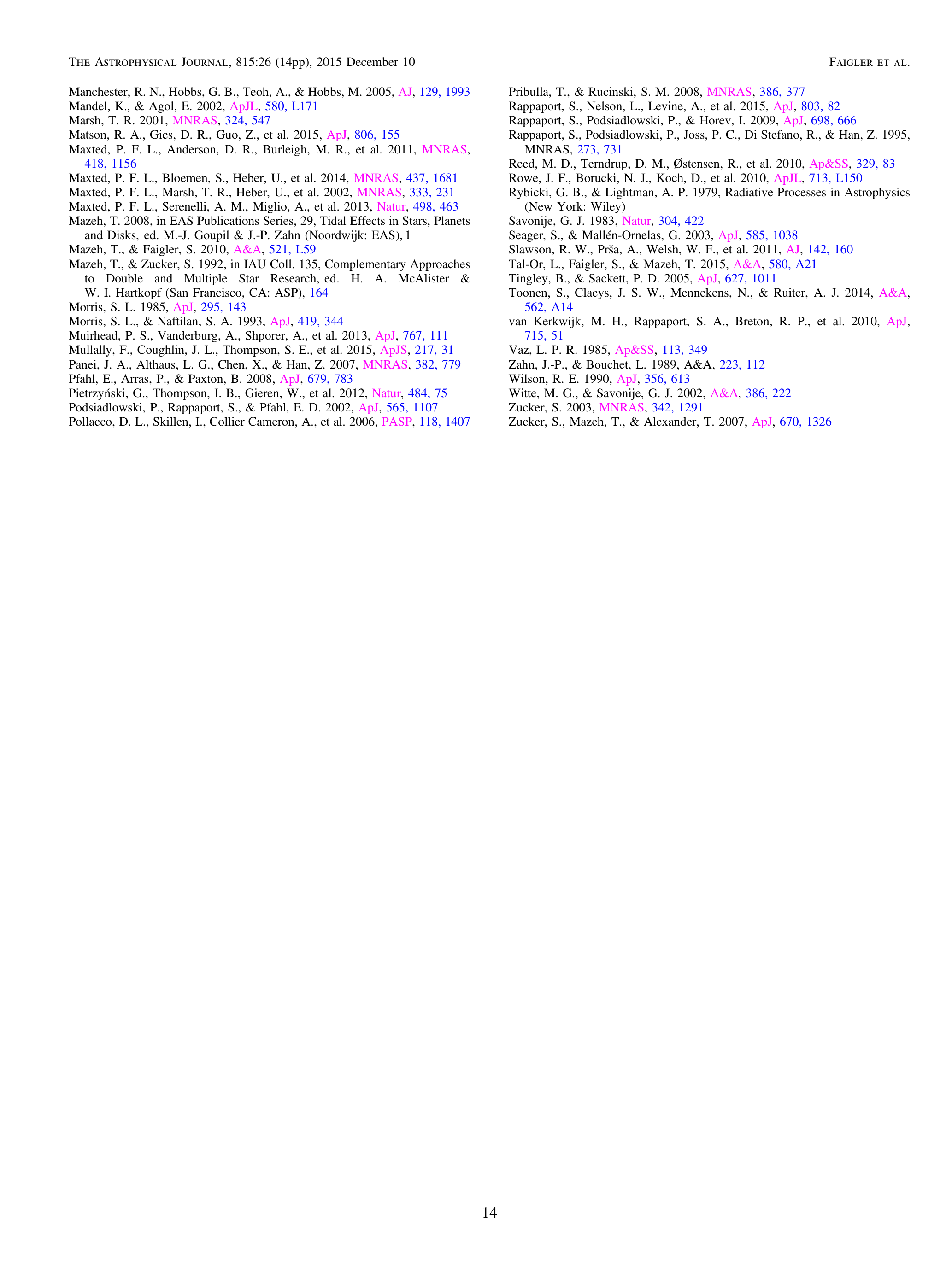,height=10.0in}}

\newpage
\addtolength{\headsep}{0.9cm} 
} 

\pagestyle{plain}
\newpage
\thispagestyle{empty}


\chapter{Discussion}
\label{discussion}

In this chapter I summarize the findings of the seven papers and briefly discuss their astrophysical implications. 

\section{The Papers}
Paper I \citep{mazeh10} presents the detection of the beaming and the ellipsoidal modulations in the light curve of CoRoT-3, a system of a $22$$M_{\rm Jup}$ brown dwarf orbiting an F3 star in an orbital period of $4.3$ days \citep{deleuil08}. This was the first time the beaming effect was detected for a substellar companion. The paper suggests that had this analysis been performed immediately after the discovery of the light-curve transits, the mass of the companion could have been estimated from the observed beaming and ellipsoidal amplitudes, thus reducing the number of RV points needed to confirm the mass of the transiting object. The detection of the effects also means that the same effects can be detected in the {\it CoRoT} and {\it Kepler} light curves of stars with massive-planet/brown-dwarf companions, even without any transits, as suggested by \citet{loeb03} and \citet{zucker07}. The ability to detect some combination of the three effects in space-missions light curves, even for non-eclipsing systems, lays the grounds for the BEER search algorithm for non-eclipsing binaries.

As a natural development following Paper I, Paper II \citep{faigler11} presents the BEER algorithm for detection of {\it non-transiting} short-period low-mass companions, through the beaming, ellipsoidal and reflection effects, in {\it Kepler} and {\it CoRoT} light curves. After describing the search algorithm and analyzing its expected performance, the paper predicts that when more {\it Kepler} data is available it may be possible to find in it candidates for planets of mass as small as $5$--$10$$M_{\rm Jup}$. This claim was later confirmed by the discovery of Kepler-76b (Paper IV). The paper also suggests that the long time span of the {\it Kepler} mission enables using not only the period, but also the phase of the BEER modulations, so that very few RV follow-up observations are needed to confirm the existence of a companion. This method was later used in the RV confirmations presented in Papers III, IV and VII.

To illustrate the effectiveness of the BEER method, Paper III \citep{faigler12} presents the discovery and RV confirmation of seven non-eclipsing binaries, with minimum secondary masses of $0.07$--$0.4 M_{\odot}$, in the {\it Kepler} data. 
The paper emphasizes that unlike eclipses searches, the BEER algorithm searches for {\it non-eclipsing} companions, and therefore can detect additional systems with lower inclination angles. It is thus effectively equivalent to performing an RV survey on hundreds of thousands of stars through scanning their {\it Kepler} and {\it CoRoT} light curves. Therefore, the paper predicts that BEER can discover hundreds of new short-period binaries.
As also stated by Paper I, Paper III predicts that once the full {\it Kepler} data is available, we should be able to detect in it brown-dwarf secondaries and even massive planets.

As envisioned by the previous papers, Paper IV \citep{faigler13} presents the discovery of Kepler-76b, a 2$M_{\rm Jup}$ transiting hot Jupiter orbiting a 13.3 mag F star with orbital period of $1.54$ days. This planet was first identified by the BEER algorithm, and then confirmed by RV follow-up observations.
Interestingly, the paper finds inconsistency between the beaming amplitude and the spectroscopically measured RV. This apparent discrepancy is explained by a phase shift of the planetary thermal modulation, due to equatorial superrotating jets in the planet atmosphere. This phenomenon was predicted by \citet{showman02} and later observed by \citet{knutson07,knutson09} in the infrared for HD 189733. The discovery illustrates that atmospheric phenomena, such as thermal winds or reflective clouds, can be probed not only by IR observations, but also by the high-precision visual-band light curves of {\it Kepler}. 

Next, Paper V \citep{faigler15} extends the discovery of superrotation in Kepler-76b to two additional known hot Jupiters, HAT-P-7b and KOI-13b. The paper presents the Lambert superrotation BEER model and shows that its planet-mass estimates are highly consistent with the planetary masses derived or constrained by RV observations. It then discusses and shows that close-in hot Jupiters are expected to show the most apparent beaming amplitude inconsistency. It is then not a surprise that this phenomenon was initially discovered in hot Jupiters, and is in agreement with the phase-shifted reflection/emission modulations of Kepler-76, HAT-P-7 and KOI-13 reported by this study. 
The paper concludes that detailed phase-curve studies of the precise light curve of {\it CoRoT}, {\it Kepler}, and future space missions, as this paper and alike \citep[e.g.,][]{esteves14}, open the opportunity to estimate the mass, and investigate the atmospheric properties of multiple close-in exoplanets.

To further establish the effectiveness of the BEER method, Paper VI \citep{tal-or15} describes the discovery and RV confirmation of {\it seventy non-eclipsing binaries} in the {\it CoRoT} data, using the AAOmega multi-object spectrograph \citep{smith04,saunders04}. Medium-resolution spectra of $281$ BEER candidates were obtained in a seven-night AAOmega RV campaign, with a precision of $\sim$$1$\,km/s. The measured RVs confirmed the binarity of seventy of the candidates, with periods of $0.3$--$10$\,days. The mass ratio of the confirmed binaries spanned a range of $0.03$--$1$. The paper also shows that red giants introduce a major source of false candidates, and suggests a method to improve BEER's performance in extracting higher-fidelity samples from future searches of {\it CoRoT} light curves.

The last study, Paper VII \citep{faigler15b}, demonstrates a different strength and utility of the BEER search algorithm. 
It presents the discovery of four short-period eclipsing dA+WD systems in the {\it Kepler} light curves.
The systems show BEER phase modulations together with primary and secondary eclipses.
These add to the 6 {\it Kepler}, and 18 WASP, previously known short-period eclipsing dA+WD binaries.
The paper shows that the new, and the previously known systems, are consistent with the mass-period relation expects for such binaries, that have gone through a mostly stable mass transfer from the WD progenitor to the current primary star.
It then shows that three of the new systems harbor the smallest WDs detected so far in short-period eclipsing dA+WD binaries. These  three binaries extend the previously known population to older systems with cooler and smaller WD secondaries.
In addition the paper points to the interesting KOI-3818 system that displays evidence for a fast-rotating primary and a minute but significant eccentricity. These features are probably the outcome of the mass-transfer process.

\section{Astrophysical contributions}
Within the scope of this thesis I can mark three main astrophysical contributions.

First, we have developed the BEER algorithm for detection of {\it non-eclipsing} short-period companions, through the beaming, ellipsoidal and reflection effects, in high-precision light curves.
The effectiveness of this new method was shown through the detection and RV confirmation of a large number of common stellar binaries (Paper III,VI). The algorithm also allowed the detection and confirmation of rare objects, the hot Jupiter Kepler-76b (Paper IV) and four eclipsing dA+WD binaries (Paper VII). 
On one hand, the discovery of many stellar binaries shows that the BEER method can serve as a high-throughput photometric detection tool of non-eclipsing binaries, with numbers comparable to those achieved by the eclipse method.  
On the other hand, the discovery of a planet, and four systems harboring WDs, 
illustrate that the new algorithm complements the known eclipse/transit method, and enables the discovery of important astrophysical systems missed by other methods. 

Next, we detected the superrotation phenomenon in the visual light curves of  Kepler-76, HAT-P-7 and KOI-13 (Paper IV, V). 
 This phenomenon, that involves a phase shift of the planetary thermal modulation, due to equatorial superrotating winds in the planet atmosphere \citep{showman02}, was previously observed only in the infrared \citep{knutson07,knutson09}.
These discoveries illustrate that detailed phase-curve studies of precise space-surveys light curves \citep[e.g., Paper V;][]{esteves14} open the opportunity to investigate atmospheric phenomena, such as thermal winds or reflective clouds, in multiple close-in exoplanets.

Finally, Using the BEER algorithm, we detected four short-period eclipsing dA+WD systems in the {\it Kepler} data,  three of which harbor the smallest WDs detected so far in such binaries (Paper VII).
The three binaries extend the previously known population of such systems, to older systems with cooler and smaller WD secondaries. The new discoveries allow comparing the small WDs properties to those predicted by binary evolution models, in a parameter region not observed before.

\section{Into the future}
In recent years, a few large photometric surveys were initiated, and more are planned for, motivated by various astrophysical research fields such as exoplanets, supernovae, gravitational-lensing and more.
Looking into the future, we can use the BEER algorithm to search the data of such large ground and space based photometric surveys. To illustrate the potential of this approach, Table \ref{tab:surveys} lists current and future large photometric surveys that are applicable for our purposes.

\begin{deluxetable}{lcccccccc}
\rotate
\tablecaption{Main properties of large photometric surveys}
\tablewidth{0pt}
\tablehead{
& \colhead{OGLE-IV$^1$}  & \colhead{WASP$^2$} &  \colhead{\it CoRoT$^3$} & \colhead{\it Kepler$^4$} & \colhead{GAIA$^5$} & \colhead{TESS$^6$} & \colhead{PLATO$^7$} & \colhead{LSST$^8$}
}
\startdata
Stars ($\times 10^6$) & $1000$ & $18$ & $0.17$  & $0.17$ & $1000$ & $0.2$ & $1000$ & $\sim$$40000$ \\
Cadence  & 1-3 d & $\sim$$10$ min. & 8.5 min.$^b$ & 30 min.$^b$& N/A & 2 min. & 25 sec.$^b$ & $\sim$4 d\\
Points/star & 300-700 & 6700 & 3000--25000$^b$ & 60000$^b$  & 70 & 19000$^c$& $2.5\times10^6$$^b$ & $825$ \\
Time span (days) & 1500 & 830 & 20--150 & 1400 & 1800 & 27 & 730 & 3650 \\
Precision$^a$ & 1\% & 1\% & 0.1\% & 0.01\% & 0.3\% & 0.02\%& 0.001\% & 1\% \\
Band (nanometer)  & 700--900 (I) & 400--700 & 450--900  & 400--900  & 330--1050 & 600--1000  & 500--1000 & 320--1050 (5 bands)\\
Magnitude range & 12--22 &7--15 & 11--16 & 7--17 & 6--20 & 4--13 & 4--16 & 16--27\\
Measurements & Ground & Ground & Space & Space & Space & Space & Space & Ground\\
Availability & at present & at present & at present & at present & 2016--22 & 2017--19 & 2024--29 & 2023--2033 \\
\tableline
\enddata
\tablenotetext{\space}{ $^1$  \cite{udalski15}}
\tablenotetext{\space}{ $^2$  \cite{butters10}}
\tablenotetext{\space}{ $^3$  \cite{auvergne09}}
\tablenotetext{\space}{ $^4$  \cite{koch10}}
\tablenotetext{\space}{ $^5$  \cite{lindegren10}}
\tablenotetext{\space}{ $^6$  \cite{ricker15}}
\tablenotetext{\space}{ $^7$  \cite{rauer14}}
\tablenotetext{\space}{ $^8$  \cite{lsst09}}
\tablenotetext{\space}{ $^a$  Typical precision per long-cadence data point.}
\tablenotetext{\space}{ $^b$ Typical for long-cadence light curves.}
\tablenotetext{\space}{ $^c$ Typical for 2 min. cadence {\it TESS} light curves.}
\label{tab:surveys}
\end{deluxetable}

Similarly to the {\it CoRoT} binary catalog constructed in Paper VI, we can search the data of each space survey and build a catalog of non-eclipsing stellar binaries, of up to {\it several million systems} for the largest surveys --- GAIA and PLATO. Such a uniform catalog, with well-defined observational biases, allows studying the binary population and its statistical characteristics, which in turn can serve as an important observational reference for binary formation and evolution theories.

In addition, the BEER method allows for searching the surveys for rare companions such as planets, brown dwarfs, white dwarfs, neutron stars and black holes. As demonstrated in Papers IV and VII, the combination of the BEER algorithm with the eclipse method can be used for finding transiting planets, brown dwarfs and white dwarfs, in the data of precise future surveys such as TESS and PLATO. 
Even more interesting will be to search for neutron stars and black holes. Such searches can be conducted on both ground and space photometry, as the ellipsoidal, and maybe even the beaming, caused by such massive companions, may be detectable even by ground surveys.

With the vast amount of astronomical data made publicly available in recent years, and the increased importance of virtual astronomy, robust and efficient search algorithms are critical for the astronomical community.
This thesis presents the BEER algorithm for searching the data of large photometric surveys, and illustrates its effectiveness in finding both common stellar binaries and rare astrophysical objects.
 I believe that the BEER tool can become an important component in the virtual astronomy toolbox for mining the vast astronomical data produced by current and future photometric surveys. 

\newpage

\singlespacing


\newpage
\thispagestyle{empty}
\

\newpage
\thispagestyle{empty}

\newpage

\doublespacing

\addcontentsline{toc}{chapter}{Acknowledgments}
\chapter*{Acknowledgments}
\vspace{-8mm}
First and foremost I wish to deeply thank my thesis advisor, Professor Tsevi Mazeh,
for his guidance and support in the course of this work.

During the years of my Ph.D. study I have benefited from the collaboration and 
assistance of many faculty members, postdocs, students and other astronomers (in alphabetical order): 
Suzanne Aigrain, Steven Bloemen, Lars Buchhave, Joshua Carter, Michael Engel, Avishay Gal-Yam, Tomer Holczer, Flavien Kiefer, Ilya Kull, Donald Kurtz, Dave Latham, Dan Maoz, Gil Nachmani, Ehud Nakar, Dovi Poznanski, Sahar Shahaf, Avi Shporer, Gil Sokol, Amiel Sternberg, Lev Tal-Or, Sam Quinn, Ignasi Ribas, William Welsh \& Shay Zucker.



The research leading to these results has received funding from the European Research Council under the EU's Seventh Framework Programme (FP7/(2007-2013)/ ERC Grant Agreement No.~291352), the ISRAEL SCIENCE FOUNDATION (grant No.~1423/11) and the Israeli Centers Of Research Excellence (I-CORE, grant No.~1829/12).

\pagestyle{plain}

\newpage
\
\pagestyle{empty}
\newpage




\setlength{\topmargin}{-2.4cm}
\begin{figure}[t]
\centerline{\psfig{figure=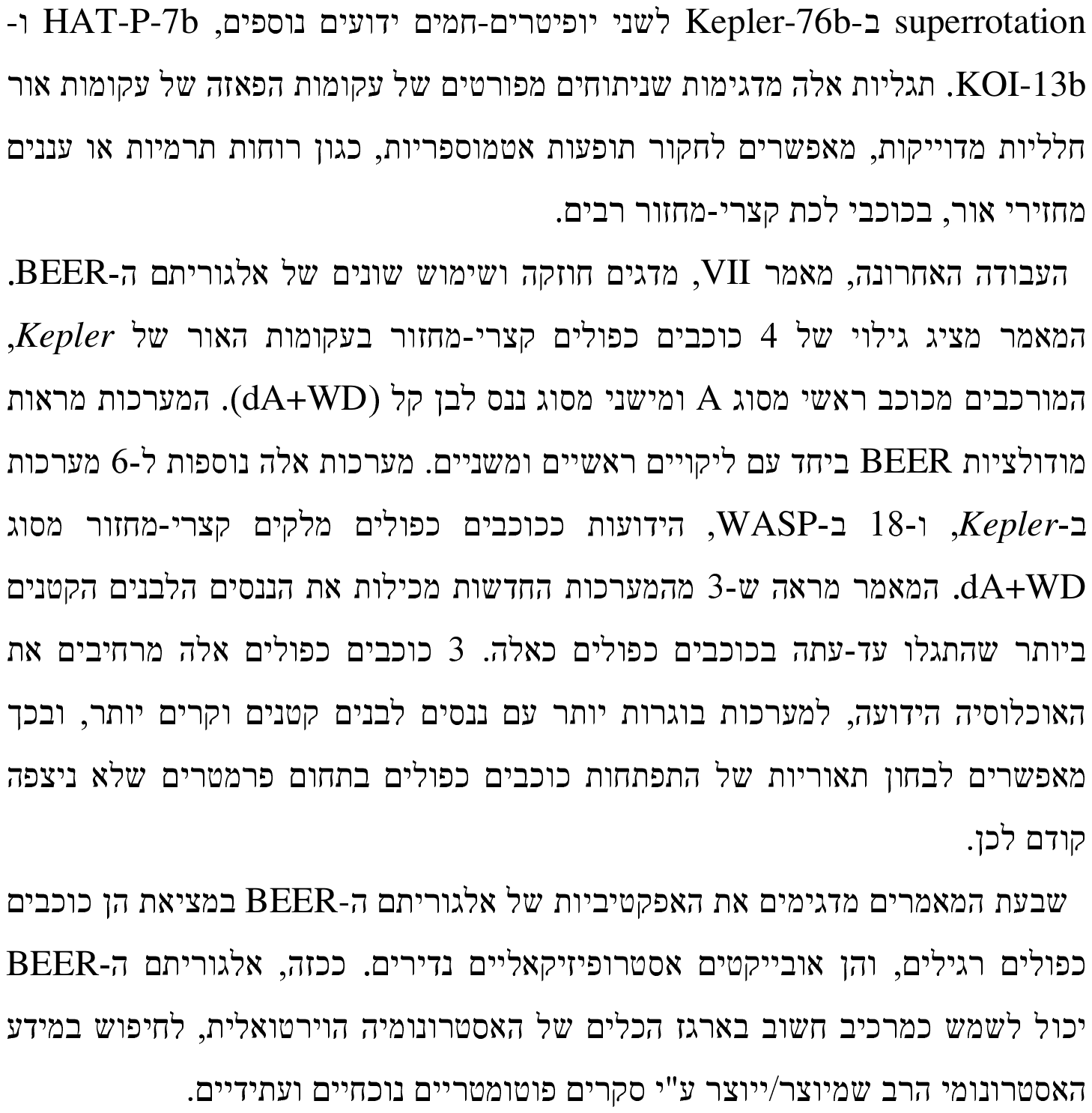}}
\end{figure}

\clearpage 

\setlength{\topmargin}{-4.5cm}
\begin{figure}[htb]
\centerline{\psfig{figure=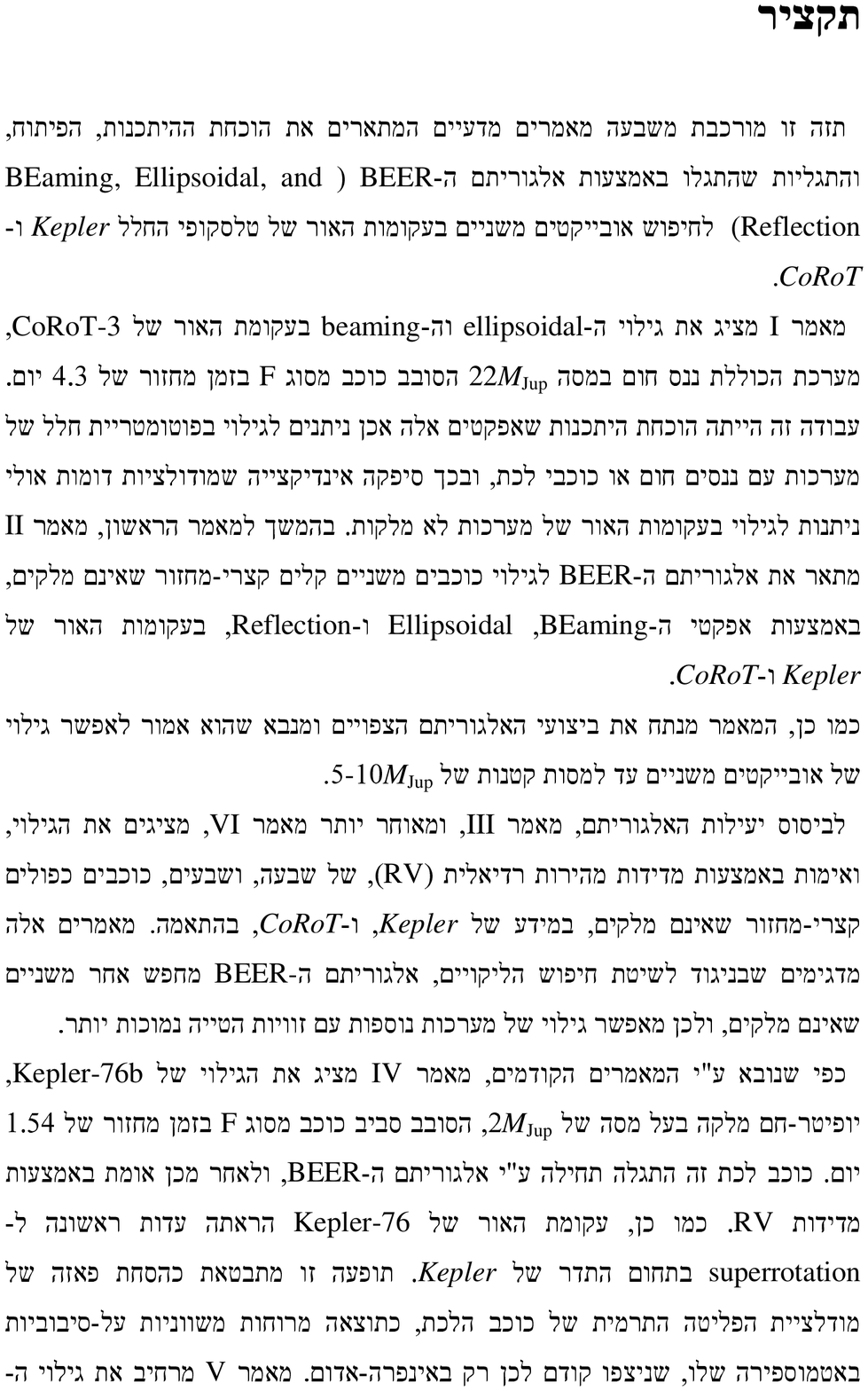}}
\end{figure}

\clearpage 

\newpage
\vline

\setlength{\topmargin}{-2.5cm}
\begin{figure}[htb]
\centerline{\psfig{figure=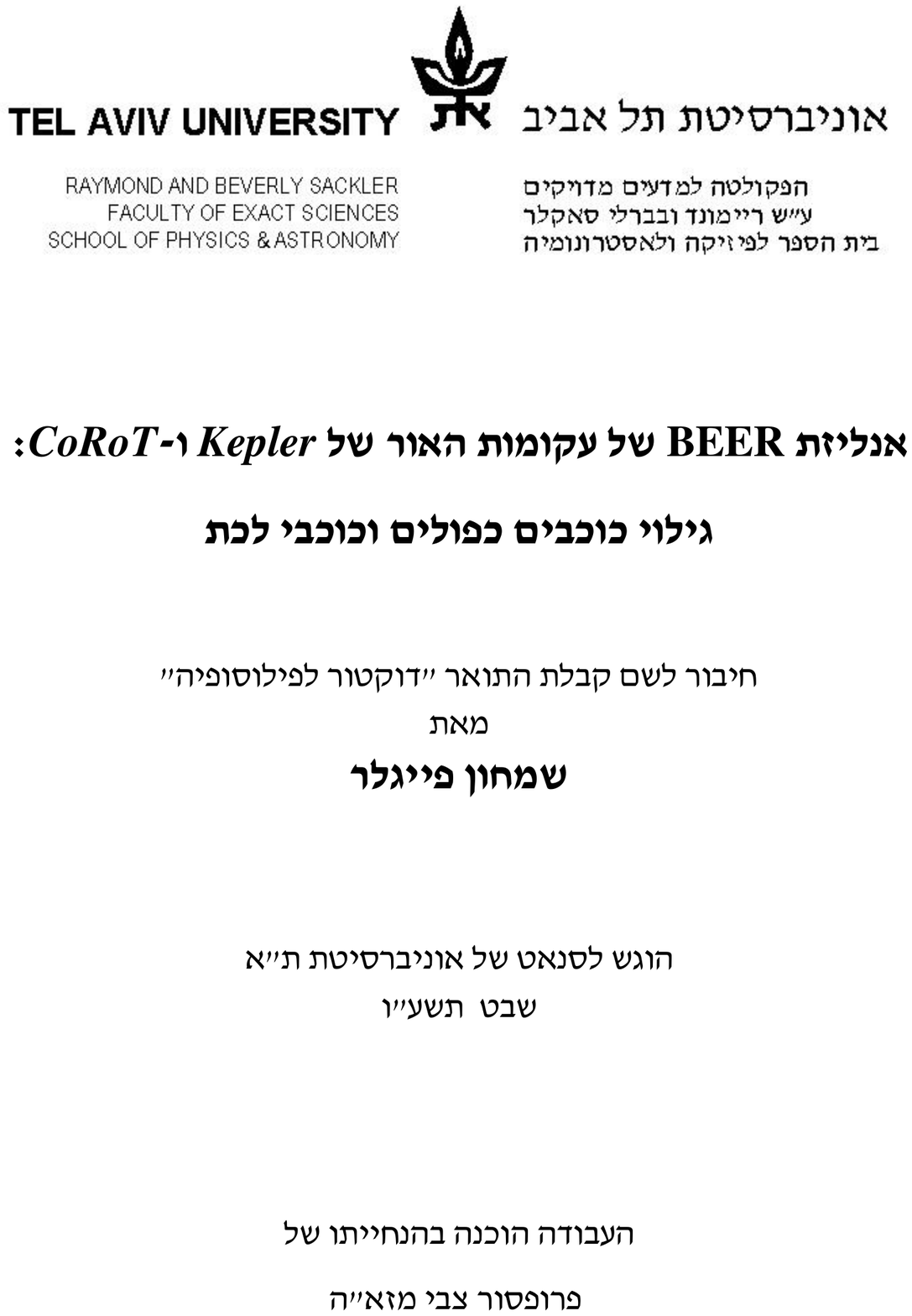,height=10in}}
\end{figure}

\end{document}